\documentclass[preprint]{elsarticle}

\usepackage{amsmath}
\usepackage{graphicx}
\usepackage[boxed]{algorithm}
\usepackage{algorithmic}
\graphicspath{{figs/}{figs-eps/}{figs-pdfnative/}}

\newcommand{\D}{\mathrm{d}}
\newcommand{\fp}[2]{\frac{\partial #1}{\partial #2}}
\newcommand{\fd}[2]{\frac{\mathrm{d} #1}{\mathrm{d} #2}}
\newcommand{\ud}{\underline{d}}
\newcommand{\dL}{d} 
\newcommand{\ctd}[1]{\frac{\tilde{\partial}#1}{\tilde{\partial}t}}
\newcommand{\csd}[1]{\frac{\tilde{\mathrm{d}}#1}{\tilde{\mathrm{d}}\sigma}}

\DeclareMathOperator{\diag}{diag}

\journal{Journal of Computational Physics}

\begin{document}


\begin{frontmatter}

    \title{A discrete geometric approach\\ for simulating the dynamics of thin viscous threads}
    
    \author[ijlrda]{B.~Audoly\corref{cor1}}
    \author[ijlrda,rutgers]{N.~Clauvelin}
    \author[ijlrda,fast]{P.-T. Brun}
    \author[col,adobe]{M.~Bergou}
    \author[col]{E.~Grinspun}
    \author[goet]{M.~Wardetzky}
    
    \cortext[cor1]{Corresponding author}
    
    \address[ijlrda]{Institut Jean Le Rond 
    d'Alembert, UMR 7190, UPMC Univ. Paris 06 and CNRS, F-75005 Paris, France}
    \address[rutgers]{BioMaPS Institute, Rutgers, the
    State University of New Jersey, Piscataway, NJ, USA}
    \address[fast]{Laboratoire FAST, UPMC Univ.~Paris 06, Universit\'e
    Paris-Sud and CNRS, B\^atiment 502, Campus Universitaire, Orsay
    91405, France}
    \address[col]{Computer Science, Columbia University, New York, NY, USA}
    \address[adobe]{Adobe Systems Incorporated}
    \address[goet]{Institute for Numerical and Applied Mathematics, University of G\"ottingen,
    37083 G\"ottingen, Germany}
    
    \begin{abstract}
	We present a numerical model for the dynamics of thin viscous
	threads based on a discrete, Lagrangian formulation of the
	smooth equations.  The model makes use of a condensed set of
	coordinates, called the centerline/spin representation: the
	kinematical constraints linking the centerline's tangent to
	the orientation of the material frame is used to eliminate two
	out of three degrees of freedom associated with rotations.
	Based on a description of twist inspired from discrete
	differential geometry and from variational principles, we
	build a full-fledged discrete viscous thread model, which
	includes in particular a discrete representation of the
	internal viscous stress.  Consistency of the discrete model
	with the classical, smooth equations is established formally
	in the limit of a vanishing discretization length.  The
	discrete models lends itself naturally to numerical
	implementation.  Our numerical method is validated against
	reference solutions for steady coiling.  The method makes it
	possible to simulate the unsteady behavior of thin viscous
	jets in a robust and efficient way, including the combined
	effects of inertia, stretching, bending, twisting, large
	rotations and surface tension.
    \end{abstract}
    

\end{frontmatter}

\section{Introduction}

\subsection{Context}

The flow of thin viscous filaments is relevant to a variety of
industrial processes such as the drawing and spinning of polymer and
glass
fibers~\cite{Forest-Zhou-Unsteady-analyses-of-thermal-2001,%
Pearson-Mechanics-of-polymer-processing-1985,%
Andreassen-Gunderson-EtAl-A-mathematical-model-for-the-melt-1997}, and
to natural phenomena such as formation of Pele's hair by lava ejected
at high speed by
volcanoes~\cite{Shimozuru-Physical-parameters-governing-1994}.  In
art, Jackson Pollock took advantage of the coiling instability of a
thin viscous fluid, the paint, impinging a surface, the canvas, to
produce a variety of decorative patterns by a fluid-mechanical
process~\cite{Herczynski-Cernuschi-EtAl-Painting-with-drops-2011}.  A
commonplace version of the same coiling instability is observed
when a thin thread of honey is poured on a morning's toast.  This
steady coiling problem is prototypical of the dynamics of thin threads.
Its apparent simplicity has made it appealing to fluid mechanicians
for a long time~\cite{Barnes-Woodcock-Liquid-rope-coil-effect-1958,%
Taylor-Instability-of-jets-threads-1968};
however the various regimes of steady coiling and its non-linear
features, such as multistability, have been understood in full details
only recently~\cite{Mahadevan-Ryu-EtAl-Fluid-rope-trick-1998,%
Ribe-Coiling-of-viscous-jets-2004,%
Ribe-Huppert-EtAl-Multiple-coexisting-states-2006,%
Ribe-Habibi-EtAl-Stability-of-liquid-rope-2006}.  To a large extent,
the analysis of steady coiling has been made possible by the
availability of numerical simulation: the shape of the thread in the
co-rotating frame is stationary, and is given by a non-linear
boundary-value problem~\cite{Ribe-Coiling-of-viscous-jets-2004} which
has been solved using numerical continuation.

In this paper we are interested in the simulation of the unsteady
behavior of thin threads, which is far less advanced.  As an
illustration, consider a recently proposed variant of the coiling
problem, similar to Pollock's painting technique, whereby the
target surface moves horizontally at a constant velocity.  The
relative motion suppresses steady coiling solutions and forces the
flow to become unsteady.
More than ten different patterns can be produced by varying the
lateral velocity of the surface and the fall
height~\cite{Chiu-Webster-Lister-The-fall-of-a-viscous-thread-2006,%
Morris-Dawes-EtAl-Meandering-instability-of-a-viscous-2008}, a number
of which have convoluted and intriguing shapes.  The patterns are
reminiscent of stitch patterns, and the experiment has been coined the
`fluid-mechanical sewing machine'.  This experiment nicely illustrates
the complex behavior that can result from the dynamics of a thin,
perfectly viscous filament.
%
%
Existing numerical methods are unable to reproduce this behavior, even
though the principle of the experiment is simple.  This highlights the
need for a robust and efficient method for simulating the dynamics of
thin viscous threads.

The dynamics of thin viscous threads is governed by the interplay of
three local modes of deformation, namely stretching, bending and
twisting modes~\cite{Trouton-On-the-coefficient-of-viscous-traction-1906,%
Buckmaster-Nachman-EtAl-The-buckling-and-stretching-of-a-viscida-1975}.
At the global scale, these modes are coupled by geometrically
non-linear terms accounting for finite rotations.  This coupling makes
the resulting dynamics remarkably rich.  Another, unfortunate
consequence of the nonlinearity is the absence of analytical solutions
to the dynamical equations.  This makes the development of robust
simulation methods even more desirable.  The main difficulty in
developing such a method is that the underlying equations are
numerically stiff, as the governing equations are non-linear partial
differential equations of fourth order in space.  This paper tackles
this difficulty by introducing a careful and well-controlled space
discretization.  In fact, we introduce a full-fledged
\emph{discrete} viscous thread model by extending all the relevant
physical quantities, such as strain rates and internal stress, to the
discrete setting.

Fluid mechanical problems involving free boundary conditions can be
simulated using refined variants of the marker and cells method,
namely the method
of~\cite{Hirt-Shannon-Free-surface-stress-conditions-1968,%
Nichols-Hirt-Improved-free-surface-1971} for 2d viscous flows, and the GENSMAC
method~\cite{Tome-McKee-GENSMAC:-A-Computational-Marker-1994,%
Tome-McKee-Numerical-simulation-of-visous-1999} for 3d viscous flows;
more recently implicit schemes coupled with projection methods have
been proposed, see
e.~g.~\cite{Oishi-Tome-EtAl-An-implicit-technique-for-solving-2008}.
The present paper is concerned with thin filaments, for which the
above methods are not efficient: when the thickness is small compared
to the longitudinal length scale, it is beneficial to use
dimensionally reduced equations as a starting point for simulations.
Thanks to dimensional reduction, the structure of the flow at small
scale is solved analytically, which makes it possible to use a
simulation grid much coarser than the thickness.

While our simulation method addresses the general non-steady dynamics
of a thin thread governed by the combined effects of twist, bending
and stretching forces, inertia and large rotations, a number of
particular cases have been simulated in the literature.  Steadily
rotating viscous threads are described by time-independent equations
in the co-rotating frame, which have been solved numerically using
methods for two-point
boundary value problems~\cite{Ribe-Coiling-of-viscous-jets-2004,%
Ribe-Habibi-EtAl-Stability-of-liquid-rope-2006,%
Arne-Marheineke-EtAl-Numerical-analysis-of-Cosserat-2010}.  The
dynamics of a viscous string, where both the bending and twisting
modes are neglected, has been
considered~\cite{Marheineke-Wegener-Asymptotic-model-for-the-dynamics-2009}.
The periodic folding of a viscous thread or sheet has been considered
in a 2d geometry~\cite{Skorobogatiy-Mahadevan-Folding-of-viscous-sheets-2000,%
Ribe-Periodic-folding-of-viscous-2003} where twist does not play any
role.  By combining the simulation of steady solutions with analytical
expansions describing oscillatory perturbations of small amplitude,
the stability of both the steady coiling
solution~\cite{Ribe-Habibi-EtAl-Stability-of-liquid-rope-2006} and of
the catenary-like profile of a dragged
thread~\cite{Blount-Lister-The-asymptotic-structure-of-a-slender-2009}
have been calculated.  Many other problems, such as the existence of
rotatory folding, the competition between folding and
coiling~\cite{Habibi-Rahmani-EtAl-Buckling-of-liquid-columns-2010},
the stitch patterns produced by the fluid-mechanical sewing
machine~\cite{Chiu-Webster-Lister-The-fall-of-a-viscous-thread-2006}
and the destabilization of steady coiling by
precession~\cite{Habibi-Moller-EtAl-Spontaneous-generation-of-spiral-2008}
remain inaccessible to those simulation methods that are based on
restrictive assumptions.

In comparison to viscous threads, elastic rods have received a lot of
attention, both from the perspective of analysis%
~\cite{MaddocDichma-Conservation-laws-in-the-dynamics-of-rods-1994,%
GorielyTabor-New-Amplitude-Equations-for-Thin-Elastic-Rods-1996,%
Goriely-Tabor-Nonlinear-dynamics-of-filaments-2-NonlinearAnalysis-1997,%
Clauvelin-Audoly-EtAl-Matched-asymptotic-expansions-2009} and
simulation~\cite{HouKlappeSi-Removing-the-Stiffness-of-Curvature-in-Computing-1998,%
Weiss-Dynamics-of-geometrically-nonlinear-rods2-2002,%
Goyal-Perkins-EtAl-Nonlinear-dynamics-and-loop-2005,%
Bergou-Wardetzky-EtAl-Discrete-Elastic-Rods-2008,%
Lim-Ferent-EtAl-Dynamics-of-a-Closed-Rod-with-2008,%
Jung-Leyendecker-EtAl-A-discrete-mechanics-approach-2010}.  By the
Rayleigh-Taylor
analogy~\cite{Taylor-Instability-of-jets-threads-1968}, the stress in
a viscous fluid is identical the stress in an elastic solid having the
same geometry, when the strain \emph{rate} relevant to the viscous
case is replaced with the \emph{current} strain relevant to the
elastic case.  Stated differently, the main difference between the
elastic and viscous problems is the presence of an additional time
derivative in the right-hand side of the viscous constitutive laws.
This analogy explains the buckling of viscous
sheets~\cite{Benjamin-Mullin-Buckling-instabilities-in-layers-1988,%
Silveira-Chaieb-EtAl-Rippling-instability-of-a-collapsing-2000}, a
phenomenon usually associated with elastic structures.  One can
take advantage of this analogy to simulate the dynamics
of viscous threads using a simulation tool written for elastic
rods~\cite{Radovitzky-Ortiz-Error-estimation-and-adaptative-1999}; we
explored this approach in a conference
paper~\cite{Bergou-Audoly-EtAl-Discrete-Viscous-Threads-2010}.  The
possibility to recycle an existing elastic code for viscous
simulations with minimal additional work is very attractive.
However, the initial effort associated with implementing and
validating an elastic code is high.  In situations where no
elastic code is available, it is simpler to implement a viscous
simulator directly.  In the present paper we follow this approach and
propose a numerical method that does not rely on an external library
for solving the dynamics of elastic rods.

While thin elastic rods are usually considered inextensible, the
stretching mode has to be retained in viscous threads, in addition to
the usual bending and twisting modes.  In both the elastic and viscous
cases, dimensional analysis shows that the strain, or strain rate,
associated with the stretching mode is small compared to that
associated with the bending and twisting modes, as the corresponding
modulus is larger by a factor proportional to the inverse of the small
aspect-ratio squared, a very large number.  A specificity of the
viscous case is that the stretching mode cannot be neglected, even
though its strain rate is small.  This is well illustrated by the
phenomenon of helical coiling: a thin thread poured from a container
onto a fixed obstacle gets stretched by gravity and remains straight
over almost the entire fall height, but it bends and twist severely in
a small boundary layer near the bottom.  Even though the stretching is
very mild, its effect is cumulated over the entire time of descent,
unlike the bending and twisting modes that come into play only near
the surface.  For this reason, the stretching mode has to be
considered in simulations of viscous threads.  By the
incompressibility condition, variations of the thread's radius along
its centerline need be considered as well.  Another difference with
the elastic case is that capillary forces can have a strong effect on
the motion, and must be taken into account.

\subsection{Model}

The derivation of dimensionally reduced models for thin viscous
filaments has a long history and is still a research topic.  The
equations for thin viscous threads were derived by asymptotic
expansion from the equations for a 3d viscous fluids by Entov and
Yarin~\cite{Entov-Yarin-The-dynamics-of-thin-liquid-1984}.  Their work
builds upon the previous analyses of viscous stretching by
Trouton~\cite{Trouton-On-the-coefficient-of-viscous-traction-1906},
and of viscous bending by Buckmaster and
co-workers~\cite{Buckmaster-The-buckling-of-thin-viscous-1973,%
Buckmaster-Nachman-The-Buckling-and-stretching-of-a-viscida-1978}.  
In
the case of elliptical cross-sections, the dynamics of the centerline
and the evolution of the geometry of the cross-section are coupled;
the corresponding equations have been derived by dimensional reduction
in 1d by Dewynne and
collaborators~\cite{Dewynne-Ockendon-EtAl-A-systematic-derivation-of-the-leading-order-1992},
and later extended to a capillary fluid
by~\cite{Cummings-Howell-On-the-evolution-of-non-axisymmetric-viscous-1999}.
Recent derivations of the equations for thin threads benefit from a
clear identification of the mechanical quantities in the 1d model
model~\cite{Panda-Marheineke-EtAl-Systematic-derivation-of-an-asymptotic-2008},
and of the systematic use of Lagrangian
coordinates~\cite{Arne-Marheineke-EtAl-Numerical-analysis-of-Cosserat-2010}.
Asymptotic models accounting for more general constitutive laws have
been proposed: the case of a visco-elastic fluid is treated
in~\cite{Bechtel-Forest-EtAl-A-one-dimensional-theory-for-viscoelastic-1986},
and a general framework is considered
in~\cite{Bechtel-Cao-EtAl-Practical-application-of-a-higher-1992}
which can produce a variety of asymptotic models when a specific set
of physical effects is considered.

Here we consider the dynamics of a thin filament of an incompressible,
purely viscous fluid having circular cross-section, under the action
of external forces such as gravity, and internal forces (viscous
stretching, bending, twisting, and capillary tension).  We consider
the 3d problem, and the curvature and kinematical twist of the thread
can be comparable to, or smaller than the inverse of the thread's
length.  Even though the fluid is very viscous, the effect of
inertia is considered.  The role of inertia is well illustrated by the
classical analysis of the pendulum modes of a viscous string, see
e.~g.~\cite{Ribe-Huppert-EtAl-Multiple-coexisting-states-2006}: in
this almost straight geometry, the flow in the axial direction is
typically governed by a small Reynolds number and dominated by
viscosity, although the flow in the transverse direction, which is
characterized by different length and time scales, is associated with a
much larger Reynolds number and dominated by inertia.  In the general
case, the local axial and transverse modes get coupled by the
curvature of the filament.  For this reason, we retain the inertial
term in the balance of momentum even though the constitutive laws are 
dominated by viscosity (Stokes' fluid).

We assume that the cross-section of the filament is a disk, and
remains so as the filament deforms.  Even when a viscous filament is
extruded from a non-circular opening, surface tension tends to
round off the shape of the cross-sections; this happens over a
time scale which we assume is short compared that of the flow.  This
is not always the case, and the possibility to account for
non-axisymmetric cross-sections has been demonstrated in the elastic
case using our numerical
method~\cite{Bergou-Wardetzky-EtAl-Discrete-Elastic-Rods-2008,%
Bergou-Audoly-EtAl-Discrete-Viscous-Threads-2010}.  While the general
case raises no fundamental difficulty, our presentation is limited to
the case of axisymmetric cross-sections, which is simpler: all the
directions in the cross-sections are then equivalent and there is no
need to keep track of their absolute orientations.  The case of tubes,
i.~e.\ of non simply-connected
cross-sections~\cite{Griffiths-Howell-Mathematical-modelling-of-non-axisymmetric-2008},
is not considered here but
this extension may be considered in future work as well.

\subsection{Proposed approach}

The main features of our numerical method are the following.  It is
based on a 1d model obtained by dimensional reduction, which makes it
much more efficient than a general-purpose model for 3d viscous flows.
We use a Lagrangian grid, making simulation vertices flow along with
the fluid; this simplifies the computation of viscous forces which, by
the constitutive law, are proportional to the comoving time derivative
of the kinematical twist and curvature.  We use a reduced set of
coordinates and eliminate two out of three degrees of freedom
associated with rotations, by making use of the fact that
cross-sections initially perpendicular to the centerline remain so
during motion: rotations are represented using a single degree of
freedom.  In addition, the absolute angle of twist of the
cross-sections is eliminated from the equations, using the fact that
the cross-sections are circular.  This results in an effective
description of rotations which only makes use of the instantaneous
angular twist velocity, denoted $v$.  Internal viscous forces include
the three physically relevant contributions of stretching, bending and
twisting.  The discrete expression of these forces is derived based on
a Rayleigh potential using variational methods.  This leads naturally
to discrete viscous forces.  The Rayleigh potential plays a similar
role in the viscous setting as the elastic energy in the elastic
setting.  The viscous forces are computed in a linear implicit manner,
which means that the expression for the viscous forces is extrapolated
to their value at the end of the time step, based on a linearization
carried out at the beginning of the time step.  This provides a good
compromise between stability and ease of implementation.  A fully
implicit evaluation of these forces is also possible, as discussed in
section~\ref{ssec:limitationsSemiImpliciteVsFullyImplicit}.

Our main contribution is to propose an elaborate space discretization
of a viscous thread by considering a fully discrete model.  All the
relevant physical quantities, such as the strain rates and the viscous
forces, are defined in the discrete setting.  In this discrete view,
bending is represented by the turning angles of the centerline at the
vertices, and we do not need to assume that the turning angles remain
small at all times.  As a consequence, the simulation of the discrete
model is stable even if the mesh size is comparable to the smallest
radius of curvature of the thread.  By contrast, stability of a
numerical scheme based on an \emph{ad hoc} discretization of the
smooth equations usual requires that the grid size is much smaller
than the smallest radius of curvature; in practice, this requirement
is severe as thin threads tend to spontaneously form strongly curved
region near the endpoints (such as the tiny rotating coils at the
bottom of a thin thread stretched by gravity and hitting a hard
surface).  The ability to run simulations with a relatively coarse
mesh is an important advantage of the discrete model, as the maximum
time step allowed by Nyquist stability criterion decreases rapidly
with mesh size in the presence of fourth-order space derivatives: the
simulation of coarser meshes is much more efficient.  Given the
numerical stiffness of the underlying problem, robustness is a central
issue in the simulation of viscous threads.  We address this issue by
keeping full control of the space discretization.

The discretization of bending in elastic rods is routinely done using
flexural springs at
hinges~\cite{Nguyen-Stability-and-nonlinear-solid-2000}, and the
extension to viscous bending is straightforward.  Enforcing the twist
forces is much less common.  Our discretization of twist is based on a
discrete notion of twist, which is directly borrowed from our previous
work on elastic
rods~\cite{Bergou-Wardetzky-EtAl-Discrete-Elastic-Rods-2008}.  This
discrete notion of twist is based on concepts from discrete
differential geometry, namely the holonomy of a discrete curve.  This
representation of twist builds upon previous work highlighting the
geometrical origin of
twist~\cite{Bishop-There-is-more-than-on-way-to-frame-1975,%
LangerSinger-Lagrangian-Aspects-of-the-Kirchhoff-Elastic-1996,%
GoldstPowersWiggin-Viscous-Nonlinear-Dynamics-of-Twist-1998}, and on
related numerical methods used in mechanical
engineering~\cite{Boyer-Primault-Finite-element-of-slender-2004}.

This paper is organized as follows.  In
section~\ref{sec:mathematicalToolbox} we derive useful identities of
geometry and differential calculus.  In section~\ref{sec:smooth} the
equations for thin viscous threads are formulated in a way that
prepares the extension to the discrete the setting, by making use of a
Lagrangian description of motion and by deriving the internal viscous
forces and moments from variational principles.
Section~\ref{sec:equivalenceWithKirchhoff} establishes the equivalence
with the formulation of the equations for thin threads classically
used by fluid mechanicians.  In section~\ref{sec:discreteModel} the
discrete model is presented in close analogy with the smooth case of
section~\ref{sec:smooth}; this section is at the core of the paper.
Section~\ref{sec:timeDiscrete} considers time discretization; formulae
that are required in the implementation are summarized, and we discuss
the treatment of boundary conditions and adaptive mesh refinement.
In section~\ref{sec:interactionWithEnvironment}, we consider the
coupling of the thread with external bodies, such as a fluid container
or hard surface.  In section~\ref{sec:validation}, the code is
validated against reference solutions for steady coiling.  In
section~\ref{sec:final}, we discuss limitations and perspectives.

In this paper, an effort is made to establish the equivalence of
different formulations.  The formulation of the equation of motion due
to Kirchhoff is popular among fluid mechanicians as it has a clear
intuitive meaning, while that based on the Rayleigh potential is
rarely if ever used in this community but lends itself to a natural
discretization.  Working out the connection between these equivalent
formulations will hopefully help to make this paper accessible to
different communities.  In addition, presenting the discrete model in
parallel with the smooth model provides mutual insights into them.
The reader should keep in mind that only a small fraction of the
formulae presented in the paper are required for the purpose of
implementing the method.  These formulae are recapitulated in
section~\ref{eq:summaryOfImplementation}.

\section{Mathematical toolbox}
\label{sec:mathematicalToolbox}
 
In this section we introduce some notations and
mathematical identities relevant to the mechanics of thin viscous
threads, such as infinitesimal rotations and the covariant derivative.

\subsection{Perpendicular projection}

We use underlines for vectors, and double underlines for rank-two
tensors (matrices).
For any unit vector $\underline{q}$ and for any vector
$\underline{a}$, the projection $\underline{P}_{\perp}$ in the
direction perpendicular to $\underline{q}$ is defined by
\begin{equation}
    \underline{P}_{\perp}(\underline{q},\underline{a}) = (\underline{\underline{1}}
    -
    \underline{q}\otimes \underline{q})\cdot \underline{a}
    =
    \underline{a} - (\underline{q}\cdot \underline{a})\,\underline{q}
	\textrm{,}
    \label{eq:PerpProjector}
\end{equation}
where the last term in the right-hand side is the longitudinal 
projection.

\subsection{Infinitesimal rotations}
\label{ssec:InfinitesimalRotations}

Consider an orthonormal frame $\ud_{i}(\sigma)$, $i=1,2,3$, which is
function of a continuous parameter $\sigma$: for any pair of integer
indices $1\leq i,j\leq 3$, and for every $\sigma$, we have
\begin{equation}
    \ud_{i}(\sigma)\cdot \ud_{j}(\sigma)
    = \delta_{ij}
	\textrm{,}
    \label{eq:diOrthonormal}
\end{equation}
where $\delta_{ij}$ is Kronecker's symbol, equal to $1$ if $i=j$ and
to $0$ otherwise.  We shall assume that the frame is $\mathcal{C}^1$
smooth with respect to the parameter $\sigma$.  

Infinitesimal rotations can be expressed by means of a skew-symmetric
matrix, or equivalently as the multiplication by a vector
$\underline{\Gamma}(\sigma)$ using the cross product.  More
accurately, for any value of the parameter $\sigma$, the Darboux vector
$\underline{\Gamma}(\sigma)$ is defined as the unique vector such that
for any $i=1,2,3$:
\begin{equation}
    \fd{\ud_{i}(\sigma)}{\sigma}
    =
    \underline{\Gamma}(\sigma)
    \times \ud_{i}(\sigma)
    \textrm{.}
    \label{eq:DarbouxGeneral}
\end{equation}
This definition will be used both when $\sigma$ is the time $t$, or
the arc length $S$.  This leads to the notions of instantaneous
angular velocity $\underline{\Gamma}(t) = \underline{\omega}$, or
twist-curvature vector $\underline{\Gamma}(S) = \underline{\pi}$,
respectively.

An explicit expression for the Darboux vector can be found by singling
out any particular vector $\ud_{i}$ in the triad, say $\ud_{3}$:
\begin{equation}
    \underline{\Gamma}(\sigma) = 
    \ud_{3}\times
    \fd{ \ud_{3}}{\sigma}
    + \Gamma_{3}\,\ud_{3}
	\textrm{,}
    \label{eq:DarbouxFromD3}
\end{equation}
where $\Gamma_{3} = \underline{\Gamma}\cdot \ud_{3}$ is 
defined by
\begin{equation}
    \Gamma_{3}(\sigma) = 
    \fd{\ud_{1}(\sigma)}{\sigma}
    \cdot
    \ud_{2}(\sigma)
    \textrm{.}
    \label{eq:Gammat}
\end{equation}
This can be checked by inserting equation~(\ref{eq:Gammat}) into
equation~(\ref{eq:DarbouxFromD3}) and then into
equation~(\ref{eq:DarbouxGeneral}).

\subsection{Covariant derivative}
\label{ssec:comovingDerivative}

Let us consider an orthonormal triad $\underline{d}_{i}(\sigma)$ and
the associated Darboux vector $\underline{\Gamma}(\sigma)$.  For any
vector field $\underline{a}(\sigma)$, we define the covariant
derivative as
\begin{equation}
    \csd{\underline{a}(\sigma)} =
    \fd{\underline{a}(\sigma)}{\sigma}
    - \underline{\Gamma}(\sigma) \times \underline{a}(\sigma)
    \textrm{.}
    \label{eq:ComovingTimeDerivative}
\end{equation}
It can be interpreted as the derivative measured in the frame moving
with the triad $\underline{d}_{i}$.  In agreement with this
interpretation, we note that equation~(\ref{eq:DarbouxGeneral}) can be
rewritten as
\begin{equation}
    \csd{\ud_{i}} = 0
    \textrm{.}
    \label{eq:cmtTIsZero}
\end{equation}

The covariant derivative satisfies the following identities, the proof
of which is left to the reader.  The general Leibniz rule for the
product of a scalar function $f(\sigma)$ by a vector
$\underline{a}(\sigma)$ reads
\begin{equation}
    \csd{(f(\sigma)\,\underline{a}(\sigma))} = 
    \fd{f(\sigma)}{\sigma}\,\underline{a}(\sigma)
    +
    f(\sigma)\,\csd{\underline{a}(\sigma)}
    \textrm{.}
    \label{eq:ctdLeibnitz}
\end{equation}
In the first term of the right-hand side above, the regular derivative
of a scalar function is used.  In the case of the scalar product of two
vectors we have
\begin{equation}
    \fd{(\underline{a}\cdot \underline{b})}{\sigma}
    =
    \csd{\underline{a}}\cdot \underline{b}
    +
    \underline{a}\cdot \csd{\underline{b}}
    \textrm{.}
    \label{eq:ctdDotProductExpansion}
\end{equation}

In the context of thin viscous threads, an important property of the
covariant derivative is its compatibility with the tangent and the
normal projections.  For any vector $\underline{a}$ we define the
tangent and normal projections, using the last triad vector
$\underline{d}_{3}$ as the tangent direction:
\begin{equation}
    \underline{a} = a_{3}\,\underline{d}_{3}+ 
    \underline{a}^\perp,
    \qquad
	\textrm{where }
    a_{3} = 
    \underline{a}\cdot \underline{d}_{3}
    \quad
	\textrm{and}
	\quad
    \underline{a}^\perp = 
    \underline{P}_\perp(\underline{d}_{3},\underline{a})
    \textrm{.}
    \label{eq:tangentTransverseNotations}
\end{equation}
Here $\underline{P}_\perp$ is the perpendicular projection operator
defined in equation~(\ref{eq:PerpProjector}).  A consequence of
equations~(\ref{eq:cmtTIsZero}--\ref{eq:ctdDotProductExpansion}) is
that the covariant derivative lives in the plane orthogonal to
$\underline{d}_{3}$ if $\underline{a}(\sigma)$ does so for all values
of $\sigma$, and is aligned with the tangent if
$\underline{a}(\sigma)$ is aligned with the tangent for all $\sigma$.
In other words, both the tangential and normal projections commute
with the covariant time derivative,
\begin{equation}
    \csd{\underline{a}}\cdot \underline{d}_{3}
    =
    \fd{a_{3}}{\sigma}
    ,\qquad
    \left(
    \csd{\underline{a}}
    \right)^\perp
    =
    \csd{(\underline{a}^\perp)}
    \textrm{.}
    \label{eq:ctdCommutesWithProjections}
\end{equation}
By contrast the regular derivative $\mathrm{d}/\mathrm{d}\sigma$ does
not commute with the projection operators.  Note that in the first
equation above, we use a regular derivative for the scalar quantity
$a_{3}$, hence the absence of a tilde.

\section{Smooth setting: a Lagrangian description of viscous threads}
\label{sec:smooth}

The equations for the dynamics of thin viscous
threads are usually expressed in Eulerian variables, see for instance
references~\cite{Ribe-Habibi-EtAl-Stability-of-liquid-rope-2006}.
In the present section we reformulate these equations in Lagrangian
variables.  This will make it possible to extend the geometrical
discretization of twist of Bergou et
al.~\cite{Bergou-Wardetzky-EtAl-Discrete-Elastic-Rods-2008}, and then
to derive a discrete model of viscous threads in a natural way.  A
Lagrangian formulation of the equations can be found in
reference~\cite{Arne-Marheineke-EtAl-Numerical-analysis-of-Cosserat-2010}
but we are not aware of any numerical method that actually makes use
of it; most numerical papers consider steady problems, for which an
Eulerian grid is sufficient.

Below, we introduce the smooth quantities that are relevant to the
kinematics and dynamics of viscous threads.  They are introduced in a
way that makes it straightforward to identify their discrete
counterparts later on.  Since the viscous thread can stretch, we make
a careful distinction between the arc length measured in reference
configuration, which is denoted $S$, and the arc length measured in
actual configuration, denoted $s$.

\subsection{Reference configuration}
\label{eq:referenceConfiguration}

Our Lagrangian description requires the definition of a reference
configuration.  This reference configuration can be imaginary, and
does not need to be the configuration of the thread at any particular
time.  A convenient choice is to define the reference configuration to
be an infinite, circular cylinder of constant radius $a_{0}$, as
illustrated in figure~\ref{fig:ThreadGeom}.  Obviously the equations
of motion will be independent of the choice of the reference
configuration, and of the value of the reference radius $a_{0}$ in 
particular.

The fluid is considered incompressible.  It therefore simpler,
although it is not required strictly speaking, to assume that the
mapping between the reference and actual configurations preserves
volume.  This is what we do in
equations~(\ref{eq:VolumeConservation}).

\subsection{Kinematics of centerline}
\label{ssec:kinematicsCenterline}

Let $S$ be the Lagrangian coordinate, and $t$ the time.  For any
function $f(S,t)$, we denote its spatial derivative using a prime,
\begin{equation}
    f'(S,t) = \fp{f(S,t)}{S}
    \textrm{,}
    \label{eq:SDerivative}
\end{equation}
and its time derivative using a dot,
\begin{equation}
    \dot{f}(S,t)
    = \fp{f(S,t)}{t}
    \textrm{.}
    \label{eq:ConvectiveDerivative}
\end{equation}
By definition, the Lagrangian coordinate $S$ follows fixed fluid
particles.  The time derivative introduced above is known as a
\emph{convective} derivative, and is often written $\dot{f} =
\fp{f}{t} = \frac{\mathrm{D}f}{\mathrm{D}t}$ in the Eulerian context.

\begin{figure}
    \centering
    \includegraphics{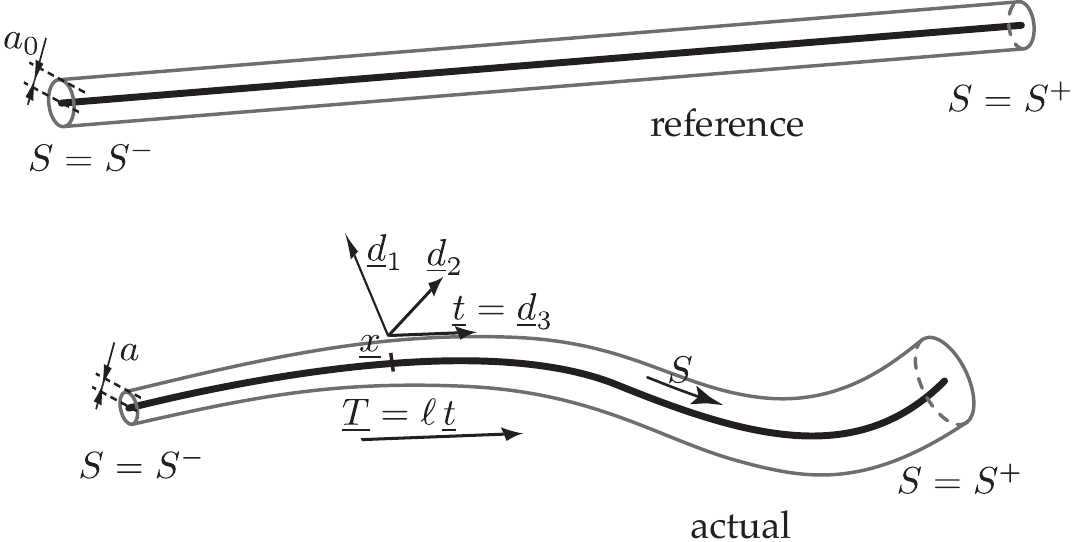}
    \caption{Reference and actual configuration of the thread.}
    \label{fig:ThreadGeom}
\end{figure}

At time $t$ the centerline of the thread is given by the function
$\underline{x}(S,t)$, see figure~\ref{fig:ThreadGeom}.
The material tangent of the thread is denoted $\underline{T}(S,t)$ and
defined by
\begin{equation}
    \underline{T}(S,t) = \underline{x}'(S,t)
    \textrm{.}
    \label{eq:MaterialTangent}
\end{equation}
Note that this is not a unit vector in general.  

The norm of $\underline{T}(S,t)$, denoted $\ell(S,t)$, measures the
amount of stretching of the centerline with respect to the reference
configuration:
\begin{equation}
    \ell(S,t) = |\underline{T}(S,t)| = \left|
    \fp{\underline{x}(S,t)}{S}
    \right|
    \textrm{.}
    \label{eq:ell}
\end{equation}
The unit tangent to the centerline is then defined as
\begin{equation}
    \underline{t}(S,t) = \frac{\underline{T}(S,t)}{\ell(S,t)}
    \textrm{.}
    \label{eq:UnitTangent}
\end{equation}

In our Lagrangian description, the arc length $s$ in actual
configuration is viewed as a secondary quantity.  It can be
reconstructed by integration of the differential equation expressing
the identity $\mathrm{d}s = |\mathrm{d}\underline{x}|$, namely
\begin{equation}
    s'(S,t) = \fp{s(S,t)}{S} = \ell(S,t)
    \textrm{.}
    \nonumber
\end{equation}

The Lagrangian axial strain rate ${\dL}$ is defined by
\begin{equation}
    {\dL}(S,t) = \fp{\ell(S,t)}{t}
    \textrm{,}
    \label{eq:AxialStrainRate}
\end{equation}
and measures the rate of stretching per unit time.  Note that this
quantity differs from the Eulerian strain rate, noted $d^{\mathrm{E}}$
and defined later in equation~(\ref{eq:EulerianStrainRate}).

In the Lagrangian framework the velocity $\underline{u}$ is simply the
time derivative of position,
\begin{equation}
    \underline{u}(S,t) = \fp{\underline{x}(S,t)}{t}
    \textrm{,}
    \label{eq:Velocity}
\end{equation}
and the acceleration its second time derivative,
\begin{equation}
    \underline{\gamma}(S,t) = \fp{\underline{u}(S,t)}{t} = 
    \fp{{}^2 \underline{x}(S,t)}{t^2}
    \textrm{.}
    \label{eq:Acceleration}
\end{equation}

A kinematical relation between the strain rate $d(S,t)$ and the
velocity $\underline{u}(S,t)$ can be established as follows:
\begin{equation}
    {\dL}(S,t) = \fp{\ell}{t} 
    = \frac{1}{2\,\ell}\,\fp{(\ell^2)}{t} 
    = 
    \frac{1}{2\,\ell}\,
    \fp{(\underline{T}^2)}{t}
    =
    \frac{1}{\ell}\,\underline{T}
    \cdot\fp{\underline{T}}{t}
    =
    \underline{t}(S,t)
    \cdot
    \fp{\underline{u}(S,t)}{S}
    \textrm{.}
    \label{eq:AxialStrainRateAsVelocityGradient}
\end{equation}
In the last equality, we have used the following identity, coming 
from the permutation of derivatives with respect of $t$ and $S$:
\begin{equation}
    \fp{\underline{T}}{t}
    =
    \fp{}{t}\left(
    \fp{\underline{x}}{S}
    \right)
    =
    \fp{}{S}\left(
    \fp{\underline{x}}{t}
    \right)
    =
    \fp{\underline{u}}{S}
    \textrm{.}
    \nonumber
\end{equation}

Another useful identity follows from inserting the definition of
$\ell$ in equation~(\ref{eq:UnitTangent}), $\underline{T} = \ell\,
\underline{t}$, into the left-hand side of the equation above.  Expanding
the time derivative, we find $\dot{\ell}\,\underline{t} +
\ell\,\dot{\underline{t}} = \underline{u}'(S)$.  Applying the
perpendicular projection operator
$\underline{P}_\perp(\underline{t},\cdot)$ on both sides and using
$\underline{P}_\perp(\underline{t},\underline{t}) = \underline{0}$ and
$\underline{P}_\perp(\underline{t},\dot{\underline{t}}) =
\dot{\underline{t}}$ (the other term cancels since $\underline{t}\cdot
\dot{\underline{t}} = \frac{1}{2}\fp{\underline{t}^2}{t} = 0$), we have
\begin{equation}
    \fp{\underline{t}(S,t)}{t} = 
    \frac{1}{\ell(S,t)}\,
    \underline{P}_\perp\left(
    \underline{t}(S,t),\fp{\underline{u}(S,t)}{S}
    \right)
    \textrm{.}
    \label{eq:tDotInTermsOfU}
\end{equation}
This equation yields the time derivative of unit tangent as a 
function of centerline geometry and of the  velocity $\underline{u}$.

\subsection{Incompressibility: radius and related quantities}
\label{ssec:incompressibility}

The viscous fluid is considered incompressible, as explained in
section~\ref{eq:referenceConfiguration}, and we assume that the
mapping from the reference configuration to the actual configuration
preserves volume.  The reference configuration is a cylinder with a
uniform radius $a_{0}$ and a cross sectional area $A_{0}=
\pi\,a_{0}^2$.  The radius of the thread in the actual configuration
is denoted $a(S,t)$, the area is $A(S,t)$, and $I(S,t)$ stands for the
moment of inertia.  We assume that the thread has locally a cylindrical
geometry, in which case
\begin{equation}
    A(S,t)=\pi\,a^2(S,t),
    \qquad
    I(S,t) = \frac{\pi\,a^4(S,t)}{4}
    \textrm{.}
    \label{eq:Area-MomentOfInertia}
\end{equation}
The fluid volume enclosed in an infinitesimal length of the thread is
$A(S,t)\,\mathrm{d}s = A(S,t)\,\ell(S,t)\,\mathrm{d}S$ in the actual
configuration, and $A_{0}(S,t)\,\mathrm{d}S$ in reference
configuration, see figure~\ref{fig:ThreadGeom}.
As a result, the incompressibility of the fluid is expressed by
\begin{equation}
    a(S,t) = \frac{a_{0}}{\sqrt{\ell(S,t)}},
    \quad
    A(S,t) = \frac{A_{0}}{\ell(S,t)},
    \quad
    I(S,t) = \frac{I_{0}}{\ell^2(S,t)}
    \label{eq:VolumeConservation}
\end{equation}
where the subscript naught refers to the reference configuration, for
which we have $\ell_{0} = 1$ by convention.  In particular, the moment
of inertia in the reference configuration reads $I_{0} =
\pi\,a_{0}^4/4$.

\subsection{Material frame, angular velocity}
\label{ssec:materialFrame}

\begin{figure}
    \centering
    \includegraphics{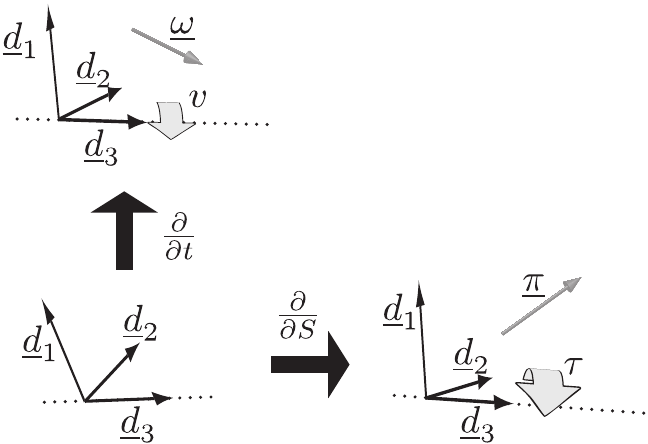}
    \caption{Darboux vectors (infinitesimal rotations) associated with
    spatial or temporal derivative of the material frame:
    instantaneous rotation velocity $\underline{\omega}$ and
    twist-curvature vector $\underline{\pi}$.  A compatibility
    condition for $\underline{\omega}$ and $\underline{\pi}$ is
    derived in section~\ref{sec:smoothCompatibilityOmegaPi}, 
    see equation~(\ref{eq:CompatibilityOmegaPi}).}
    \label{fig:SpaceAndTimeRotations}
\end{figure}

To complete the description of the motion, we need to keep track of
twist, defined as the rotation of the cross sections about the
tangent.  Indeed twist gives rise to viscous shear stress in the plane
of a cross-section, which affects the dynamics of the thread.  With
the aim to measure twist, we introduce a triad, denoted
$(\ud_{1}(S,t),\ud_{2}(S,t),\ud_{3}(S,t))$, which is rigidly attached
to the cross-sections.  This triad follows the motion of the
surrounding fluid particles and is called the material frame.

The flow inside the thread is shearless in the limit of very thin
thread, as explained for instance in the work of Ribe and
collaborators~\cite{Ribe-Habibi-EtAl-Stability-of-liquid-rope-2006}.
A similar argument holds in the case of elastic rods, see for instance
reference~\cite{Audoly-Pomeau-Elasticity-and-geometry:-from-2010}.  As
a result, material cross sections cannot slide with respect to another
and remain perpendicular to the local tangent to the centerline.  This
is known as Kirchhoff kinematical hypothesis, but it can be justified
rigorously by asymptotic analysis.  Therefore we impose the following
conditions: (\emph{i}) the triad satisfies the orthonormality
condition~(\ref{eq:diOrthonormal}), and (\emph{ii}) it is compatible
with the centerline in the sense that
\begin{equation}
    \ud_{3}(S,t) = \underline{t}(S,t)
    \textrm{.}
    \label{eq:TangentCompatibility}
\end{equation}
This kinematical condition is at the core of the mechanics of thin
threads; it couples the rotations of the material frame, measured
using the triad $\ud_{i}$ in the left-hand side of the equation, with
the motion of the centerline whose tangent appears in the right-hand
side.

The definition of the Darboux vector in
equation~(\ref{eq:DarbouxGeneral}) yields, after replacing the general
parameter $\sigma$ with time, $\sigma=t$,
\begin{equation}
    \fp{\ud_{i}(S,t)}{t} = 
    \underline{\omega}(S,t) \times \ud_{i}(S,t)
    \label{eq:DefMaterialRotationVelocity}
\end{equation}
for any value of the index $i=1,2,3$.  In the context of time
derivation, the Darboux vector $\underline{\Gamma}(t)$ is denoted
$\underline{\omega}(S,t)$ and is called the angular velocity vector;
its tangential component is called the axial spin velocity $\Gamma_{3}
= v(S,t)$,
\begin{equation}
    v(S,t) = \underline{\omega}(S,t)\cdot 
    \underline{t}(S,t)
    \textrm{.}
    \label{eq:DefOmegaT}
\end{equation}
By equation~(\ref{eq:Gammat}), $v$ is given by $v = \dot{\ud}_{1}\cdot
\ud_{2}$. An explicit expression for the angular velocity is
obtained from equation~(\ref{eq:DarbouxFromD3}):
\begin{equation}
    \underline{\omega}(S,t) = \underline{t}(S,t) \times
 	\dot{\underline{t}}(S,t)Ê+ 
    v(S,t)\,\underline{t}(S,t)
    \textrm{.}
    \label{eq:OmegaDecomposition}
\end{equation}

A second Darboux vector is obtained in the case of spatial
derivatives, inserting $\sigma = S$ in
equation~(\ref{eq:DarbouxGeneral}).  This vector is associated with
infinitesimal changes of the Lagrangian coordinate $S$, and is called
the twist-curvature vector.  It is denoted $\underline{\pi}(S,t)$ and
satisfies the fundamental relation~(\ref{eq:DarbouxGeneral}):
\begin{equation}
    \fp{\ud_{i}(S,t)}{S}  = 
    \underline{\pi}(S,t) \times \ud_{i}(S,t)
    \textrm{,}
    \label{eq:DefMaterialCurvature}
\end{equation}
for any value of the index $1\leq i\leq 3$.  Its tangential component
is called the (Lagrangian) kinematical twist and noted $\tau(S,t)$
instead of the generic notation $\Gamma_{3}$:
\begin{equation}
    \tau (S,t) = \underline{\pi}(S,t)\cdot \underline{t}(S,t)
    \textrm{.}
    \label{eq:DefMaterialTwist}
\end{equation}
This scalar $\tau(S,t)$ measures the rate of rotation of the material
frame about the tangent: $\tau(s,t) = \ud_{1}'\cdot \ud_{2}$.  It is
different from the Fr\'enet-Serret notion of torsion for a
three-dimensional curve which does not make any reference to the
material frame $\underline{d}_{i}$.  The Fr\'enet-Serret torsion
measures the non-planarity of a curve, although the kinematical twist
$\tau$ can be non-zero even though the centerline is planar, as
happens for instance in the case of a straight but twisted
configuration.

The normal projection of $\underline{\pi}$ is given by the general
expression of the Darboux vector in equation~(\ref{eq:DarbouxFromD3}),
combined with the condition of
compatibility~(\ref{eq:TangentCompatibility}),
\begin{equation}
    \underline{\pi}(S,t) = \underline{K}(S,t) + 
	\tau(S,t)\,\underline{t}(S,t)
    \textrm{,}
    \label{eq:BinormalLagrangianCurvatureAndMaterialCurvature}
\end{equation}
where we have introduced the Lagrangian binormal curvature
\begin{equation}
    \underline{K}(S,t) = 
    \underline{t}(S,t)
    \times \fp{\underline{t}(S,t)}{S}
    \textrm{.}
    \label{eq:LagrangianBinormal}
\end{equation}
Note $\underline{K}(S,t)$ depends only on the centerline, not on the
material frame.  Consistently with our Lagrangian approach, both the
kinematical twist $\tau$ and the binormal curvature $\underline{K}$
refer to a unit increment of the arc length $S$ in the reference
configuration, not to the arc length $s$ in the actual configuration.
As a result, these quantities differ from the usual twist and
curvature used in the Eulerian framework and some care is required
when comparing our equations with their Eulerian variants.
Equation~(\ref{eq:BinormalLagrangianCurvatureAndMaterialCurvature})
shows that the tangential component of $\underline{\pi}$ encodes the
twist while its normal projection encodes for the curvature, hence the
name twist-curvature vector.

The measure of strain relevant to the stretching mode is provided by
the extension $\ell(S,t)$ defined earlier in
equation~(\ref{eq:UnitTangent}).  This vector $\underline{\pi}(S,t)$
just introduced provides a measure of strain relevant to the two other
modes, twisting and bending.  For viscous threads, we need to evaluate
the \emph{rates of strain}.  This is the goal of the next sections: in
section~\ref{sec:smoothCompatibilityOmegaPi} we derive an identity for
the covariant time derivative of $\underline{\pi}(S,t)$, and in
section~\ref{ssec:LagrangianStrainRates} we identify this quantity as
the rate of strain, which is needed in the constitutive relations.

\subsection{Compatibility of rotations}
\label{sec:smoothCompatibilityOmegaPi}

The rotation velocity $\underline{\omega}$ and the twist-curvature
vector $\underline{\pi}$ characterize infinitesimal changes of the
material frame $\underline{d}_{i}(S,t)$ corresponding to increments of
time $t$ and of arc length $S$, respectively.  A compatibility
condition relating the vectors $\underline{\omega}'$ and
$\underline{\dot{\pi}}$ can be derived from the identity of the
cross-derivatives in Lagrangian coordinates $(t,S)$:

\begin{equation}
    \fp{}{t}
    \left(
    \fp{\ud_{i}(t,S)}{S}
    \right)
    =
    \fp{}{S}
    \left(
    \fp{\ud_{i}(t,S)}{t}
    \right)
    \textrm{.}
    \nonumber
\end{equation}
In this expression, we express the innermost derivatives with the
help of equations~(\ref{eq:DefMaterialRotationVelocity})
and~(\ref{eq:DefMaterialCurvature}) and write
\begin{equation}
    \fp{(\underline{\pi}\times \ud_{i})}{t}
    =
    \fp{(\underline{\omega}\times \ud_{i})}{S}
    \textrm{.}
    \nonumber
\end{equation}
We use equations~(\ref{eq:DefMaterialRotationVelocity})
and~(\ref{eq:DefMaterialCurvature}) again to expand the derivatives,
and find
\begin{equation}
    \left(
    \fp{\underline{\pi}}{t}
    -
    \fp{\underline{\omega}}{S}
    \right)\times \ud_{i}
    =
    - \underline{\pi} \times (\underline{\omega}\times \ud_{i})
    + \underline{\omega} \times (\underline{\pi}\times \ud_{i})
    \textrm{.}
    \nonumber
\end{equation}
The right-hand side can be simplified using Jacobi's identity, $
\underline{a}\times(\underline{b}\times \underline{c})+
\underline{b}\times(\underline{c}\times \underline{a})+
\underline{c}\times(\underline{a}\times \underline{b}) =
\underline{0}$, valid for any set of vectors $\underline{a}$,
$\underline{b}$, $\underline{c}$.  With $\underline{a} =
\underline{\pi}$, $\underline{b} = \underline{\omega}$ and
$\underline{c} = \underline{d}_{i}$, this yields
\begin{equation}
    \fp{\underline{\omega}}{S} = 
    \fp{\underline{\pi}}{t} - \underline{\omega}\times \underline{\pi}
    \nonumber
    \textrm{,}
\end{equation}
which is also known as the Maurer-Cartan identity.
Here we have simplified by $\underline{d}_{i}$, since the relations
hold for any value of the index $i=1,2,3$.

In the right-hand side we can
identify the covariant derivative defined in
section~\ref{ssec:comovingDerivative}, here with $\sigma=t$ and 
$\underline{t} = \underline{d}_{3}$. It is defined for an arbitrary 
vector field $\underline{a}(S,t)$ by
\begin{equation}
    \ctd{\underline{a}(S,t)} = \fp{\underline{a}(S,t)}{t}
    - \underline{\omega}(S,t) \times \underline{a}(S,t)
    \textrm{.}
    \label{eq:timeComovingDerivative}
\end{equation}
We can then rewrite the compatibility condition in a compact form,
\begin{equation}
    \fp{\underline{\omega}(S,t)}{S} = 
    \ctd{\underline{\pi}(S,t)}
    \textrm{.}
    \label{eq:CompatibilityOmegaPi}
\end{equation}

Note that the role of $\underline{\omega}$ and $\underline{\pi}$ is
symmetric and it is possible to rewrite this equation the other way
around,
\begin{equation}
    \fp{\underline{\pi}(S,t)}{S} = 
    \ctd{\underline{\omega}(S,t)}
    \textrm{.}
    \nonumber
\end{equation}
However, we shall only make use of
equation~(\ref{eq:CompatibilityOmegaPi}) in the following, as it
provides the relevant measure of rate of strain.

\subsection{Strain rate vector}
\label{ssec:LagrangianStrainRates}

The quantities appearing in equation~(\ref{eq:CompatibilityOmegaPi}) are
fundamental for the dynamics of viscous rods, and are related to the strain
rates of the different modes of deformation. The left-hand side is denoted
$\underline{e}$:
\begin{equation}
    \underline{e} = \fp{\underline{\omega}(S,t)}{S}
    \label{eq:RateOfStrainTwistCurvature}
    \textrm{.}
\end{equation}
The vector $\underline{e}$, called the \emph{strain rate vector},
provides the measure of the rates of deformation relevant to the twist
and curvature modes.  As such it appears in the right-hand sides of the
constitutive laws obtained by dimensional reduction, see
equations~(\ref{eq:constitutiveEqnsRaw-moment}).  In particular, when
the thread undergoes a rigid-body motion, $\underline{\omega}$ is
independent of $S$, the strain rate vector $\underline{e}$ vanishes
and so does the viscous stress, as expected.

Let us introduce the projections of $\underline{e}(S,t)$ in the
tangential and normal directions, denoted $e_{\mathrm{t}}$ and
$\underline{e}_{\mathrm{b}}$ respectively:
\begin{equation}
    \underline{e}(S,t) = e_{\mathrm{t}}(S,t)\,\underline{t}(S,t) + 
    \underline{e}_{\mathrm{b}}(S,t),
    \qquad
   \underline{e}_{\mathrm{b}}(S,t) \cdot  \underline{t}(S,t)  = 0
    \textrm{.}
    \label{eq:strainRateVectorDecomposition}
\end{equation}
The subscripts are motivated by the fact that the projections are
associated with the twisting and bending deformations, respectively,
as we show below.  The projections are defined by the following
explicit formulas,
\begin{subequations}
    \label{eq:strainRateVectorDecomposition-2}
    \begin{align}
    e_{\mathrm{t}}(S,t) & = \underline{e}(S,t)\cdot \underline{t}(S,t)
    \label{eq:strainRateVectorDecomposition-et} \\
    \underline{e}_{\mathrm{b}}(S,t) & = 
    \underline{P}_{\perp}\left(\underline{t}(S,t),\underline{e}(S,t)\right)
    \label{eq:strainRateVectorDecomposition-eb}
\end{align}
\end{subequations}
 
The covariant derivative in the right-hand side of the compatibility
condition~(\ref{eq:CompatibilityOmegaPi}) commutes with the projection
by equation~(\ref{eq:ctdCommutesWithProjections}).  As a result, the
decomposition of the twist-curvature vector $\underline{\pi}$ in
equation~(\ref{eq:BinormalLagrangianCurvatureAndMaterialCurvature})
implies
\begin{subequations}
    \label{eq:StrainRateVectorComponentsInTermsOfTauKappa}
    \begin{align}
        e_{\mathrm{t}}(S,t) & =\fp{\tau(S,t)}{t}
        \label{eq:StrainRateVectorComponentsInTermsOfTauKappa-T}\\
        \underline{e}_{\mathrm{b}}(S,t) & = \ctd{\underline{K}(S,t)}
        \label{eq:StrainRateVectorComponentsInTermsOfTauKappa-S}
	\textrm{.}
    \end{align}
Here $e_{\mathrm{t}}$ and $\underline{e}_{b}$ appear to be given by
the time-derivative of the strain measures --- the kinematical twist
$\tau$ and the curvature binormal $\underline{K}$.  These derivatives
are evaluated in a frame following the material (notice the presence
of the covariant derivative in the second equation, and recall that
the time derivative in the first equation is actually a convective
derivative as we use Lagrangian variables).
Equations~(\ref{eq:StrainRateVectorComponentsInTermsOfTauKappa})
confirms that $e_{\mathrm{t}}$ and $\underline{e}_{\mathrm{b}}$
measure the rates of change of the twist strain $\tau$ and of the
bending strain $\underline{K}$, respectively.  For the sake of
completeness, we recall the definition of the rate of strain
associated with the stretching mode of deformation, given earlier in
equation~(\ref{eq:AxialStrainRateAsVelocityGradient}):
\begin{equation}
    d(S,t) = \fp{\ell(S,t)}{t}
    \textrm{.}
    \label{eq:StrainRateVectorComponentsInTermsOfTauKappa-RecallD}
\end{equation}
\end{subequations}

The decomposition of $\underline{e}$ in
equation~(\ref{eq:strainRateVectorDecomposition}), and the
interpretations of its projections in
equations~(\ref{eq:StrainRateVectorComponentsInTermsOfTauKappa}) raise
some difficulties in the discrete setting.  Indeed, the definition of
$\underline{e}$ as the gradient of rotation, given in
equation~(\ref{eq:RateOfStrainTwistCurvature}), cannot be extended to
the discrete case, as we shall see.  However it is possible to propose
discrete equivalents to its tangent and normal projections
$e_{\mathrm{t}}$ and $\underline{e}_{\mathrm{b}}$ separately.

\subsection{A geometrical identity for the twist's rate of strain}

We derive in this section a geometric identity which is central to the
mechanics of thin elastic rods and viscous threads.  It explains the
coupling between the motion of the centerline $\underline{x}(S,t)$ and
the the twist $\tau(S,t)$.  In a previous
work~\cite{Bergou-Wardetzky-EtAl-Discrete-Elastic-Rods-2008} focusing
on the case of elastic rods, this equation was used to obtain a
natural discretization of the twist.  A similar discretization
strategy is followed here.

We start by projecting the compatibility
condition~(\ref{eq:CompatibilityOmegaPi}) along the tangent 
direction, using the product rule:
\begin{equation}
    e_{\mathrm{t}} = 
    \fp{\tau(S,t)}{t} 
    =
    \underline{t} \cdot \fp{\underline{\omega}}{S}
    =
    \fp{(\underline{t}\cdot \underline{\omega})}{S} - 
    \fp{\underline{t}}{S}\cdot \underline{\omega}
    \textrm{.}
    \nonumber
\end{equation}
By equation~(\ref{eq:OmegaDecomposition}), we can identify the
quantity $(\underline{t}\cdot \underline{\omega})$ appearing in the
right-hand side as the angular twist velocity $v$.  In the second
term, the time derivative of the tangent is given by combining
equations~(\ref{eq:DefMaterialCurvature})
and~(\ref{eq:TangentCompatibility}) as
\begin{equation}
    e_{\mathrm{t}} =  \fp{v(S,t)}{S}
    - (\underline{\pi}\times \underline{t})\cdot \underline{\omega}
    \textrm{.}
    \nonumber
\end{equation}
In the right-hand side, we can replace $\underline{\pi}$ by its
transverse projection $\underline{K}$ inside the cross product
$\underline{\pi}\times\underline{t}$.  We then permute the mixed
product and use $\underline{\omega}\times \underline{t} =
\dot{\underline{t}}$ by
equation~(\ref{eq:DefMaterialRotationVelocity}).  Then we find
\begin{equation}
    e_{\mathrm{t}} =  \fp{v(S,t)}{S} +
    \underline{K}(S,t)\cdot \fp{\underline{t}(S,t)}{t}
    \textrm{.}
    \label{eq:reconstructTwistFromOmegaT}
\end{equation}
We shall now comment a little on this equation which is important both at a
practical and at a fundamental level.

On a practical side, equation~(\ref{eq:reconstructTwistFromOmegaT})
makes it possible to use the centerline position $\underline{x}(S,t)$
and the spin velocity $v(S,t)$ as the primary variables for
parameterizing the thread.  When combined the definitions
$\underline{K} = \underline{t}\times \underline{t}'$ and
$\underline{t} = \underline{x}'$,
equation~(\ref{eq:reconstructTwistFromOmegaT}) provides the value of
the strain rate for the twisting mode, $e_{\mathrm{t}} = \dot{\tau}$,
a quantity that is required in the constitutive
law~(\ref{eq:constitutiveEqnsRaw-moment}).  This centerline/spin
parameterization has important benefits, as we shall argue later on.

More fundamentally, equation~(\ref{eq:reconstructTwistFromOmegaT})
explains the coupling between the centerline motion and the twisting
motion of the material frame: the twisting moment is proportional to
the rate of strain $e_{\mathrm{t}}$, which not only depends on the
rotational degree of freedom $v$ (through the first term in the
right-hand side) but also on the centerline motion (through the
second term).
The geometry underlying equation~(\ref{eq:reconstructTwistFromOmegaT}) has
been discussed by several authors but has never been used as a starting
point for setting up simulations of viscous threads. This equation can be
seen as an incremental version of the C\u{a}lug\u{a}reanu-White-F\"uller
(CWF) theorem~\cite{Calugareanu-Lintegrale-de-Gauss-et-lanlyse-1959,%
Pohl:The-Self-Linking-Number-o:1968,%
White-Self-Linking-and-the-Gauss-Integral-1969,%
Fuller-The-Writhing-Number-of-a-Space-1971,%
Fuller-Decomposition-of-the-linking-number-of-a-closed-ribbon:-1978} which
defines the notion of writhing for a closed curve --- for a short review on
this theorem, see
references~\cite{Aldinger:Formulae-for-the-calculat:1995,%
LangerSinger-Lagrangian-Aspects-of-the-Kirchhoff-Elastic-1996}. The CWF
theorem has become very well known in the context of supercoiled
DNA~\cite{Tanaka-Takahashi-Elastic-theory-of-supercoiled-1985,%
White:Calculation-of-the-twist-:1986,%
Tobias-Coleman-EtAl-The-dependence-of-DNA-tertiary-structure-1994,%
Klenin-Langowski-Computation-of-writhe-in-modeling-2000} or
polymers~\cite{Maggs:Writhing-geometry-at-fini:2001}, and has been used in
other contexts such as the dynamics of elastic filaments in a viscous
fluid~\cite{GoldstPowersWiggin-Viscous-Nonlinear-Dynamics-of-Twist-1998,%
Wolgemuth-Powers-EtAl-Twirling-and-Whirling:-Viscous-2000} or the
mechanics of
proteins~\cite{Levitt-Protein-folding-by-restrained-1983}. 
%
%
The binormal curvature $\underline{K}(S,t)$ in the right-hand side of
equation~(\ref{eq:reconstructTwistFromOmegaT}) above is directly
responsible for the geometrically non-linear term $\underline{T}\times
\underline{n}$ of the equations of motion, see
equation~(\ref{eq:KirchhoffEqnsLagrangian}).  There is no obvious way
to discretize this term of the equations of motion, but a
natural discretization will become apparent when this term is
connected to the geometric context of
equation~(\ref{eq:reconstructTwistFromOmegaT}), a variant of which
will appear in the discrete setting, see
equation~(\ref{eq:reconstructDiscreteTwistRate}) below.

\subsection{Centerline/spin representation}

The initial parameterization of the thread introduced in
sections~\ref{ssec:kinematicsCenterline} and~\ref{ssec:materialFrame}
is based on the centerline position $\underline{x}(S,t)$ and the
orthonormal frame $(\underline{d}_{i}(S,t))$ with $1\leq i\leq 3$.
This parameterization is subjected to the kinematical constraint of
compatibility expressed by equation~(\ref{eq:TangentCompatibility}).
From a numerical viewpoint it is desirable to eliminate this
constraint, and use a reduced set of variables instead.  To this end
we introduce the centerline/spin representation $(\underline{x},v)$:
the centerline is described using $\underline{x}(S,t)$ as earlier but
the material frame is described in an incremental way using the spin
velocity $v(S,t)$ introduced in equation~(\ref{eq:DefOmegaT}).  By
`incremental', we mean that we do not keep track of the absolute
direction of the transverse material vectors $\underline{d}_{1}$ and
$\underline{d}_{2}$ but only of the spin velocity $v =
\dot{\underline{d}}_{1}\cdot \underline{d}_{2}$.  Indeed, in the case
of isotropic cross-sections which we consider here, the absolute
direction of the material frame can be eliminated from the equations
of motion.  Our centerline/\emph{spin} representation is inspired from
the centerline/\emph{angle} representation introduced by Langer and
Singer in the context of elastic
rods~\cite{LangerSinger-Lagrangian-Aspects-of-the-Kirchhoff-Elastic-1996};
while they define the orientation of the cross-section incrementally
with respect to arc length $S$ using a twist angle $\theta$, we define
it incrementally with respect to time $t$ using a spinning velocity
$v$.  The benefit of our representation is that it makes the matrix
governing the dynamics of the thread sparse, as explained later.  The
initial $(\underline{x},\underline{d}_{i})$ representation uses 3
degrees of freedom, such as the Euler angles, for the orientation of
the material frame, which are subjected to 2 scalar constraints for
the compatibility.  By comparison, the centerline/spin representation
uses a single degree of freedom $v(S,t)$ for the description of the
twisting degree of freedom, which is free of kinematical constraint.
This has the important benefit of decreasing the number of degrees of
freedom and removing constraints.

In this section and in the next ones, we expose the principle of the
time-stepping algorithm in the centerline/spin representation.  This
algorithm is used at a particular time $t$ to derive positions and
velocities at the next time step.  In the absence of any ambiguity, we
shall often omit the time argument $t$ from now on.  The equations of
motion being second-order in time, we assume that the actual position,
as well as the linear and angular velocities as prescribed,
\begin{equation}
    \underline{x}(S) = \underline{x}(S,t),\quad
    \underline{u}(S) = \dot{\underline{x}}(S,t),\quad
    \underline{v}(S) = v(S,t)
    \textrm{.}
    \label{eq:SmoothRestrictions}
\end{equation}
We show how the linear and angular accelerations
$\underline{\gamma}(S) = \ddot{\underline{x}}(S)$ and $\dot{v}(S)$ can
be computed.  This requires reconstructing a number of intermediate
quantities.

To begin with, we can readily compute the axial strain $\ell(S)$, the
unit tangent $\underline{t}$ and the binormal curvature
$\underline{K}$ defined respectively by equations~(\ref{eq:ell}),
(\ref{eq:UnitTangent}), and~(\ref{eq:LagrangianBinormal}) 
in terms of the centerline, without even using the velocities:
\begin{subequations}
    \label{eq:smoothReconsructSpatialQuantities}
    \begin{align}
        \ell(S) & = 
	\left|
    \underline{x}'(S)
    \right|
        \label{eq:smoothReconsructSpatialQuantities-ell}\\
        \underline{t}(S) & =
	\frac{
    \underline{x}'(S)
    }{\ell(S)}
        \label{eq:smoothReconsructSpatialQuantities-unitTangent}\\
        \underline{K}(S) & =
	\underline{t}(S)\times \underline{t}'(S)
		\textrm{.}
        \label{eq:smoothReconsructSpatialQuantities-binormal}
    \end{align}
\end{subequations}

\subsection{Reconstruction of strain rates from velocities}

We now proceed to the reconstruction of the strain rates $d$,
$e_{\mathrm{t}}$ and $\underline{e}_{\mathrm{b}}$ defined in
equations~(\ref{eq:StrainRateVectorComponentsInTermsOfTauKappa}).  We
introduce a reconstruction scheme valid for arbitrary velocities,
denoted $\hat{\underline{u}}$ and $\hat{v}$, that are not necessarily
equal to the velocities $\underline{u}$ and $v$ of the actual motion.
These velocities $\hat{\underline{u}}$ and $\hat{v}$ will be called
virtual.  Working out the dependence of the strain rates on arbitrary
(virtual) velocities will enable us to put the constitutive laws in a
variational framework, using dissipation potentials.  This approach
will allow for a natural discretization of the constitutive laws.

We start by reconstructing the time derivative of the tangent,
$\dot{\underline{t}}$.  By equation~(\ref{eq:tDotInTermsOfU}) this
involves the operator $\underline{\mathcal{V}}$
\begin{subequations}
    \label{eq:VLinearFormSmooth}
\begin{equation}
    \underline{\mathcal{V}}(
    \underline{x};\hat{\underline{u}};S
    ) =
    \frac{1}{\ell(S)}\,
    \underline{P}_\perp\left(
    \underline{t}(S),
    \hat{\underline{u}}'(S)
    \right)
    \textrm{.}
    \label{eq:VLinearFormSmooth-Def}
\end{equation}
Note that $\underline{\mathcal{V}}$ depends on the \emph{functions}
$\underline{x}$ and $\hat{\underline{u}}$, and not just on their
values at $S$.  In the right-hand side, $\ell(S)$ and
$\underline{t}(S)$ are reconstructed from the centerline
$\underline{x}(S)$ passed as the first argument of
$\underline{\mathcal{V}}$ using
equations~(\ref{eq:smoothReconsructSpatialQuantities}).

In the particular case of the \textrm{real motion}, when the operator is
evaluated with the actual velocity $\hat{\underline{u}}(S) = \underline{u}(S)
=\dot{\underline{x}}(S,t)$ defined in
equation~(\ref{eq:SmoothRestrictions}), $\underline{\mathcal{V}}$
yields by construction
\begin{equation} 
    \underline{\mathcal{V}}(\underline{x};\underline{u};S) 
    =
    \fp{\underline{t}(S,t)}{t}
    \textrm{.}
    \label{eq:VLinearFormSmooth-Prop}
\end{equation}
\end{subequations}

Before considering the strain rates, we also need to reconstruct the
material rotation $\underline{\omega}$. For this purpose,
equation~(\ref{eq:OmegaDecomposition}) suggests introducing the operator
\begin{subequations}
    \label{eq:WLinearFormSmooth}
\begin{equation}
    \underline{\mathcal{W}}(\underline{x};
    \hat{\underline{u}},\hat{v};
    S) = 
    \hat{v}(S)\,\underline{t}(S)
    +
    \underline{t}(S)
    \times
    \underline{\mathcal{V}}(
    \underline{x};
    \hat{\underline{u}};
    S
    )
    \textrm{.}
    \label{eq:WLinearFormSmooth-Def}
\end{equation}
Here again, the tangent $\underline{t}$ is implicitly a function
of the first argument $\underline{x}$.  This operator 
$\underline{\mathcal{W}}$ yields the
material rotation in the case of a real motion $\hat{\underline{u}} 
= \underline{u}$:
\begin{equation}
    \underline{\mathcal{W}}(\underline{x};\underline{u},v;S) 
    =
    \underline{\omega}(S,t)
    \textrm{.}
    \label{eq:WLinearFormSmooth-Prop}
\end{equation}
\end{subequations}

The strain rates $d$, $\underline{e}_{\mathrm{t}}$ and
$\underline{e}_{\mathrm{b}}$ are reconstructed by means of three
operators, noted $\mathcal{L}_{\mathrm{s}}$, $\mathcal{L}^\mathrm{t}$
and $\underline{\mathcal{L}}^\mathrm{b}$ respectively.  The definition
of the operator associated with the stretching mode is motivated by 
equation~(\ref{eq:AxialStrainRateAsVelocityGradient}):
\begin{equation}
    \mathcal{L}_{\mathrm{s}}(\underline{x};
    \hat{\underline{u}};
    S)
    = \underline{t}(S) \cdot \hat{\underline{u}}'(S)
    \label{eq:SmoothFormsAxialStrain-Def}
	\textrm{.}
\end{equation}
When evaluated with a real motion it yields the strain rate,
\begin{equation}
    \mathcal{L}_{\mathrm{s}}(\underline{x};\underline{u};S)
    =
    d(S,t)
    \textrm{.}
    \label{eq:SmoothFormsAxialStrain-RealMotion}
\end{equation}

The twisting and bending strains are introduced through a decomposition
equivalent to the one used in
equation~(\ref{eq:strainRateVectorDecomposition}):
\begin{equation}
    \fd{\underline{\mathcal{W}}(
    \underline{x};
    \hat{\underline{u}},\hat{v};
    S
    )}{S}
    =
    \mathcal{L}^\mathrm{t}(\underline{x};
    \hat{\underline{u}},\hat{v};
    S)\,
    \underline{t}(S)
    +
    \underline{\mathcal{L}}^\mathrm{b}(\underline{x};
    \hat{\underline{u}},\hat{v};
    S)
    \textrm{,}
    \label{eq:GradientOfRotationReconstruction}
\end{equation}
where the right-hand sides are defined by the following projections:
\begin{subequations}
    \label{eq:LtbByProjection}
    \begin{align}
	\mathcal{L}^\mathrm{t}( 
	\underline{x}; \hat{\underline{u}},\hat{v}; S
	) 
	& =
	\underline{t}(S)
	\cdot
	\fd{\underline{\mathcal{W}}(
	\underline{x}; \hat{\underline{u}},\hat{v}; S
	)}{S} 
	\label{eq:LtbByProjection-t}\\
	\underline{\mathcal{L}}^\mathrm{b}( 
	\underline{x}; \hat{\underline{u}},\hat{v}; S
	) 
	& =
	\underline{P}_{\perp}\left(
	\underline{t}(S),
	\fd{\underline{\mathcal{W}}(
	\underline{x}; \hat{\underline{u}},\hat{v}; S
	)}{S}
	\right)
	\textrm{.}
        \label{eq:LtbByProjection-s}
    \end{align}
\end{subequations}
Note that the total derivative of $\underline{\mathcal{W}}$ in
equation~(\ref{eq:GradientOfRotationReconstruction})
and~(\ref{eq:LtbByProjection}) produce terms proportional to
$\hat{\underline{u}}'(S)$ and $\hat{v}'(S)$ according to the
definition of $\mathcal{W}$ in
equation~(\ref{eq:WLinearFormSmooth-Def}).  As a result, the operators
$\mathcal{L}^\mathrm{t}$ and $\underline{\mathcal{L}}^\mathrm{b}$ also
depend on the local values of $\hat{\underline{u}}'$ and $\hat{v}'$.

For a real motion, $\underline{\mathcal{W}}$ evaluates to
$\underline{\omega}$ by equation~(\ref{eq:WLinearFormSmooth-Prop}),
and so the left-hand side of equation~(\ref{eq:GradientOfRotationReconstruction}) evaluates to the strain rate
vector $\underline{e}$.  Comparison of the
decompositions~(\ref{eq:strainRateVectorDecomposition})
and~(\ref{eq:LtbByProjection}) shows that, by construction, the operators
operators $\underline{\mathcal{L}}^\mathrm{t}$ and
$\underline{\mathcal{L}}^\mathrm{b}$ yield in the case of a real motion
\begin{subequations}
    \label{eq:Lt+bLinearFormSmooth-RealMotion}
    \begin{align}
	\mathcal{L}^\mathrm{t}(\underline{x};\underline{u},v;S)
	& =
	e_{\mathrm{t}}(S,t)
	\label{eq:Lt+bLinearFormSmooth-RealMotion-T}\\
	\underline{\mathcal{L}}^\mathrm{b}(\underline{x};\underline{u},v;S)
	&=
	\underline{e}_{\mathrm{b}}(S,t)
	\textrm{.}
        \label{eq:Lt+bLinearFormSmooth-RealMotion-B}
    \end{align}
\end{subequations}

A more explicit form of the operator $\mathcal{L}^\mathrm{t}$ defined 
in equation~(\ref{eq:LtbByProjection-t}) can be found by inserting 
the definition of $\underline{\mathcal{W}}$ into 
equation~(\ref{eq:WLinearFormSmooth-Def}). After some algebra, we find
\begin{subequations}
    \label{eq:LtLinearFormSmooth}
\begin{equation}
    \mathcal{L}^\mathrm{t}(\underline{x};
    \hat{\underline{u}},\hat{v};
    S)=
    \hat{v}'(S) + 
    \underline{K}(S)
    \cdot
    \underline{\mathcal{V}}(
    \underline{x};\hat{\underline{u}};S
    )
    \textrm{,}
    \label{eq:LtLinearFormSmooth-Def}
\end{equation}
where $\underline{K}$ is the curvature binormal computed from the
centerline configuration $\underline{x}$ using
equation~(\ref{eq:smoothReconsructSpatialQuantities-binormal}). This 
is an extension of equation~(\ref{eq:reconstructTwistFromOmegaT}) to 
virtual velocities.
\end{subequations}

All the operators $\underline{\mathcal{V}}$,
$\underline{\mathcal{W}}$, $\mathcal{L}_{\mathrm{s}}$,
$\mathcal{L}^\mathrm{t}$, $\underline{\mathcal{L}}^{\mathrm{b}}$
introduced above depend linearly on the virtual velocities
$\hat{\underline{u}}$ and $\hat{v}$.  Their discrete equivalents shall
allow us to calculate the viscous stress.

\subsection{Dissipation potentials}

In the Lagrangian framework, internal viscous stress can be described
by a Rayleigh potential, see for instance
reference~\cite{Torby-Advanced-Dynamics-for-Engineers-1984}.  This
potential, which plays a role similar to the potential energy for
elastic rods, expresses the power dissipated by viscosity during a
virtual motion prescribed by the velocities $\underline{\hat{u}}(S)$
and $\hat{v}(S)$.  This potential has three contributions,
corresponding to the stretching, twisting and bending modes of
deformation:
\begin{equation}
    \mathcal{D} (
    \underline{x};
    \hat{\underline{u}},\hat{v}
    ) 
    =
    \mathcal{D}_{\mathrm{s}} (\underline{x};\underline{\hat{u}})
    + \mathcal{D}_{\mathrm{t}} (\underline{x};\underline{\hat{u}},\hat{v})
    + \mathcal{D}_{\mathrm{b}} (\underline{x};\underline{\hat{u}},\hat{v})
    \textrm{.}
    \label{eq:SplitDissipation}
\end{equation}

The stretching contribution is proportional to the stretching
modulus whose expression is due to
Trouton~\cite{Trouton-On-the-coefficient-of-viscous-traction-1906}:
\begin{equation}
    D(\ell) = 3\,\mu\, A(\ell) = 
    \frac{D_{0}}{\ell^2}
    \textrm{,}
    \label{eq:StretchingModulus}
\end{equation}
where $\mu$ is the fluid's dynamic viscosity.  Here $D_{0} =
3\mu\,A_{0} = 3\mu\,(\pi\,a_{0}^2)$ is the value of the stretching
modulus in reference configuration.  The twist modulus $C$ and the
bending modulus $B$ read~\cite{Yarin-Free-Liquid-Jets-1993,Ribe-Coiling-of-viscous-jets-2004}
\begin{equation}
    C(\ell) = 2\mu\,I(\ell) = \frac{C_{0}}{\ell^2}
    ,\qquad
    B(\ell) = 3\mu\,I(\ell) = \frac{B_{0}}{\ell^2}
    \textrm{,}
    \label{eq:BendingAndTwistModuli}
\end{equation}
where $C_{0} = 2\mu\,I_{0}$ and $B_{0} = 3\,\mu\,I_{0}$ are the moduli
in reference configurations, and the moment of inertia in reference
configuration $I_{0}$ is defined in
section~\ref{ssec:incompressibility}.

We propose the following expressions for the stretching, twisting and
bending contributions to the Rayleigh potential:
\begin{subequations}
    \label{eq:smoothDissipationInTermsOfLForms}
    \begin{align}
	\label{eq:Dissipation-Stretch}
	\mathcal{D}_{\mathrm{s}}(
	\underline{x};
	\underline{\hat{u}}
	)
	& = \int_{S_-}^{S^+} \frac{D(\ell(S))}{2\,\ell(S)}
	\,\Big(
	 \mathcal{L}_{\mathrm{s}}(\underline{x};\underline{\hat{u}};S)
	\Big)^2\,\mathrm{d}S
	\textrm{,}
	\\
	\label{eq:Dissipation-Twist}
	\mathcal{D}_{\mathrm{t}}(
	\underline{x};
	\underline{\hat{u}},\hat{v}
	)
	& =
	\int_{S_-}^{S^+} 
	\frac{C(\ell(S))}{2\,\ell(S)}\,
	\Big(
	 \mathcal{L}^{\mathrm{t}}(
	 \underline{x};
	 \underline{\hat{u}},\hat{v};S)
	\Big)^2\,\mathrm{d}S
	\textrm{,}
	\\
	\label{eq:Dissipation-Bending}
	\mathcal{D}_{\mathrm{b}}(
	\underline{x};
	\underline{\hat{u}},\hat{v}
	)
	& =
	\int_{S_-}^{S^+} 
	\frac{B(\ell(S))}{2\,\ell(S)}\,
	\Big(
	 \underline{\mathcal{L}}^{\mathrm{b}}(
	 \underline{x};
	 \underline{\hat{u}},\hat{v};S)
	\Big)^2\,\mathrm{d}S
	\textrm{.}
    \end{align}
\end{subequations}
Here $S^-$ and $S^+$ denotes the Lagrangian coordinates of the
endpoints of the thread.  Both $S^-$ and $S^+$ may depend on time
even though this time dependence is implicit for the sake of
readability.  In all the expressions above, $\ell$ is a function of
the first argument $\underline{x}$ by
equation~(\ref{eq:smoothReconsructSpatialQuantities}).  Note that the
stretching contribution $\mathcal{D}_{\mathrm{s}}$ does not depend on
the rotational degree of freedom $\hat{v}$ but solely on the
centerline velocity $\underline{\hat{u}}$.  Since
$\mathcal{L}_{\mathrm{s}}$, $\mathcal{L}^\mathrm{t}$ and
$\underline{\mathcal{L}}^\mathrm{b}$ are linear forms, all
contributions $\mathcal{D}_{\mathrm{s}}$, $\mathcal{D}_{\mathrm{b}}$
and $\mathcal{D}_{\mathrm{t}}$ and the total Rayleigh potential
$\mathcal{D}$ are quadratic forms of their velocity arguments
$\underline{\hat{u}}$ and $\hat{v}$.  This quadratic dependence
reflects the linear character of the viscous constitutive laws.

The expressions introduced in
equations~(\ref{eq:smoothDissipationInTermsOfLForms}) for
$\mathcal{D}_{\mathrm{s}}$, $\mathcal{D}_{\mathrm{t}}$ and
$\mathcal{D}_{\mathrm{b}}$ will be justified \emph{a posteriori} in
section~\ref{sec:equivalenceOfConstitutiveLaws}, by checking that they
lead to the correct constitutive laws.  In particular, we will explain
the reason for the factor $1/\ell$ in the integrands above.

The Rayleigh potential $\mathcal{D}$ allows the equations of motion
for a thin viscous thread to be put in variational form.  As noticed
by Batty and Bridson~\cite{Batty-Bridson-Accurate-Viscous-Free-2008}
in the context of 3D fluids with free boundaries, this variational
setting provides a natural discretization of these equations.  We
follow this general approach, and start by exposing the variational
structure of the smooth equations.

\subsection{Equations of motion}

The main property of the Rayleigh potential is that is gives the
viscous force by derivation with respect to the virtual velocity: the
resultant of the internal viscous stress on the centerline is given by
\begin{subequations}
    \label{eq:smoothFunctionalDerivativeOfRayleighPotential}
    \begin{equation}
        \underline{P}_{\mathrm{v}}(S,t) = 
	-\left.
	\frac{\partial \mathcal{D}(\underline{x};
	\underline{\hat{u}},\hat{v})}{\partial 
	\underline{\hat{u}}(S)}
	\right|_{(\underline{\hat{u}},\hat{v}) = (\underline{u},v)}
        \label{eq:smoothFunctionalDerivativeOfRayleighPotential-resultant}
    \end{equation}
    The notation in the right-hand side must be understood as follows:
    we first take the functional derivative of the potential with
    respect to its argument $\underline{\hat{u}}$, and later
    substitute the velocity arguments with their real values,
    $\underline{\hat{u}} = \underline{u}$ and $\hat{v}=v$.  This
    quantity $\underline{P}_{\mathrm{v}}$ is the resultant of the
    internal viscous forces, per unit length $\mathrm{d}S$ in
    reference configuration.  It includes the
    stretching, twisting and bending forces, each contribution being
    listed in equation~(\ref{eq:SplitDissipation}).  
    
    The stress conjugated to the spin velocity $v$ is the twisting 
    moment
    due to the internal viscous stress in the thread.  It is given by
    a similar formula,
    \begin{equation}
        Q_{\mathrm{v}}(S,t) = 
	-\left.
	\frac{\partial \mathcal{D}(\underline{x};
	\underline{\hat{u}},\hat{v})}{\partial \hat{v}(S)}
	\right|_{(\underline{\hat{u}},\hat{v}) = (\underline{u},v)}
        \label{eq:smoothFunctionalDerivativeOfRayleighPotential-twist}
    \end{equation}
\end{subequations}

In equations~(\ref{eq:smoothFunctionalDerivativeOfRayleighPotential})
the functional derivatives in the right-hand sides are calculated
practically by computing the first variation $\D\mathcal{D}$ of the
Rayleigh potential for small increments of the virtual velocities,
denoted $\delta \underline{u}$ and $\delta v$, and by identifying the
result with
\begin{equation}
    -\D\mathcal{D}(\underline{x};
     \underline{u},v;
    \delta \underline{u}, \delta v) =
    \int_{S_-}^{S^+}
    (\underline{P}_{\mathrm{v}}\,\delta \underline{u}
    + Q_{\mathrm{v}}\,\delta v)
    \;\mathrm{d}S
    \textrm{.}
    \label{eq:RayleighPotentialFirstVariation}
\end{equation}
This calculation will be carried out later in
section~\ref{sec:netViscousForceAndTwistingMoment}, when we work out
explicit expressions for the net viscous force
$\underline{P}_{\mathrm{v}}$ and moment $Q_{\mathrm{v}}$, and check
that they are equivalent with the classical Kirchhoff equations.

In terms of the net viscous force $\underline{P}_{\mathrm{v}}$ and 
twisting moment $Q_{\mathrm{v}}$, the balance of
linear and angular momentum can be written
\begin{subequations}
    \label{eq:equationsOfMotionStrong}
    \begin{align}
        \rho\,A_{0}\,\ddot{\underline{x}}(S,t) & = 
	\underline{P}_{\mathrm{v}}(S,t) + \underline{P}(S,t)
	\label{eq:equationsOfMotion-Linear}
	\\
	\ell\,J\,\dot{v}(S,t) & = Q_{\mathrm{v}}(S,t) + Q(S,t)
	\textrm{.}
	\label{eq:equationsOfMotion-Angular}
    \end{align}
\end{subequations}
Here $\underline{P}(S,t)$ is the density of external force and
$Q(S,t)$ the density of external twisting moment.  These balance laws
are for an infinitesimal segment, and per unit length $\mathrm{d}S$ in
reference configuration.  The factors $(\rho\,A_{0})$ and $(\ell\,J)$
are its mass and moment of inertia about the tangent, measured per
unit reference length $\mathrm{d}S$.  Here, $J$ is the moment of
inertia per unit length $\mathrm{d}s$ in \emph{actual} configuration,
given by the usual formula
\begin{equation}
    J(\ell) =
    \iint_{|r|<a} r^2\,r\,\mathrm{d}r\,\mathrm{d}\theta = 
    2\rho\,I(\ell)
    \textrm{.}
    \label{eq:momentOfInertia}
\end{equation}
The factor $\ell$ in the left-hand side of 
equation~(\ref{eq:equationsOfMotion-Angular}) is because an element 
of reference length $\mathrm{d}S$ has a moment of inertia 
$J\,\mathrm{d}s = \ell\,J\,\mathrm{d}S$.

By multiplying both terms of the first
equation~(\ref{eq:equationsOfMotion-Linear}) by a vector-valued test
function $\delta\underline{u}(S)$ and both terms of the second
equation~(\ref{eq:equationsOfMotion-Angular}) by a scalar-valued test
function $\delta v(S)$, we can write the equations of
motion~(\ref{eq:equationsOfMotionStrong}) in weak form.  The motion is
such that, for any choice of the functions $\delta\underline{u}(S)$
and $\delta v(S)$ and at any time $t$,
\begin{equation}
    \int_{S_-}^{S^+}
    (
    \rho\,A_{0}\,\dot{\underline{u}}\cdot\delta \underline{u}
    +
    J\,\dot{v}\,\delta v
    )\,\mathrm{d}S = 
    -\D\mathcal{D}(\underline{x};\underline{u},v;
    \delta\underline{u},\delta v)
    + \int_{S_-}^{S^+}
    (
    \underline{P}\cdot \delta \underline{u}
    +
    Q\,\delta v
    )\,\mathrm{d}S
    \label{eq:equationsOfMotionWeak}
\end{equation}
where $\underline{u}(S,t) = \dot{\underline{x}}(S,t)$ is the actual
velocity.  In continuum mechanics, the left-hand side is called the
virtual work of acceleration, the first term in the right-hand side is
the internal virtual work, and the last term is the external virtual
work.  This weak form of the equations of motion will be useful for
going to the discrete case.

\subsection{External loading}

In equations~(\ref{eq:equationsOfMotionStrong}), $\underline{P}$ and
$Q$ denote the the force resultant and the twisting moment due to
external forces --- as opposed to the internal, viscous forces in the
thread.  Those forces are given per unit length $\mathrm{d}S$ in
reference configuration.  In our validation examples, we consider a
thread moving under the action of gravity and surface tension.  Contact
with the ground is not handled by applying forces but instead by
freezing the motion of particles, as explained later in
section~\ref{ssec:descriptionOfFloor}: we do not need an explicit 
expression for these contact forces.

Gravity is represented by a force
\begin{equation}
    \underline{P}(S,t) = \underline{P}_{\mathrm{g}}(S,t) = 
    \rho\,A_{0}\,\underline{g},\qquad
    Q(S,t) = Q_{\mathrm{g}}(S,t) = 0
    \textrm{.}
    \label{eq:gravityForce}
\end{equation}

We now work out the expression of the effective forces acting on the
centerline as a result of surface tension on the lateral boundaries of
the thread.  Those forces are derived by variation from an energy
proportional to the lateral area of these boundaries.  In the case of
a slender thread which we consider, the capillary energy is written
using the following approximation of the lateral area:
\begin{equation}
    \mathcal{E}_{\gamma}(\underline{x}) = \int_{S_-}^{S^+} 
    \gamma\,2\pi\,a(\ell(S))\,\ell(S)
    \;\mathrm{d}S
    \textrm{,}
    \label{eq:smoothCapillaryEnergy}
\end{equation}
where $\gamma$ is the surface tension, possibly depending on time and
position along centerline, and $2\pi\,a(\ell)\,(\ell\,\mathrm{d}S)$ is
the lateral area of a cylinder of radius $a$ and length $\mathrm{d}s =
\ell\,\mathrm{d}S$.  We neglect the small conical angle of the lateral
surface, which has a negligible influence on the capillary forces for a
thin thread.  

The effective capillary force acting on the centerline can be obtained
by variation, using the definition of $a(\ell)$ in
equation~(\ref{eq:VolumeConservation}) and the definition of $\ell$ in
terms of $\underline{x}(S)$ in equation~(\ref{eq:ell}).  This yields
\begin{subequations}
    \label{eq:smoothSurfaceTension}
\begin{equation}
    \D\mathcal{E}(\underline{x};\delta \underline{x})
    =
    \Big[\underline{n}_{\gamma}\,\delta \underline{x}\Big]_{S^-}^{S^+}
    -\int_{S_-}^{S^+}  \underline{P}_{\gamma}(S)\cdot\delta \underline{x}\,\mathrm{d}S
    \mathrm{,}
    \label{eq:smoothSurfaceTension-ForceDistribution}
\end{equation}
where the bracket denotes the boundary term coming from the 
integration by parts, and
\begin{align}
    \underline{P}_{\gamma}(S,t) & = 
    \fp{\underline{n}_{\gamma}(S,t)}{S}
    \\
    Q_{\gamma}(S,t) &= 0 \\
    \underline{n}_{\gamma}(S,t) & = 
    \gamma\,\pi\,a(S,t)\,\underline{t}(S,t)
    \textrm{.}
\end{align}
\end{subequations}

\subsection{Neglecting rotational inertia}
\label{ssec:smoothQuasiStatic}

For thin elastic rods or viscous threads, a classical approximation,
proposed by Kirchhoff himself, is to neglect the rotational inertia,
that is to set $J=0$ in the equation of
motion~(\ref{eq:equationsOfMotion-Angular}).  This approximation can
be justified by the fact that the kinetic energy associated with
rotational inertia scales like $(\ell\,J)\,v^2\sim
\ell\,\rho\,a^4\,(1/t^*)^2$ for a motion happening on a typical
time-scale $t^*$.  By contrast the kinetic energy associated with
translation of the centerline scales like
$\ell\,\rho\,A\,\underline{u}^2\sim \ell\,\rho\,a^2\,(L/t^*)^2$, where
$L$ is the typical length-scale of the motion.  The energy of the
rotational mode is therefore negligible for slender threads, for which
$L\gg a$.  Rotational inertia is always dominated by translational
inertia --- except for a straight viscous thread moving in pure twist,
a particular case where the kinetic energy in translation is exactly
zero and the above argument is inapplicable.  Therefore, we set $J=0$
in the following.  The longitudinal balance of angular
momentum~(\ref{eq:equationsOfMotion-Angular}) becomes a condition for
the quasi-static equilibrium of the twisting mode:
\begin{equation}
    0 = Q_{\mathrm{v}}(S,t) + Q(S,t)
    \textrm{.}
    \label{eq:smoothQuasiStaticTwist}
\end{equation}
This equation expresses the fact that the typical time associated with
the damping of twist waves, which is much shorter than that associated
with the damping of bending waves by the above scaling argument, is
considered to be shorter than the time step of the simulation.

\section{Equivalence with Kirchhoff equations for a thin viscous thread}
\label{sec:equivalenceWithKirchhoff}

This section aims at demonstrating the equivalence of the Lagrangian
description of threads based on the centerline/spin representation
$(\underline{x},v)$ exposed in the previous section, and the
classical Kirchhoff equations for thin viscous threads.  The goal is
to bridge the gap with classical formulations, and to identify
important stress variables that underlie the dynamic equations
derived in the previous section.  No new ingredient required in the
numerical model will be introduced, and the reader interested only in
the implementation can skip ahead to the derivation of the discrete
model in section~\ref{sec:discreteModel}.

\subsection{Constitutive laws underlying the dissipation 
potential}
\label{sec:underlyingConstitutiveRelations}

To establish the connection between our formalism and the Kirchhoff
equations, we shall start by calculating the force
$\underline{P}_{\mathrm{v}}$ and twisting moment $Q_{\mathrm{v}}$
arising from viscous stress.  By
equation~(\ref{eq:RayleighPotentialFirstVariation}), this requires
working out the first variation of the dissipation potential with
respect to the velocities $\underline{\hat{u}}$ and $\hat{v}$, near
the real motion $\underline{\hat{u}} = \underline{u}$ and $\hat{v} =
v$.  Combining equations~(\ref{eq:SplitDissipation})
and~(\ref{eq:smoothDissipationInTermsOfLForms}), we can write this
first variation as
\begin{multline}
    -\D\mathcal{D}( \underline{x}; \underline{u},v;
    \delta\underline{u},\delta v ) 
    =
    \\
    -\int_{S_-}^{S^+}\Bigg(
    \frac{D\,\mathcal{L}_{\mathrm{s}}}{\ell}
    \,
    \D\mathcal{L}_{\mathrm{s}}(\delta\underline{u})
    +
    \frac{C\,\mathcal{L}^{\mathrm{t}}}{\ell}
    \,
    \D\mathcal{L}^{\mathrm{t}}(\delta\underline{u},\delta v)
    + 
    \frac{B\,\underline{\mathcal{L}}^{\mathrm{b}}}{\ell}
    \cdot
    \D\underline{\mathcal{L}}^{\mathrm{b}}(\delta \underline{u},\delta v)
    \Bigg)\,\mathrm{d}S
    \textrm{,}
    \nonumber
\end{multline}
where we have temporarily omitted the arguments $\underline{x}$,
$\underline{u}$, $v$ and $S$ in the integrand for better readability.
Since the latter are the real velocities, we can make use of
equations~(\ref{eq:SmoothFormsAxialStrain-RealMotion})
and~(\ref{eq:Lt+bLinearFormSmooth-RealMotion}), and write
\begin{multline}
    -\D\mathcal{D}( \underline{x}; \underline{u},v;
    \delta\underline{u},\delta v ) 
    =
    \\
    -\int_{S_-}^{S^+}\Bigg(
    \frac{D\,d}{\ell}
    \,
    \D\mathcal{L}_{\mathrm{s}}(\delta\underline{u})
    +
    \frac{C\,e_{\mathrm{t}}}{\ell}
    \,
    \D\mathcal{L}^{\mathrm{t}}(\delta\underline{u},\delta v)
    + 
    \frac{B\,\underline{e}_{\mathrm{b}}}{\ell}
    \cdot
    \D\underline{\mathcal{L}}^{\mathrm{b}}(\delta \underline{u},\delta v)
    \Bigg)\,\mathrm{d}S
    \textrm{,}
    \label{eq:FirstVariationOfDissipationPotential-step1}
\end{multline}
where the omitted arguments $\underline{x}$, $\underline{u}$ and $v$
again refer to the real motion.

Let us denote $n_{\mathrm{s}}$ the first coefficient appearing in the
integrand, and assemble the two other coefficients into a vector
denoted $\underline{m}$:
\begin{subequations}
    \label{eq:constitutiveEqnsRaw}
\begin{align}
    n_{\mathrm{s}}(S,t) & = 
    \frac{1}{\ell}\,(D\,d)
    \label{eq:constitutiveEqnsRaw-tension} \\
    \underline{m}(S,t) & = 
    \frac{1}{\ell}\,(
    C\,e_{\mathrm{t}} \,\underline{t}+ B\,\underline{e}_{\mathrm{b}}
    )
    \textrm{.}
    \label{eq:constitutiveEqnsRaw-moment}
\end{align}
\end{subequations}
Note that these coefficients do not depend on the virtual motion but
only on the actual one.  
The quantities introduced in equation~(\ref{eq:constitutiveEqnsRaw})
will be identified as the viscous stress in the thread.  More
accurately, $n_{\mathrm{s}}$ is the scalar tension resisting
stretching and $\underline{m}$ the internal moment arising from 
twisting
and bending.

The \emph{vector} tension will also be useful later on,
\begin{equation}
    \underline{n}_{\mathrm{s}}(S,t) = n_{\mathrm{s}}(S,t)\,
    \underline{t}(S,t)
    \label{tensionForce} 
     \textrm{.}
\end{equation}
It can be interpreted as the internal force that arises in response 
to stretching deformations.

\subsection{Equivalence with Eulerian constitutive laws}
\label{sec:equivalenceOfConstitutiveLaws}

The constitutive laws can be rewritten in Eulerian form, which is how
they are classically presented in the literature.  From
equations~(\ref{eq:constitutiveEqnsRaw}), we have
\begin{subequations}
    \label{eq:constitutiveEqnsEulerian}
\begin{align}
    n_{\mathrm{s}}(S,t) & = D\,d^\mathrm{E}
    \label{eq:constitutiveEqnsEulerian-tension} \\
    \underline{m}(S,t) & = 
    \big[
    C\,(\underline{t}\otimes 
	\underline{t})
	+
	B\,(\underline{\underline{1}} - \underline{t}\otimes 
	\underline{t})
    \big]\cdot
    \underline{e}^\mathrm{E}
    \textrm{,}
    \label{eq:constitutiveEqnsEulerian-moment}
\end{align}
where we have introduced the Eulerian strain rate,
\begin{equation}
    d^\mathrm{E} = \frac{d}{\ell},\qquad
    \underline{e}^\mathrm{E} = \frac{1}{\ell}\,\underline{e} = 
    \frac{1}{\ell}\,
    \fp{\underline{\omega}}{S}
    =
    \fp{\underline{\omega}}{s}
    \textrm{.}
    \label{eq:EulerianStrainRate}
\end{equation}
\end{subequations}
In the square brackets of
equation~(\ref{eq:constitutiveEqnsEulerian-moment}), we have
introduced the tensor of viscous moduli.  The operators in parentheses
inside these square brackets are the tangential and perpendicular
projection operators.

In~\ref{app:RibeCompatibility}, we show that the constitutive laws
used in the classical work of
Ribe~\cite{Ribe-Coiling-of-viscous-jets-2004}, and derived by him from
the Stokes equations in 3D, are equivalent to
equations~(\ref{eq:constitutiveEqnsEulerian}) above.

We note that, according to the Rayleigh-Taylor analogy, the case of an
elastic rod is described by very similar equations, namely by
replacing the strain rate $\underline{e}^\mathrm{E}$ in the
constitutive law~(\ref{eq:constitutiveEqnsEulerian-moment}) by the
Eulerian twist-curvature vector $\underline{\pi}^\mathrm{E}$, and the
constitutive law~(\ref{eq:constitutiveEqnsEulerian-tension}) for the
internal tension by the condition of inextensibility, $\ell=1$.

%
%
%
%

\subsection{Canonical form of the internal virtual work}

The viscous introduced stress introduced in
equations~(\ref{eq:constitutiveEqnsRaw}) allows
equation~(\ref{eq:FirstVariationOfDissipationPotential-step1}) to be 
written as
\begin{subequations}
    \label{eq:FirstVariationOfDissipationPotential-step2}
\begin{equation}
    -\D\mathcal{D}( \underline{x}; \underline{u},v;
    \delta\underline{u},\delta v ) 
    =
    -\int_{S_-}^{S^+}\Big(
    n_{\mathrm{s}}
    \,\D\mathcal{L}_{\mathrm{s}}(\delta\underline{u})
    +
    \underline{m}\cdot
    \D\underline{\mathcal{M}}(\delta \underline{u},\delta v)
    \Big)\,\mathrm{d}S
    \textrm{,}
    \label{eq:FirstVariationOfDissipationPotential-step2-dD}
\end{equation}
where 
\begin{equation}
    \D\underline{\mathcal{M}}(\underline{x};\delta \underline{u},\delta v;S)
    =
     \underline{t}(S)\,
    \D\mathcal{L}^{\mathrm{t}}(\underline{x};\delta\underline{u},\delta v;S)
    +
    \D\underline{\mathcal{L}}^{\mathrm{b}}(\underline{x};\delta 
    \underline{u},\delta v;S)
    \textrm{.}
    \label{eq:FirstVariationOfDissipationPotential-step2-defMtb}
\end{equation}
\end{subequations}

In this definition of the operator $\D\underline{\mathcal{M}}$, we
recognize the first variation of the right-hand side of
equation~(\ref{eq:GradientOfRotationReconstruction}) --- this can be
shown by noting that its right-hand side is linear with respect to the
virtual motion $\delta \underline{u}$ and $\delta v$.  Inserting into
the above equation, we find a compact expression for the first
variation of the dissipation potential:
\begin{equation}
    -\D\mathcal{D}(\delta \underline{u},\delta v) = -
    \int_{S_-}^{S^+}
    \Big(
    \underline{n}_{\mathrm{s}}(S)\cdot \fd{[\delta \underline{u}]}{S}
    +\underline{m}(S)\cdot 
    \fd{[\D\underline{\mathcal{W}}(\delta \underline{u},\delta v)]}{S}
    \Big)\,\mathrm{d}S
    \textrm{.}
    \label{eq:virtualWorkInternalForceCanonical}
\end{equation}
Here, the first term in the integrand has been rewritten using
$n_{\mathrm{s}}\,\D\mathcal{L}_\mathrm{s} =
n_{\mathrm{s}}\,(\underline{t}\cdot \delta \underline{u}'(S)) =
\underline{n}_{\mathrm{s}}\cdot \delta \underline{u}'(S)$ by
equations~(\ref{eq:SmoothFormsAxialStrain-Def})
and~(\ref{tensionForce}).

For any real motion, the quantity $\underline{\mathcal{W}}$ evaluates
to the angular velocity $\underline{\omega}$ of a cross-section by
equation~(\ref{eq:WLinearFormSmooth-Def}).  Therefore
$\D\underline{\mathcal{W}}$ can be interpreted as the virtual rotation
associated with the virtual motion $\delta \underline{u}$ and $\delta
v$.  In view of this,
equation~(\ref{eq:virtualWorkInternalForceCanonical}) is similar to
the classical expression for the internal virtual work in a 3D
continuum, $-\iiint
\underline{\underline{\sigma}}:\underline{\underline{\nabla}}(\delta
\underline{x})\,\mathrm{d}^3x$.  In the context of thin viscous
threads, the 3D stress $\underline{\underline{\sigma}}$ is replaced by
the stretching force $\underline{n}_{s}$ in the first term and by the
bending and twisting moment $\underline{m}$ in the second term, the
gradient is replaced by the spatial derivative
$\mathrm{d}/\mathrm{d}S$, and the virtual displacement $(\delta
\underline{x})$ is replaced by the virtual velocity $\delta
\underline{u}$ and virtual spin velocity $\delta v$.

\subsection{Identification of the net viscous force and twisting moment}
\label{sec:netViscousForceAndTwistingMoment}

We shall now derive explicit expressions for the effective force and
moment acting on the centerline as a result of the internal viscous
stress.  Inserting the definition of $\mathcal{V}$ in
equation~(\ref{eq:VLinearFormSmooth-Def}) into the definition of
$\mathcal{W}$ in equation~(\ref{eq:WLinearFormSmooth-Def}) and
computing the first variation, we have
\begin{equation}
    \D\mathcal{W}(\delta \underline{u},\delta v) 
    = \underline{t}\,\delta v + 
    \frac{\underline{t}\times \delta \underline{u}'(S)}{\ell}
    \textrm{.}
    \nonumber
\end{equation}
Inserting into the
expression~(\ref{eq:virtualWorkInternalForceCanonical}) for the
virtual internal work, the first variation of the dissipation 
potential reads
\begin{equation}
    -\D\mathcal{D}(\delta \underline{u},\delta v) = 
    -\int_{S_-}^{S^+} \left(
    \underline{n}_{\mathrm{s}}\cdot \fd{\delta \underline{u}}{S}
    +
    \underline{m}\cdot
    \fd{(\underline{t}\,\delta v + 
    \frac{1}{\ell}\,\underline{t}\times \delta \underline{u}'(S))}{S}
    \right)\;\mathrm{d}S
    \textrm{.}
    \nonumber
\end{equation}
We can integrate by parts to cast the right-hand side into a form
similar to equation~(\ref{eq:RayleighPotentialFirstVariation}):
\begin{equation}
    -\D\mathcal{D}(\delta \underline{u},\delta v) = 
    \mathrm{BT}+
    \int_{S_-}^{S^+}\left(
    \underline{t}\cdot \fp{\underline{m}}{S}\,\delta v
    +
    \fp{(\underline{n}_{\mathrm{s}} +
    \frac{1}{\ell}\,\underline{t}\times 
    \underline{m}'(S))}{S}\cdot \delta \underline{u}
    \right)
    \;\mathrm{d}S
    \label{eq:firstVariationOfRayleighPot-tmp1}
\end{equation}
where $\mathrm{BT}$ stands for boundary terms.  Those boundary terms
are omitted in the smooth case, and will be readily obtained from our
variational formulation in the discrete case.

Identifying equations~(\ref{eq:RayleighPotentialFirstVariation})
and~(\ref{eq:firstVariationOfRayleighPot-tmp1}), we find an explicit 
expression for the net force $\underline{P}_{\mathrm{v}}$ acting 
on the centerline and for the net twisting moment $Q_{\mathrm{v}}$:
\begin{subequations}
    \label{eq:smoothPvQv}
    \begin{align}
        \underline{P}_{\mathrm{v}}(S,t) & =
	\fp{\underline{n}(S,t)}{S}
        \label{eq:smoothPvQv-P} \\
	Q_{\mathrm{v}}(S,t) & = \underline{t}(S,t)\cdot \fp{\underline{m}(S,t)}{S}
        \label{eq:smoothPvQv-Q}
	\intertext{where}
	\underline{n}(S,t) & = \underline{n}_{\mathrm{s}}
     +\frac{1}{\ell(S,t)}\,\underline{t}(S,t)\times 
     \fp{\underline{m}(S,t)}{S}
     \textrm{.}
     \label{eq:smoothPvQv-internalForce}
    \end{align}    
\end{subequations}
This net force results from the combination of viscous stretching,
twisting and bending stresses inside the thread, and has been computed
using the Rayleigh dissipation potentials.  In these expressions we
have omitted pointwise forces and moments at the endpoints $S=S^\pm$
coming from the boundary terms.  They will be restored in the discrete
model.

\subsection{Equivalence with Kirchhoff equations}

We proceed to show that the expressions~(\ref{eq:smoothPvQv}) for the
net force $\underline{P}_{\mathrm{v}}$ and twisting moment
$Q_{\mathrm{v}}$ acting on the centerline are equivalent to the
equation of motion for a thin thread due to Kirchhoff.

The Kirchhoff equations are usually written in Eulerian variables as
\begin{subequations}
    \label{eq:KirchhoffEqnsEulerian}
\begin{align}
    \fp{\underline{n}}{s} + \underline{P}^{\mathrm{E}} & = 
    \rho\,A\,\ddot{\underline{x}}
    \label{eq:KirchhoffEqnsEulerian-Transl} \\
    \fp{\underline{m}}{s} + \underline{t}\times \underline{n} 
    + Q^{\mathrm{E}}\,\underline{t}
    &=
    J\,\dot{v}\,\underline{t}
    \label{eq:KirchhoffEqnsEulerian-Rot}
\end{align}
\end{subequations}
Here, $\underline{P}^{\mathrm{E}}$ and $Q^\mathrm{E}$ are density of
applied force and twist per unit length $\textrm{d}s$ in actual
configuration, respectively.

When these loads are multiplied by $\ell = \fp{s}{S}$, we obtain our
Lagrangian densities of load $\underline{P} =
\ell\,\underline{P}^\mathrm{E}$, $Q = \ell\,Q^\mathrm{E}$.
Multiplying both equations~(\ref{eq:KirchhoffEqnsEulerian-Transl})
and~(\ref{eq:KirchhoffEqnsEulerian-Rot}) by $\ell$, we have
\begin{subequations}
    \label{eq:KirchhoffEqnsLagrangian}
\begin{align}
    \fp{\underline{n}}{S} + \underline{P} & = 
    \rho\,A_{0}\,\ddot{\underline{x}}
    \label{eq:KirchhoffEqnsLagrangian-Transl} \\
    \fp{\underline{m}}{S} + \underline{T}\times \underline{n} +
    Q\,\underline{t}&=
    \ell\,J\,\dot{v}\,\underline{t}
    \label{eq:KirchhoffEqnsLagrangian-Rot}
\end{align}
where $A_{0} = \ell\,A$ is the area of the cross-section in reference
configuration, and $\underline{T} = \ell \, \underline{t}$ is the
deformed (non-unit) material tangent defined in
equation~(\ref{eq:MaterialTangent}).

Projecting equation~(\ref{eq:KirchhoffEqnsLagrangian-Rot}) along the
tangent and normal directions successively, we have
    \label{eq:KirchhoffEqnsLagrangian-Rot-Proj}
    \begin{align}
	\underline{t}\cdot \fp{\underline{m}}{S} + Q &
	= \ell\,J\,\dot{v} 
        \label{eq:KirchhoffEqnsLagrangian-Rot-Proj-Tang}\\
	\underline{P}_{\perp}(\underline{t},\underline{n}) & =
	\frac{1}{\ell}\,
	\underline{t}\times \fp{\underline{m}}{S}
        \label{eq:KirchhoffEqnsLagrangian-Rot-Proj-Norm}
    \end{align}
\end{subequations}

The tangential component of the internal force $\underline{n}$ can be
interpreted as that resisting stretching of the centerline.  It is
called the tension and is denoted $\underline{n}_{\mathrm{s}}$.  The
full internal force $\underline{n}$ can then be reconstructed by
combining this tangential component $\underline{n}_{\mathrm{s}} =
n_{\mathrm{s}}\,\underline{t}$ with its normal component, given by
equation~(\ref{eq:KirchhoffEqnsLagrangian-Rot-Proj-Norm}):
\begin{subequations}
    \label{eq:netForceAndTwistingMomentFromKirchhoff}
\begin{equation}
    \underline{n} = \underline{n}_{\mathrm{s}} 
    +
    \frac{1}{\ell}\,
    \underline{t}\times \fp{\underline{m}}{S}
    \label{eq:netForceAndTwistingMomentFromKirchhoff-N}
\end{equation}
Comparison of equations~(\ref{eq:KirchhoffEqnsLagrangian-Transl})
and~(\ref{eq:equationsOfMotion-Linear}) reveals that the viscous
stress, described by the quantities $\underline{n}$ and
$\underline{m}$, produces an net force
\begin{equation}
    \underline{P}_{\mathrm{v}} = \fp{\underline{n}}{S}
    \label{eq:netForceAndTwistingMomentFromKirchhoff-Force}
\end{equation}
on the centerline.  Similarly, comparison of
equations~(\ref{eq:equationsOfMotion-Angular})
and~(\ref{eq:KirchhoffEqnsLagrangian-Rot-Proj-Tang}) reveals that the 
net twisting moment due to viscous forces reads
\begin{equation}
    Q_{\mathrm{v}} = \underline{t}\cdot \fp{\underline{m}}{S}
    \label{eq:netForceAndTwistingMomentFromKirchhoff-Moment}
\end{equation}
\end{subequations}
The
expressions~(\ref{eq:netForceAndTwistingMomentFromKirchhoff-Force})
and~(\ref{eq:netForceAndTwistingMomentFromKirchhoff-Moment}) for
$\underline{P}_{\mathrm{v}}$ and $Q_{\mathrm{v}}$ derived from the
Kirchhoff equations~(\ref{eq:KirchhoffEqnsEulerian}) are identical to
those derived earlier in equations~(\ref{eq:smoothPvQv}) from our
dissipation potentials, based on the centerline/twist representation.
We have therefore established the equivalence of our formalism with the
classical, Eulerian equations for thin viscous threads.  The benefit
of our formalism is that it can be discretized in a natural way, and
leads to an efficient implementation.

\section{Space discretization: the discrete viscous thread model}
\label{sec:discreteModel}

In this section, the spatial discretization is carried out, by closely
following the smooth formalism of the previous sections.  Our
derivation of the discrete model makes use of three key ideas that
have been exposed in the smooth setting.  First, we extend the
centerline/spin representation to the discrete case; its benefit is to
eliminate two out of three rotational degrees of freedom using the
condition of compatibility of the tangent.  Second, we introduce a
discrete twist using the geometrical notion of parallel transport.
Third, we derive equations of motion in the discrete setting by
variational principles, starting from discrete dissipation potentials.

We start our analysis of spatial discretization by defining discrete
quantities such as centerline position, linear and angular velocities,
rate of strain, etc.  Time discretization will not be discussed until
section~\ref{sec:timeDiscrete}.

\subsection{Kinematics of centerline}
\label{ssec:discreteKinematics}

\begin{figure}[tbp]
    \centering
    \includegraphics[width=\textwidth]{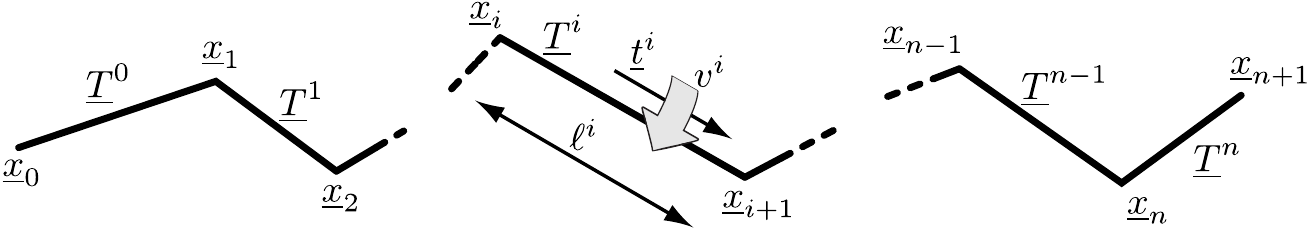}
    \caption{Discrete setting: centerline is a polygonal curve.  Note
    that we use subscripts for vertex-based quantities, such as 
    vertex positions, and superscripts for segment-based quantities, 
    such as segment length $\ell^i$.}
    \label{fig:VerticesEdges}
\end{figure}

The centerline is discretized using $(n+2)$ vertices.  Their positions
in space are noted $\underline{x}_{0}(t)$, $\underline{x}_{1}(t)$,
\dots, $\underline{x}_{n+1}(t)$, see figure~\ref{fig:VerticesEdges}.
Our numerical model involves setting up a force, assigning a mass, and
integrating the fundamental law of dynamics at each vertex
$\underline{x}_{i}(t)$.  The thin thread behavior is produced by
means of a discrete viscous force, which by design converges to the
force $\underline{P}_{\mathrm{v}}(S,t)$ defined in
equations~(\ref{eq:smoothFunctionalDerivativeOfRayleighPotential-resultant})
and~(\ref{eq:smoothPvQv}) in the smooth limit, $n\to\infty$.

The segment joining vertices $\underline{x}_{i}$ and
$\underline{x}_{i+1}$ is noted
\begin{equation}
    \underline{T}^i(t) = 
    \underline{x}_{i+1}(t) -
    \underline{x}_{i}(t)
    \textrm{,}
    \label{eq:edgeTi}
\end{equation}
as shown in figure~\ref{fig:VerticesEdges}.  Following classical
conventions, we use subscripts for indices $0\leq i\leq n+1$
associated with vertices, and superscripts for indices $0\leq i\leq n$
associated with segments.  Since the vertex index $i$ plays the role of
the Lagrangian coordinate $S$, the segment vector $\underline{T}^i(t)$
defined above is the discrete equivalent of the material, non-unit
tangent $\underline{T}(S,t)$ defined in
equation~(\ref{eq:MaterialTangent}).  More accurately it is, like many
other discrete quantities introduced next, an \emph{integrated}
quantity: the discrete tangent is essentially the smooth tangent
multiplied by the discretization length.

The discrete segment length $\ell^i(t)$ and unit tangent
$\underline{t}^i(t)$ are defined by formulas similar to
equations~(\ref{eq:ell}--\ref{eq:UnitTangent})
\begin{equation}
    \ell^i(t) = |\underline{T}^i(t)|
    \textrm{,}
    \label{eq:DiscreteEdgeLength}
\end{equation}
\begin{equation}
    \underline{t}^i(t) = \frac{\underline{T}^i(t)}{\ell^i(t)}
    \textrm{.}
    \label{eq:DiscreteTangent}
\end{equation}

We define the vertex velocities by
\begin{equation}
    \underline{u}_{i}(t) = \fd{\underline{x}_{i}(t)}{t}
    \textrm{.}
    \label{eq:discreteCenterlineVelocity}
\end{equation}
In terms of the velocities $\underline{u}_{j}(t)$, we define the
integrated axial strain rate for segment $\underline{T}^i$ :
\begin{equation}
    d^{i}(t) = \fd{\ell^i(t)}{t} = 
    \underline{t}^i(t)\cdot(\underline{u}_{i+1}(t) - 
    \underline{u}_{i}(t) )
    \textrm{,}
    \label{eq:discreteAxialStrainRate}
\end{equation}
in analogy with equations~(\ref{eq:AxialStrainRate})
and~(\ref{eq:AxialStrainRateAsVelocityGradient}). This is again an 
integrated form of the smooth strain rate $d(S,t)$.

The time derivative of the unit tangent is given in terms of the vertex
velocities by a geometrical formula analogous to
equation~(\ref{eq:tDotInTermsOfU}),
\begin{equation}
    \dot{\underline{t}_{i}}(t) = \frac{1}{\ell^i(t)}\,\underline{P}_\perp\left(
    \underline{t}^i(t),
    \underline{u}_{i+1}(t) - \underline{u}_{i}(t)
    \right)
    \textrm{.}
    \label{eq:discreteTimeDerivativeTangent}
\end{equation}

To define the bending strain, we shall later need vertex-based tangents.
There are several possible definitions that are equivalent in the
smooth limit, and we opt for one that preserves the unit character of 
the tangent, namely
\begin{equation}
    \tilde{\underline{t}}_{i}(\underline{x}_{i-1}, \underline{x}_{i},
    \underline{x}_{i+1}) = 
    \frac{
    \underline{t}^{i-1}
    +
    \underline{t}^i
    }{
    \left|
    \underline{t}^{i-1}
    +
    \underline{t}^i
    \right|
    } 
    \textrm{.}
    \label{eq:DiscreteDissipation-Bending-EdgeTangent}
\end{equation}
The tilde notation is used here and in several other places when we
introduce vertex-based versions of quantities that are primarily
defined at segments, and \emph{vice-versa}.

Similarly, there are several possible definitions for the discrete binormal
curvature vector.  One particular definition is:
\begin{equation}
    \underline{K}_{i}(t) = \frac{\underline{t}^{i-1}\times
    \underline{t}^i}{\frac{1}{2}\,(1+\underline{t}^{i-1}\cdot 
    \underline{t}^i)}
    \textrm{.}
    \label{eq:DiscreteHolonomy-HolonomyCurvature}
\end{equation}
The motivation for choosing this particular definition comes from the
forthcoming equation~(\ref{eq:reconstructDiscreteTwistRate}): this
particular expression $\underline{K}_{i}$ will emerge from the calculation of the
discrete twist. The vector $\underline{K}_{i}$ is an integrated measure of
the smooth binormal curvature vector $\underline{K}(S,t)$ defined in
equation~(\ref{eq:LagrangianBinormal}). Indeed the denominator in
equation~(\ref{eq:DiscreteHolonomy-HolonomyCurvature}) converges to 1 in
the smooth limit where $\underline{t}^{i-1}\sim \underline{t}^i\sim
\underline{t}(S,t)$, while the numerator is equivalent to
$\underline{t}^{i-1}\times \underline{t}^i \sim \underline{t}^{i-1}\times
(\underline{t}^i - \underline{t}^{i-1})\sim
\underline{K}(S,t)\,\tilde{\ell}_{i}$ where $\tilde{\ell}_{i}$ is the
length of the Voronoi cell around vertex $\underline{x}_{i}$, defined below in
equation~(\ref{eq:VoronoiLength}).

\subsection{Incompressibility: radius and related quantities}
\label{ssec:discreteIncompressibility}

Each segment $\underline{T}^i$ carries a volume of fluid $V^i$ and a
mass of fluid $m^i$.  Those quantities are initialized based on the
prescribed initial segment length, radius and mass density of the
fluid.  They are conserved during the simulation, except in the case
of an adaptive mesh, discussed in section~\ref{ssec:meshRefinement},
which requires segment subdivision.  As in the smooth case we use
incompressibility to reconstruct the local radius $a^i(t)$ and
cross-sectional area $A^i(t)$, assuming that each segment has a
cylindrical geometry:
\begin{equation}
    A^i(t) = \frac{V^i}{\ell^i(t)},\qquad
    a^i(t) = \left(\frac{A^i(t)}{\pi}\right)^{1/2}
    \textrm{.}
    \label{eq:discreteRadiusAndAreaReconstruction}
\end{equation}

We shall need later the length $\tilde{\ell}_{i}$ of the Voronoi 
region near a given vertex.
For an interior vertex $\underline{x}_{i}$ with $1\leq i\leq n$, it is
defined as the curvilinear distance between the midpoints of the
adjacent  segments, measured along the polygonal line traced out by 
the vertices:
\begin{equation}
    \tilde{\ell}_{i}(t) = \frac{\ell^{i-1}(t)+\ell^i(t)}{2}
    \quad\textrm{for $1\leq i\leq n$}.
    \label{eq:VoronoiLength}
\end{equation}
This is a vertex-based discretization length, as opposed to the
original segment-based discretization length $\ell^j$.  This length is
not required for the end vertices, $i=0$ and $n+1$.

\subsection{Material frame, angular velocity}

The unit tangent $\underline{t}^i$ is defined at segments.  We define the
orthonormal triads $(\underline{d}_{1}^i,\underline{d}_{2}^i,
\underline{d}_{3}^i)$ at the segments too.  This allows the condition of
compatibility in equation~(\ref{eq:TangentCompatibility}) to be easily
extended to the discrete case:
\begin{equation}
    \underline{d}_{3}^i(t) = \underline{t}^i(t)
    \textrm{.}
    \label{eq:discreteCompatibility}
\end{equation}
Repeating the argument of section~\ref{ssec:materialFrame}, one can 
show that the angular rotation $\underline{\omega}^i$ of the material 
frame can be decomposed as
\begin{equation}
   \underline{\omega}^i(t) = v^i(t)\,\underline{t}^i(t)
   +\underline{t}^i(t)\times \dot{\underline{t}}^i(t)
   \textrm{.}
    \label{eq:RotationVectorDiscrete}
\end{equation}
In the first term of the right-hand side, the quantity $v^i(t)$ is the
spin velocity, \emph{i.\ e.}\ the angular velocity of the material
frame about the tangent, as shown in figure~\ref{fig:VerticesEdges}.
The second term warrants that the time evolution of the centerline,
$\dot{\underline{t}}^i = \underline{\omega}^i\times \underline{t}^i$
remains consistent with the condition of compatibility in
equation~(\ref{eq:discreteCompatibility}).

\subsection{Revisiting the case of zero twist: parallel transport}
\label{ssec:DiscreteHolonomy}

The main difficulty in setting up a discrete model for thin viscous
thread resides in the definition of twist.  In the smooth case, twist
is defined by projecting the infinitesimal rotation vector
$\underline{\pi}$ along the tangent.  This operation is no longer
possible in the discrete case, as rotations are finite and are
represented by a matrix.  To remedy this difficulty, we introduce the
geometrical notion of parallel transport.  It enables us to revisit
the notion of twist, in a way that makes its extension to the discrete
setting natural.

For a given configuration of the centerline, parallel transport
defines a series of rotations from one segment to the next.  Those
rotations define a minimalist motion along the centerline, and will be
used to define twist-less states of the thread.  Parallel transport
has also been used in the smooth setting to define the so-called
natural or Bishop
frame~\cite{Bishop-There-is-more-than-on-way-to-frame-1975,%
LangerSinger-Lagrangian-Aspects-of-the-Kirchhoff-Elastic-1996}.

Consider the unit tangents $\underline{t}^{i-1}$ and $\underline{t}^i$
of the segments adjacent to a vertex $\underline{x}_{i}$.  
We shall
assume that these tangents are not opposite to each other,
\begin{equation}
    \underline{t}^{i-1}\neq -\underline{t}^i 
    \textrm{.}
    \label{eq:tangentsAreNotOpposite}
\end{equation}
This assumption is satisfied, except for a subset of configurations
whose measure is zero.

For a reason that will be clear in the next section, we define a
rotation $\underline{\underline{Q}}$ to be \emph{compatible} at vertex
$\underline{x}_{i}$ if it maps $\underline{t}^{i-1}$ to
$\underline{t}^i$:
\begin{equation}
    \underline{\underline{Q}}\cdot \underline{t}^{i-1}=
    \underline{t}^i
    \textrm{.}
    \label{eq:rotationCompatibility}
\end{equation}
Next, we define parallel transport across vertex $\underline{x}_{i}$ as
\emph{the minimal rotation that is compatible}; here, the word
`minimal' refers to the rotation having the smallest possible angle of
rotation about its own axis~\footnote{In geometrical terms, this
minimal rotation minimizes the distance to the identity over the Lie
group of direct rotations in Euclidean 3D space.}.  This defines a
unique rotation under the assumption of
equation~(\ref{eq:tangentsAreNotOpposite}), as we show now.

Parallel transport, defined above in geometrical terms, has an
explicit representation: it is the rotation
$\underline{\underline{T}}_{i}$ whose axis is along the binormal
$\underline{K}_{i}$ and whose angle is the turning angle $\varphi_{i}$
across vertex $\underline{x}_{i}$.  The turning angle is defined by
\begin{equation}
    \varphi_{i} = \cos^{-1}(\underline{t}^{i-1}\cdot \underline{t}^i)
    \quad
    \textrm{with $0\leq \varphi_{i} <\pi$}
    \textrm{.}
    \label{eq:turningAngle}
\end{equation}
Note that $\varphi_{i} \neq \pi$ by
equation~(\ref{eq:tangentsAreNotOpposite}).  The rotation
$\underline{\underline{T}}_{i}$ just defined satisfies
\begin{subequations}
    \label{eq:parallelTransportAsRotationAboutBinormal}
    \begin{align}
        \underline{\underline{T}}_{i}^T\cdot 
	\underline{\underline{T}}_{i} & 
	=\underline{\underline{1}}
        \label{eq:parallelTransportAsRotationAboutBinormal-IsRotation}\\
        \underline{\underline{T}}_{i}\cdot \underline{K}_{i} & 
	= \underline{K}_{i}
        \label{eq:parallelTransportAsRotationAboutBinormal-LeavesBinormalInvariant}\\
        \underline{\underline{T}}_{i}\cdot \underline{t}^{i-1} & 
	= \underline{t}^i
        \label{eq:parallelTransportAsRotationAboutBinormal-MapsTangents}
    \end{align}
These equations express the fact that $\underline{\underline{T}}_{i}$
is a rotation, that its axis is aligned with binormal
$\underline{K}_{i}$, and that is it compatible --- compare with
equation~(\ref{eq:rotationCompatibility}).  Compatibility follows from
prescribing the rotation angle to be the turning angle $\varphi_{i}$.
In the particular case $\varphi_{i}=0$, that is when the adjacent
segments are aligned, $\underline{t}^{i-1}=\underline{t}^i$, the three
equations above no longer define a unique rotation; in this case,
parallel transport is defined to be the identity,
\begin{equation}
    \underline{\underline{T}}_{i} = \underline{\underline{1}}
    \quad\textrm{if $\varphi_{i} = 0$.}
    \label{eq:parallelTransportAsRotationAboutBinormal-FlatVertex}
\end{equation}
\end{subequations}

Since a compatible rotation maps $\underline{t}^{i-1}$ and
$\underline{t}^i$, its angle of rotation has to be greater or equal to
the angle $\varphi_{i}$ between them.  The angle of rotation of the
matrix $\underline{\underline{T}}_{i}$ that we have just defined is
precisely $\varphi_{i}$.  Therefore, to show that the geometric
definition parallel transport uniquely defines the matrix
$\underline{\underline{T}}_{i}$, it remains to prove that any other
compatible rotation has an angle of rotation strictly larger than
$\varphi_{i}$.  This is what we do now.

First note that any compatible rotation $\underline{\underline{Q}}$
can be decomposed as
\begin{equation}
    \underline{\underline{Q}} =
    \underline{\underline{T}}_{i}\cdot
    \underline{\underline{R}}(\underline{t}^{i-1},\tau(\underline{\underline{Q}}))
    \label{eq:compatibleRotationDecomposition}
\end{equation}
for some angle $\tau(\underline{\underline{Q}})$.  Here
$\underline{\underline{R}}(\underline{t}^{i-1},\tau)$ denotes the
rotation about $\underline{t}^{i-1}$ with angle $\tau$.  This
decomposition follows from the remark that
$\underline{\underline{T}}_{i}^{-1}\cdot \underline{\underline{Q}}$ is
a rotation leaving $\underline{t}^{i-1}$ invariant.  Denoting
$\underline{q}_{\sigma}$ the unit vector obtained by rotating
$\underline{t}^{i-1}$ about the binormal by an angle $\sigma$, one can
compute the dot product of $\underline{q}_{\sigma}$ with its image
$\underline{q}_{\sigma}' = \underline{\underline{Q}}\cdot
\underline{q}_{\sigma}$ by the rotation $\underline{\underline{Q}}$ in
equation~(\ref{eq:compatibleRotationDecomposition}) as
\begin{equation}
    \underline{q}_{\sigma}\cdot \underline{q}_{\sigma}' = 
    \cos\varphi_{i} 
    +[\sin(\varphi_{i}-\sigma)\sin(\sigma)]\,(1-\cos\tau(\underline{\underline{Q}}))
    \textrm{.}
    \label{eq:makeCosSmaller}
\end{equation}
This equality can be established in the direct orthonormal basis whose
first and last vectors are $\underline{t}^{i-1}$ and
$\underline{K}_{i}/|\underline{K}_{i}|$, respectively; in this frame,
$\underline{q}_{\sigma} = \{\cos\sigma,\sin\sigma,0\}$ and
$\underline{t}^i = \{\cos\varphi_{i},\sin\varphi_{i},0\}$. Details of 
the calculation are left to the reader.

For any value of $\varphi_{i}$ such that $0\leq \varphi_{i}<\pi$,
there exists at least a value of $\sigma$ that makes the function
inside the square bracket of equation~(\ref{eq:makeCosSmaller})
negative and non-zero.  If $\cos\tau(\underline{\underline{Q}})\neq
1$, this implies
\begin{equation}
    \underline{q}_{\sigma}\cdot \underline{q}_{\sigma}' < \cos\varphi_{i}
    \textrm{.}
    \label{eq:rotationHasLargerAngle}
\end{equation}
If the rotation $\underline{\underline{Q}}$ of
equation~(\ref{eq:compatibleRotationDecomposition}) is different from
$\underline{\underline{T}}_{i}$,
$\cos\tau(\underline{\underline{Q}})\neq 1$.  Then
equation~(\ref{eq:rotationHasLargerAngle}) shows that the angle of
rotation of $\underline{\underline{Q}}$ about its own axis is greater
or equal than $\cos^{-1}(\underline{q}_{\sigma}\cdot
\underline{q}_{\sigma}')$, and therefore strictly larger than
$\varphi_{i}$.  The proof is complete: there is a unique compatible
rotation whose angle of rotation is minimal, and this is the rotation
$\underline{\underline{T}}_{i}$ defined in
equation~(\ref{eq:parallelTransportAsRotationAboutBinormal}).  It will
be called parallel transport.

Parallel transport establishes a natural mapping between
cross-sections belonging to neighboring segments, and is similar to
the notion of a Levi-Civita connection in differential geometry, see
e.\ g.\ the book by Wald~\cite{Wald-General-relativity-1984}.  This
property will now be used to define the discrete twist.  More
accurately, we shall define twist-less configurations of the rod to 
be those obtained by parallel-transporting the material frames from 
one segment to the next, and will \emph{define discrete twist by
difference with parallel transport}.

The identification of parallel-transport with twist-less
configurations of the rod can be justified by examining compatible
rotations in the smooth case.  Infinitesimal rotation
$\underline{\pi}$ are compatible when they can be associated with
material frames satisfying the compatibility condition in
equation~(\ref{eq:TangentCompatibility}).  Such vectors $\underline{\pi}$
are of the form $\underline{\pi} = \underline{K} +
\tau\,\underline{t}$ by
equation~(\ref{eq:BinormalLagrangianCurvatureAndMaterialCurvature}).
For a given configuration of the centerline, the binormal curvature
$\underline{K}$ is prescribed but $\tau$ is a free function.  Smooth
parallel transport is defined, as in the discrete case, by minimizing
the magnitude $|\underline{\pi}|$ of the infinitesimal rotation,
keeping the centerline fixed; this minimization yields $\tau(S,t)=0$.
This result confirms that parallel transport is associated with
twist-less configurations of the rod in the smooth case.  In view of
this, it makes sense to use parallel transport to extend the notion of
twist to the discrete setting.

\subsection{Discrete twist}

Since material frames are orthonormal, there exists a unique rotation
mapping one material frame to the next.  The rotation connecting the
two material frames adjacent to the vertex $\underline{x}_{i}$ is denoted
$\underline{\underline{Q}}_{i}$,
\begin{equation}
    \underline{d}_{j}^{i} = 
    \underline{\underline{Q}}_{i}\cdot
    \underline{d}_{j}^{i-1}\quad
    \textrm{for $j=1,2,3$}
    \textrm{.}
    \label{eq:discreteRotationAcrossVertex}
\end{equation}
This finite rotation is the discrete equivalent of the infinitesimal
rotation vector $\underline{\pi}(S,t)$ defined in
equation~(\ref{eq:DefMaterialCurvature}); in the smooth case, the
kinematical twist has been extracted from $\underline{\pi}$ by
projection along the local tangent direction.  This operation is no
longer possible with the finite rotation matrix
$\underline{\underline{Q}}_{i}$.

Parallel transport provides an alternative route for defining twist in
the discrete case.  To begin with, note that setting $j=3$ in
equation~(\ref{eq:discreteRotationAcrossVertex}) shows that
$\underline{\underline{Q}}_{i}$ is compatible
 --- compare with
equation~(\ref{eq:rotationCompatibility}), using $\underline{d}^j_{3}
= \underline{t}^j$.  Being a compatible rotation,
$\underline{\underline{Q}}_{i}$ can be decomposed using parallel
transport, as earlier in 
equation~(\ref{eq:compatibleRotationDecomposition}),
\begin{equation}
    \underline{\underline{Q}}_{i} =
    \underline{\underline{T}}_{i}
    \cdot 
    \underline{\underline{R}}(\underline{t}^{i-1},\tau_{i})
    = 
    \underline{\underline{R}}(\underline{t}^i,\tau_{i})
    \cdot 
    \underline{\underline{T}}_{i}
    \textrm{.}
    \label{eq:decompositionOfCompatibleRotation}
\end{equation}
Here, we use the notation $\tau_{i} =
\tau(\underline{\underline{Q}}_{i})$.  The alternative decomposition
after the second equal sign in
equation~(\ref{eq:decompositionOfCompatibleRotation}) follows from the
following argument: the rotation $[ \underline{\underline{T}}_{i}
\cdot \underline{\underline{R}}(\underline{t}^{i-1},\tau_{i}) \cdot
\underline{\underline{T}}_{i}^T]$ is a rotation of angle $\tau_{i}$,
being conjugated with
$\underline{\underline{R}}(\underline{t}^{i-1},\tau_{i})$, and in
addition leaves $\underline{t}^i$ invariant --- it is therefore the 
rotation about $\underline{t}^i$ with angle $\tau_{i}$, which is 
noted $\underline{\underline{R}}(\underline{t}^i,\tau_{i})$.  Note that
the angle $\tau_{i}$ is uniquely defined modulo $2\pi$.

In equation~(\ref{eq:decompositionOfCompatibleRotation}), the angle
$\tau_{i}$ defines an axial rotation required to match one material
frame to the next, in complement with parallel transport
$\underline{\underline{T}}_{i}$.  This $\tau_{i}$ is called the
discrete angle of twist across vertex $\underline{x}_{i}$.  It is an
integrated version of the kinematical twist $\tau(S,t)$ appearing in
the smooth setting.

Let us check that the discrete twist $\tau_{i}$ is consistent with the
smooth twist $\tau(S,t)$ in the smooth limit.  In
equation~(\ref{eq:decompositionOfCompatibleRotation}), the rotations
$\underline{\underline{Q}}_{i}$, $\underline{\underline{T}}_{i}$ and
$\underline{\underline{R}}(\underline{t}^i,\tau_{i})$ converge in the
smooth limit towards infinitesimal rotations which are represented by
the vectors $\underline{\pi}$, $\underline{K}$ and
$\tau_{i}\,\underline{t}$ respectively; the proof of this is left to
the reader.  In this limit,
equation~(\ref{eq:decompositionOfCompatibleRotation}) becomes
$\underline{\pi} = \underline{K} + \tau_{i}\,\underline{t}$, and we
recover the decomposition in
equation~(\ref{eq:BinormalLagrangianCurvatureAndMaterialCurvature}).
This confirms that our definition of the discrete twist is consistent
in the smooth limit.

\subsection{Rate of change of twisting strain}
\label{ssec:rateOfChangeOfDiscreteTwist}

To simulate the dynamics of viscous threads, we need an expression for
the rate of strain associated with the twist mode.  By analogy with
the smooth case, it is defined as the material derivative of the angle
of twist,
\begin{equation}
    e^\mathrm{t}_{i} = \dot{\tau}_{i}
    \textrm{.}
    \label{eq:discreteEtwist}
\end{equation}
Like the quantity $\tau_{i}$, this is a spatially integrated form of
the smooth rate of strain $e_{\mathrm{t}}(S,t)$.  The goal of the
present section is compute $e^\mathrm{t}_{i}$ in a form that can be
used in our centerline/spin representation, i.\ e.\ to express
$e^\mathrm{t}_{i}$ as a function of the velocities $\underline{u}_{j}$
and $v^j$.

\begin{figure}[tbp]
    \centering
    \includegraphics[width=.99\columnwidth]{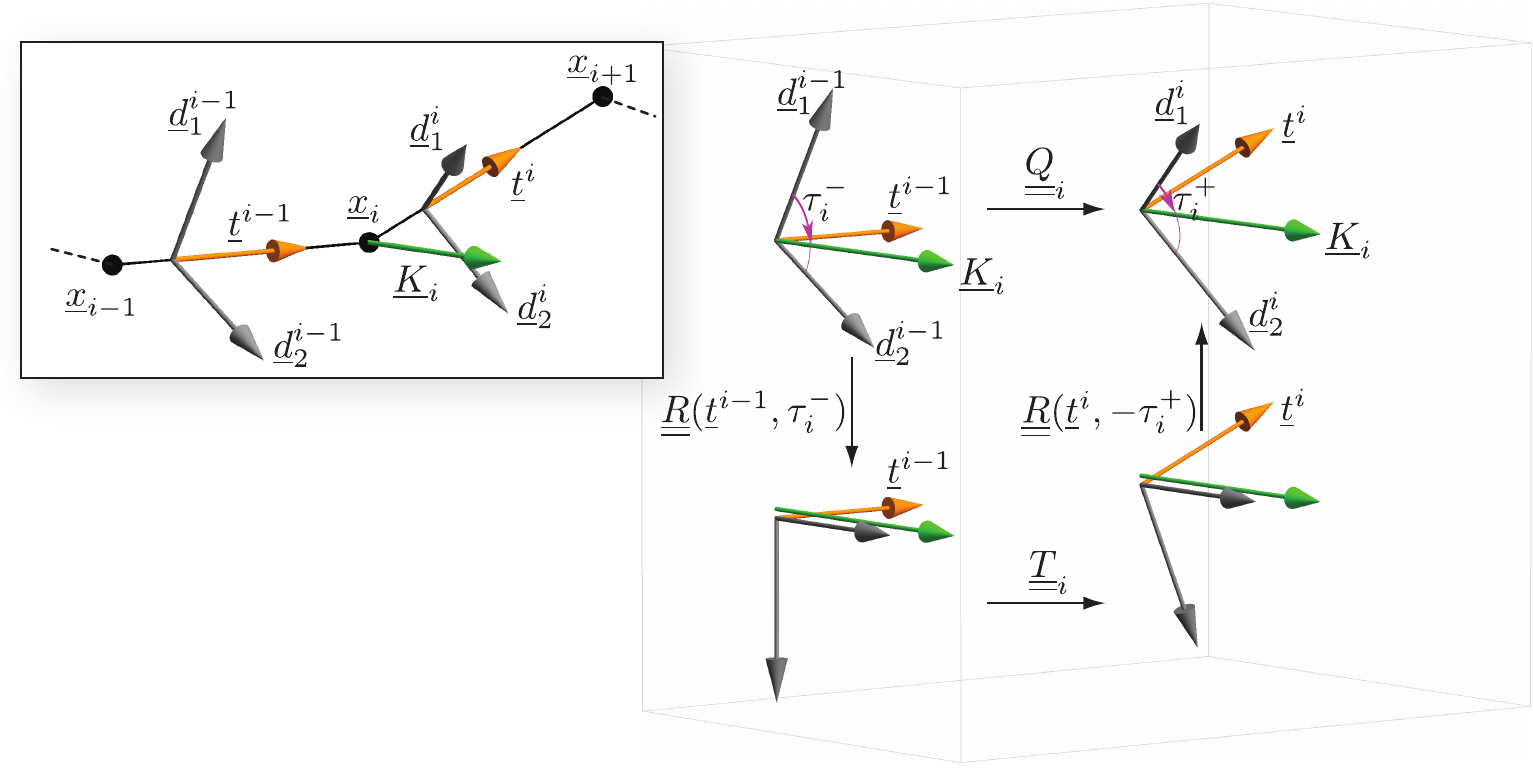}
    \caption{Illustration of
    equation~(\ref{eq:rotationDecompositionAsInFigure}).  The rotation
    mapping one material frame to the next is decomposed using
    parallel transport and the polar angles $\tau_{i}^-$ and
    $\tau_{i}^+$.}
    \label{fig:RotationDecomposition}
\end{figure}

To this end, let us start by introducing the polar angles $\tau_{i}^-$
and $\tau_{i}^+$ of the binormal $\underline{K}_{i}$ in the frames
$(\underline{d}_{1}^{i-1},\underline{d}_{2}^{i-1})$ and
$(\underline{d}_{1}^{i},\underline{d}_{2}^{i})$, respectively.  These
angles are represented in figure~\ref{fig:RotationDecomposition}.
They can be computed by taking the arctangent of the coordinates of
the vector $\underline{t}^{i-1}\times \underline{t}^i$, which is
aligned with $\underline{K}_{i}$ by definition of the latter.  These
coordinates are denoted $(\tau_{i1}^-,\tau_{i2}^-)$ in the first frame
and $(\tau_{i1}^+,\tau_{i2}^+)$ in the second frame:
\begin{subequations}
    \label{eq:tauPMDecomposition}
    \begin{equation}
	\tau_{i}^\pm =\tan^{-1}\left(\frac{\tau_{i2}^\pm}{\tau_{i1}^\pm}\right)
	\textrm{,}
	\label{eq:tauPMDecomposition-ArcTan}
    \end{equation}
    where
    \begin{equation}
	\tau_{ij}^\pm  = (\underline{t}^{i-1}\times \underline{t}^i)\cdot 
	\underline{d}_{j}^{i+(\pm 1-1)/2}
	\textrm{.}
	\label{eq:tauPMDecomposition-Coord}
    \end{equation}
\end{subequations}
In the second equation, $j$ takes on the values $1$ or $2$, and the
superscript in the last factor evaluates to the index $(i-1)$ of the
segment in the left-hand side when $\pm=-$, and to the index $(i)$ of
the segment in the right-hand side when $\pm=+$.

As shown graphically in figure~\ref{fig:RotationDecomposition}, the 
rotation $\underline{\underline{Q}}_{i}$ can be decomposed into a 
rotation about $\underline{t}^{i-1}$ with angle $\tau_{i}^-$ that 
brings $\underline{d}^{i-1}_{1}$ onto the binormal $\underline{K}_{i}$, 
composed by the parallel transport that maps $\underline{t}^{i-1}$ to 
$\underline{t}^i$ without affecting the binormal, and composed by a 
rotation about $\underline{t}^{i}$ with angle $(-\tau_{i}^+)$ that 
brings back the binormal to $\underline{d}_{1}^i$ without affecting 
the tangent:
\begin{equation}
    \underline{\underline{Q}}_{i} =
    \underline{\underline{R}}(\underline{t}^i,-\tau_{i}^+)
    \cdot
    \underline{\underline{T}}_{i}
    \cdot
    \underline{\underline{R}}(\underline{t}^{i-1},\tau_{i}^-)
    \textrm{.}
    \label{eq:rotationDecompositionAsInFigure}
\end{equation}
As earlier in equation~(\ref{eq:decompositionOfCompatibleRotation}), 
we can use conjugation to group the axial rotations in equation 
above. Identifying the result with the definition of $\tau_{i}$ in 
equation~(\ref{eq:decompositionOfCompatibleRotation}), we have
\begin{equation}
    \tau_{i} = \tau_{i}^- - \tau_{i}^+
    \textrm{.}
    \nonumber
\end{equation}
Inserting into the definition~(\ref{eq:discreteEtwist}) of the strain
rate, we have
\begin{equation}
    e^\mathrm{t}_{i} = \dot{\tau}_{i}^- - \dot{\tau}_{i}^+
    \textrm{.}
    \label{eq:discreteEtwist-tmp1}
\end{equation}
We proceed to compute the time derivatives $\dot{\tau}_{i}^\pm$ by
differentiating equations~(\ref{eq:tauPMDecomposition}).

The calculation of $\dot{\tau}_{i\alpha}^-$ is done in the frame
moving with the first material frame $(d^{i-1}_{j})_{1\leq j\leq 3}$.
There, all vectors in the right-hand side of
equation~(\ref{eq:tauPMDecomposition-Coord}) are still, except
$\underline{t}^i$ which has angular velocity
$\underline{\omega}^i-\underline{\omega}^{i-1}$.  This yields, for
$j=1,2$,
\begin{equation}
    \dot{\tau}_{ij}^- = 
    (\underline{t}^{i-1}\times 
    [(\underline{\omega}^i-\underline{\omega}^{i-1})\times\underline{t}^i])\cdot 
    \underline{d}_{j}^{i-1}
    =
    - (\underline{t}^{i-1}\times \underline{d}_{j}^{i-1})\cdot 
    [(\underline{\omega}^i-\underline{\omega}^{i-1})\times 
    \underline{t}^i]
    \textrm{,}
    \nonumber
\end{equation}
after permutation of the mixed product.  Inserting this expression
into the derivative of the arctangent in
equation~(\ref{eq:tauPMDecomposition-ArcTan}), we find
\begin{subequations}
    \label{eq:dotTauIPM}
\begin{equation}
    \dot{\tau}_{i}^- = 
    \frac{
    \tau_{i1}^-\,\dot{\tau}_{i2}^- - 
    \tau_{i2}^-\,\dot{\tau}_{i1}^-
    }{(\tau_{i1}^-)^2+(\tau_{i2}^-)^2}
    =
    \frac{(\underline{t}^{i-1}\times \underline{t}^i)\cdot[(\underline{\omega}^i-\underline{\omega}^{i-1})\times 
    \underline{t}^i]}{|\underline{t}^{i-1}\times \underline{t}^i|^2}
    \textrm{.}
    \label{eq:dotTauIPM-Minus}
\end{equation}
The time derivative of the second angle $\tau_{i}^+$ is given by the 
same formula, with the indices $i$ and $i-1$ swapped:
\begin{equation}
    \dot{\tau}_{i}^+ =
    \frac{(\underline{t}^{i-1}\times \underline{t}^i)\cdot[(\underline{\omega}^i-\underline{\omega}^{i-1})\times 
    \underline{t}^{i-1}]}{|\underline{t}^{i-1}\times \underline{t}^i|^2}
    \textrm{.}
    \label{eq:dotTauIPM-Plus}
\end{equation}
\end{subequations}
Inserting this expression into 
equation~(\ref{eq:discreteEtwist-tmp1}) and permuting the mixed 
product, we have
\begin{equation}
    e^\mathrm{t}_{i} =
    (\underline{\omega}^i-\underline{\omega}^{i-1})
    \cdot
    \frac{
    (\underline{t}^i-\underline{t}^{i-1})
    \times 
    (\underline{t}^{i-1}\times \underline{t}^i)
    }{
    |\underline{t}^{i-1}\times \underline{t}^i|^2
    }
    \nonumber
\end{equation}
In the right-hand side, the second factor can be simplified using the
fact that both $\underline{t}^{i-1}$ and $\underline{t}^i$ are unit
vectors.  This yields
\begin{equation}
    e^\mathrm{t}_{i} =
    (\underline{\omega}^i-\underline{\omega}^{i-1})
    \cdot
    \frac{\underline{t}^{i-1}+\underline{t}^i}{
    1+
    \underline{t}^{i-1}\cdot \underline{t}^i
    }
    \textrm{.}
    \label{eq:eitProjection}
\end{equation}

Inserting now the decomposition~(\ref{eq:RotationVectorDiscrete}) of the 
material velocity $\underline{\omega}^j$ and 
simplifying, we have
\begin{equation}
    e^{\mathrm{t}}_{i}=
    v^{i} -
    v^{i-1}
    +
    \underline{K}_{i} 
    \cdot
    \frac{\dot{\underline{t}}^{i-1} + \dot{\underline{t}}^i}{2}
    \textrm{,}
    \label{eq:reconstructDiscreteTwistRate}
\end{equation}
after using the
definition~(\ref{eq:DiscreteHolonomy-HolonomyCurvature}) of the
discrete binormal curvature $\underline{K}_{i}$.

This formula is fundamental as it yields the rate of strain for the
twisting mode in a form that is suitable for our centerline/twist
representation.  It closely resembles the smooth
equation~(\ref{eq:reconstructTwistFromOmegaT}).  The second term in
the right-hand side has a geometrical origin.  It captures the change
of parallel transport resulting from a change in the centerline, an
effect that was dubbed holonomy in our previous work.  
The holonomy term is responsible for the coupling of centerline
motion with the twisting mode, a phenomenon which appears to be
geometrical in essence.

The geometrical origin of this coupling has been recognized earlier
but has not been used as a starting point for dynamical simulations of
threads, to the best of our knowledge.  The role of the
binormal curvature $\underline{K}$ has been noted in the related
context of F\"uller's
theorem~\cite{Fuller-Decomposition-of-the-linking-number-of-a-closed-ribbon:-1978}
for the increment of writhe of a space curve.  Expressions similar to
those in equations~(\ref{eq:dotTauIPM}) have been derived for the
increment of discrete writhe, and have been used for the simulation of
the Brownian dynamics of DNA modelled as an elastic
rod~\cite{Vries-Evaluating-changes-of-writhe-2005}.

\subsection{Rate of change of bending strain}

In the smooth case, we have defined in
equation~(\ref{eq:RateOfStrainTwistCurvature}) the strain rate vector
$\underline{e}(S,t)$ to be the gradient of rotation.  We introduce
similarly a discrete strain rate vector $\underline{e}^i$ by
\begin{equation}
    \underline{e}^i = \underline{\omega}^i - \underline{\omega}^{i-1}
    \textrm{.}
    \label{eq:discreteStrainRateVector}
\end{equation}
In equation~(\ref{eq:strainRateVectorDecomposition}), we have shown
that the tangent and perpendicular projections of $\underline{e}$ are
the rates of strain relevant to the twisting and bending modes,
respectively.  In the discrete case, we carry out a similar
decomposition, using the vertex-based tangent
$\underline{\tilde{t}}_{i}$,
\begin{equation}
    e^\mathrm{t}_{i} = 
    h_\mathrm{t}(\varphi_{i})\,
    \underline{\tilde t}_{i}
    \cdot
    \underline{e}
    \qquad
    \underline{e}^\mathrm{b}_{i} = 
    h_\mathrm{b}(\varphi_{i})\,
    \underline{P}_{\perp}(\underline{\tilde t}_{i},
    \underline{e}_{i})
    \textrm{.}
    \label{eq:discreteStrainRateVectorDecomposition}
\end{equation}
In these projections, have introduced two normalizing functions
$h_\mathrm{t}$ and $h_\mathrm{b}$ of the turning angle $\varphi_{i}$.
These functions reflect the fact that the definition of the
vertex-based tangent $\underline{\tilde t}_{i}$ in
equation~(\ref{eq:DiscreteDissipation-Bending-EdgeTangent}) is
somewhat arbitrary.  For instance, we could use instead a different 
vertex-based tangent,
\begin{equation}
    \underline{\tilde t}_{i}' = 
    \frac{t^{i-1}+\underline{t}^i}{1+\underline{t}^{i-1}\cdot
    \underline{t}^i} = 
    h(\varphi_{i})\,\underline{\tilde t}_{i}
    \textrm{.}
    \label{eq:alternativeTiTilde}
\end{equation}
Here $h(\varphi_{i}) = |\underline{\tilde t}_{i}'| =
1/\cos(\varphi_{i}/2)$, as can be shown by using trigonometric
relations in the triangle whose sides are $\underline{t}^{i-1}$ and
$\underline{t}^i$.  For consistency with the smooth case, we require
that the functions $h_\mathrm{t}$ and $h_\mathrm{b}$ converge to one
when their argument vanishes.

Our definition of a discrete twist based on parallel transport imposes
a particular choice of the function $h_{\mathrm{t}}$.  Indeed,
identifying equations~(\ref{eq:eitProjection})
and~(\ref{eq:alternativeTiTilde}) we find $e_{i}^\mathrm{t} =
\underline{\tilde t}_{i}'\cdot\underline{e}$ and so
$h_{\mathrm{t}}(\varphi_{i}) = h(\varphi_{i}) = |\underline{\tilde
t}_{i}'| = 1/\cos(\varphi_{i}/2)$.

By contrast, any choice of the function $h_{\mathrm{b}}(\varphi_{i})$
is acceptable as long as $h_{\mathbf{b}}\to 1$ for $\varphi_{i}\to
0$. This leads to infinitely many different discrete thread models, which are 
all equivalent in the smooth limit. In this paper, we choose 
$h_{\mathrm{b}}(\varphi_{i})=1$ for simplicity: the strain rate 
associated with the bending mode then reads
\begin{equation}
    \underline{e}_{i}^\mathrm{b} = 
    \underline{P}_{\perp}(\underline{\tilde t}_{i},
    \underline{e}_{i})
    \textrm{.}
    \label{eq:ebiSimpleChoice}
\end{equation}

\subsection{Reconstruction of strain rates from velocities}
\label{ssec:discreteStrainRateOperators}

As in the smooth case, we keep track of the dependence of all
secondary quantities on the velocities, and introduce virtual vertex
velocities $\underline{\hat{u}}_{i}$ and virtual spin velocities $v^i$
at the segments.  This shall enable us to compute the discrete viscous
forces as the gradient of the dissipation potential with respect to
velocities.

The vertex positions are collected into a generalized coordinate
$\underline{X}(t)$, a vector of dimension $3(n+2)$:
\begin{equation}
    \underline{X}(t) = \{\underline{x}_{0}(t),\dots,\underline{x}_{n+1}(t)\}
    \textrm{.}
    \label{eq:positionalGeneralizedCoordinate}
\end{equation}
Note that there is no need to keep track of the orientation of the
material frame in the generalized coordinate $\underline{X}(t)$, as we
consider isotropic cross-sections.  The twisting mode will only be
included in the generalized velocity $\underline{U}(t)$, defined
later, through which it gets coupled with the centerline motion.

Given the centerline configuration $\underline{X}(t)$, we first
compute all quantities, such as $\ell^i(\underline{X})$,
$\underline{t}^i(\underline{X})$,
$\underline{\tilde{t}}_{i}(\underline{X})$ and
$\underline{K}_{i}(\underline{X})$, which do not depend on velocities.
To make the notations lighter, the argument $\underline{X}$ will often
not appear explicitly in the following.  Next, we extend the
operators $\underline{\mathcal{V}}$ and $\underline{\mathcal{W}}$
defined in equations~(\ref{eq:VLinearFormSmooth})
and~(\ref{eq:WLinearFormSmooth}) to the discrete setting.  The
operator $\underline{\mathcal{V}}^i$ is attached to segment
$\underline{T}^i$ and defined by
\begin{subequations}
    \label{eq:discreteLinearFormsVW}
\begin{equation}
   \underline{\mathcal{V}}^i(\underline{X};
   \underline{\hat{u}}_{i},\underline{\hat{u}}_{i+1}) =
    \frac{1}{\ell^i}\,
    \underline{P}_\perp
    \left(
    \underline{t}^i,
    \underline{\hat{u}}_{i+1}-\underline{\hat{u}}_{i}
    \right)
    \textrm{.}
    \label{eq:TangentDerivativeInTermsOfCenterlineVelocity-Discrete}
\end{equation}
This definition comes from
equation~(\ref{eq:discreteTimeDerivativeTangent}): by design, this
operator yields the time derivative of the tangent when applied to a
real motion, as in the smooth case.

The discrete operator $\underline{\mathcal{W}}^i$ associated with the
segment $\underline{T}^i$ is defined by
\begin{equation}
    \underline{\mathcal{W}}^i(\underline{X};
    \underline{\hat{u}}_{i},\underline{\hat{u}}_{i+1},\hat{v}^i) =
    \hat{v}^i\,\underline{t}^i
    +\underline{t}^i\times 
    \frac{\underline{\hat{u}}_{i+1}-\underline{\hat{u}}_{i}}{\ell^i}
    \textrm{.}
    \label{eq:RotationVectorDiscrete-LinearForm}
\end{equation}
\end{subequations}
This definition is motivated by
equation~(\ref{eq:RotationVectorDiscrete}): when evaluated with a 
real motion,
$\underline{\mathcal{W}}^i$ yields the angular velocity vector
$\underline{\omega}^i$.

We now propose discrete versions of the three fundamental linear forms
defining the viscous dissipation potential, which have been introduced
previously in equations~(\ref{eq:SmoothFormsAxialStrain-Def}) 
and~(\ref{eq:LtbByProjection}).

The discrete axial strain rate $d^{i}$ on segment $\underline{T}^i$ is
given by equation~(\ref{eq:discreteAxialStrainRate}).  In view of
this, we introduce the linear form
\begin{subequations}
    \label{eq:DiscreteDissipation-LinearForm}
\begin{equation}
    \mathcal{L}_\mathrm{s}^{i}(\underline{X};
   \underline{\hat{u}}_{i},\underline{\hat{u}}_{i+1})
    =
    \underline{t}^i\cdot (\underline{\hat{u}}_{i+1}- \underline{\hat{u}}_{i})
    \textrm{.}
    \label{eq:DiscreteDissipation-Stretching-LinearForm}
\end{equation}
Then, we have $
\mathcal{L}_\mathrm{s}^{i}(\underline{X};
\underline{u}_{i}, \underline{u}_{i+1}) = \dot{\ell}^i = d^{i}$ for
any real motion.

The discrete strain rates for the twisting and bending modes are given by
equations~(\ref{eq:reconstructDiscreteTwistRate})
and~(\ref{eq:ebiSimpleChoice}), respectively.  Dependence of these
strain rates on the velocities is captured by the following operators
\begin{multline}
    \mathcal{L}^\mathrm{t}_{i}
    (\underline{X}; 
    \underline{\hat{u}}_{i-1}, \underline{\hat{u}}_{i}, 
    \underline{\hat{u}}_{i+1},
    \hat{v}^{i-1},\hat{v}^i)
    \\ =
    \hat{v}^i - \hat{v}^{i-1}
    + \underline{K}_{i}\cdot
    \frac{
   \underline{\mathcal{V}}^{i-1}(\underline{X};
    \underline{\hat{u}}_{i-1},\underline{\hat{u}}_{i})
    +
   \underline{\mathcal{V}}^i(\underline{X};
    \underline{\hat{u}}_{i},\underline{\hat{u}}_{i+1})
    }{2}
    \textrm{.}
    \label{eq:DiscreteDissipation-Twist-LinearForm}
\end{multline}
and
\begin{multline}
    \underline{\mathcal{L}}^\mathrm{b}_{i}
    (\underline{X}; 
    \underline{\hat{u}}_{i-1}, \underline{\hat{u}}_{i}, \underline{\hat{u}}_{i+1},
    \hat{v}^{i-1},\hat{v}^i) 
    \\
    =
    \underline{P}_\perp\Big(
    \tilde{\underline{t}}_{i}
    ,
    \mathcal{W}^i(\underline{X};
    \underline{\hat{u}}_{i},\underline{\hat{u}}_{i+1},\hat{v}^i)
    -
    \mathcal{W}^{i-1}(\underline{X};
    \underline{\hat{u}}_{i-1},\underline{\hat{u}}_{i},\hat{v}^{i-1})
    \Big)
    \textrm{.}
    \label{eq:DiscreteDissipation-Bending-LinearForm}
\end{multline}
\end{subequations}
For any real motion, $\mathcal{L}^\mathrm{t}_{i}$ and
$\underline{\mathcal{L}}^\mathrm{b}_{i}$ are equal to the strain rates
$e_{i}^\mathrm{t} = \dot{\tau}_{i}$ and
$\underline{e}_{i}^\mathrm{b}$, respectively.  The definition
$\mathcal{L}^\mathrm{t}_{i}$ makes use of the geometrical definition
of discrete twist based on parallel transport.

\subsection{Dissipation potentials}

In our centerline/spin representation, the generalized velocity is a
vector of dimension $4n+7$ defined by collecting the linear velocities
at the vertices, and the angular velocities of spin at the segments:
\begin{equation}
    \underline{U}(t) =
    \{\underline{u}_{0}(t),v^0(t),\underline{u}_{1}(t),v^1(t),\cdots,
    v^n(t),\underline{u}_{n+1}(t)\}
    \textrm{.}
    \label{eq:generalizedVelocity}
\end{equation}
This vector $\underline{U}(t)$ is the generalized velocity of the real
motion.  The dissipation potentials are defined in terms of arbitrary
velocities, which we collect under the name of generalized virtual
velocity $\underline{\hat{U}}$:
\begin{equation}
    \underline{\hat{U}}(t) =
    \{\underline{\hat{u}}_{0}(t),\hat{v}^0(t),\underline{\hat{u}}_{1}(t),\hat{v}^1(t),\cdots,
    \hat{v}^n(t),\underline{\hat{u}}_{n+1}(t)\}
    \textrm{.}
    \label{eq:generalizedVelocity-Virtual}
\end{equation}

As in the smooth case, the viscous internal forces are introduced by means
of dissipation potentials.  The discrete potentials extend the smooth
ones defined in equations~(\ref{eq:smoothDissipationInTermsOfLForms}):
\begin{subequations}
    \label{eq:discreteDissipationInTermsOfLForms}
    \begin{align}
	\mathcal{D}_{\mathrm{s}}(\underline{X};
	\underline{\hat{U}}) & =
	\frac{1}{2}\sum_{0\leq i\leq n}D^{i}\, \Big(\mathcal{L}_\mathrm{s}^{i}
	(\underline{X}; \underline{\hat{u}}_{i},
	\underline{\hat{u}}_{i+1})\Big)^2 
        \label{eq:DiscreteDissipation-Stretching}
	\\
        \mathcal{D}_{\mathrm{t}}
	(\underline{X};
	\underline{\hat{U}})
	& =
	\frac{1}{2}\sum_{1\leq i\leq n}C_{i}\,
	\Big(\mathcal{L}^\mathrm{t}_{i}
	(\underline{X}; 
	\underline{\hat{u}}_{i-1}, \underline{\hat{u}}_{i}, \underline{\hat{u}}_{i+1},
	\hat{v}^{i-1},\hat{v}^i)
	\Big)^2
        \label{eq:DiscreteDissipation-Twist-Terms}
	\\
	\mathcal{D}_{\mathrm{b}}
	(\underline{X};
	\underline{\hat{U}})
	& = \frac{1}{2}\sum_{1\leq i\leq n}B_{i}
	 \,
	\Big(\underline{\mathcal{L}}^\mathrm{b}_{i}
	(\underline{X};
	\underline{\hat{u}}_{i-1}, \underline{\hat{u}}_{i},
	\underline{\hat{u}}_{i+1},
	\hat{v}^{i-1},\hat{v}^i)
	\Big)^2
	\textrm{.}
	\label{eq:DiscreteDissipation-Bending-Terms}
    \end{align}
\end{subequations}
Note that the stretching contribution involves a sum over all
segments, although the twisting and bending contributions involve a sum
over interior vertices.  This is because the discrete axial strain
$d^i$ is defined on segments, while the strain rate vector 
$\underline{e}_{i}$ relevant to the twisting and bending modes is 
defined on interior vertices.

In equations~(\ref{eq:discreteDissipationInTermsOfLForms}), the
discrete moduli are defined by
\begin{subequations}
    \label{eq:discreteModuli}
    \begin{align}
        D^{i}  &  = \frac{3\,\mu^i\,A^i}{\ell^i}
        \label{eq:discreteModuli-Stretching}\\
        C_{i} &
	= \frac{2\,[\widetilde{\mu\,I}]_{i}}{{\tilde \ell}_{i}} 
        \label{eq:discreteModuli-Twist} \\
	B_{i} & 
	=
	\frac{3\,[\widetilde{\mu\,I}]_{i}}{{\tilde \ell}_{i}}
	\textrm{.}
        \label{eq:discreteModuli-Bending}
    \end{align}
where $\mu^i$ is the fluid's dynamic viscosity which is stored at
segments like other fluid properties, $A^i$ is the segment's
cross-sectional area reconstructed by
equation~(\ref{eq:discreteRadiusAndAreaReconstruction}), $\ell_{i}$
the segment length given by equation~(\ref{eq:DiscreteEdgeLength}) and
$\tilde{\ell}^i$ the length of the Voronoi cell around an interior vertices given by
equation~(\ref{eq:VoronoiLength}).  The factor
$[\widetilde{\mu\,I}]_{i}$ appearing the twisting and bending moduli is
defined at vertices by linear interpolation over the adjacent segments:
\begin{equation}
   [\widetilde{\mu\,I}]_{i} = \frac{1}{2}\,
   \frac{\mu^{i-1}\,(A^{i-1})^2 + \mu^{i}\,(A^{i})^2}{4\,\pi}.
\end{equation}
\end{subequations}    
This definition is motivated by the fact that $I=A^2/(4\pi)$ in the
smooth case, as shown by equation~(\ref{eq:Area-MomentOfInertia}).
Note that the discrete moduli satisfy the same relation $B_{i}/C_{i} =
3/2$ as in the smooth case; physically, this relation is a consequence
of the fluid's incompressibility.  All moduli depend on the actual
configuration $\underline{X}(t)$ but not on velocities.

The definitions~(\ref{eq:discreteModuli}) of the discrete moduli are
identical to the definitions~(\ref{eq:StretchingModulus})
and~(\ref{eq:BendingAndTwistModuli}) in the smooth setting, up to
factors proportional to the discretization length $\ell^{i}$ or
$\tilde{\ell}_{i}$.  These factors were introduced so as to warrant
convergence of the dissipation potentials in the smooth limit.  For
instance, for the stretching contribution we have
$D^i\sim\frac{D}{\ell^i}$, and $\mathcal{L}_{\mathrm{s}}^i\sim
\mathcal{L}_{\mathrm{s}}\,\ell^i$ by
equation~(\ref{eq:DiscreteDissipation-Stretching-LinearForm}).  Formal
convergence of the corresponding dissipation potential follows:
\begin{equation}
    \frac{1}{2}\sum_{i}
    D^i\,(\mathcal{L}_\mathrm{s}^{i})^2
    \sim
    \frac{1}{2}\sum_{i}
    \frac{D}{\ell^i}\,(\mathcal{L}_\mathrm{s}\,\ell^i)^2
    \sim
    \frac{1}{2}\sum_{i}
    D\,(\mathcal{L}_\mathrm{s})^2\,\ell^i
    \sim
    \frac{1}{2}\int_{S_-}^{S^+} D\,(\mathcal{L}_\mathrm{s})^2\,\mathrm{d}S
    \sim \mathcal{D}_{\mathrm{s}}
    \textrm{.}
    \nonumber
\end{equation}

The total dissipation potential is defined by summing up the
stretching, twisting and bending contributions:
\begin{equation}
    \mathcal{D}(\underline{X};\underline{\hat{U}})
    =
    \mathcal{D}_{\mathrm{s}}(\underline{X};\underline{\hat{U}})
    + \mathcal{D}_{\mathrm{t}}(\underline{X};\underline{\hat{U}})
    + \mathcal{D}_{\mathrm{b}}(\underline{X};\underline{\hat{U}})
    \textrm{.}
    \label{eq:discreteTotalDissipationPotential}
\end{equation}

\subsection{Discrete equations of motion}

The Rayleigh dissipation potential $\mathcal{D}$ is used to derive the
discrete viscous forces and moments.  By analogy with
equations~(\ref{eq:smoothFunctionalDerivativeOfRayleighPotential-resultant})
and~(\ref{eq:smoothFunctionalDerivativeOfRayleighPotential-twist}), we
model the internal viscous stress by a net force
$\underline{P}^\mathrm{v}_{i}$ acting on the vertex
$\underline{x}_{i}$, and by a twisting moment $Q_{\mathrm{v}}^j$
acting on the segment $\underline{T}^j$, both of which are given by a
derivative of the dissipation potential:
\begin{subequations}
    \label{eq:discreteViscousForcesMom}
    \begin{align}
	\underline{P}^\mathrm{v}_{i}(\underline{X};\underline{U}) & =
	-\left.
	\fp{ \mathcal{D}(\underline{X};\underline{\hat{U}})
	}{ \underline{\hat{u}}_{i} }
	\right|_{\underline{\hat{U}} = \underline{U}}
	\label{eq:discreteViscousForcesMom-Force}\\
	Q_{\mathrm{v}}^i(\underline{X};\underline{U}) & =  
	-\left.\fp{
	\mathcal{D}(\underline{X};\underline{\hat{U}})
	}{
	\hat{v}^i
	}\right|_{\underline{\hat{U}} = \underline{U}}
        \label{eq:discreteViscousForcesMom-Moment}
    \end{align}
\end{subequations}
In these expressions, the partial derivative is with respect to the
$\underline{u}_{i}$ or $\hat{v}^i$ entry inside the generalized
coordinate vector $\underline{\hat{U}}$, see
equation~(\ref{eq:generalizedVelocity-Virtual}).

The discrete equations of motion read
\begin{subequations}
    \label{eq:discreteEqnsOfMotion}
\begin{align}
    \tilde{m}_{i}\,\dot{\underline{u}}_{i}(t) & = \underline{P}_{i}^{\mathrm{v}}
    (\underline{X}(t);\underline{U}(t))
    + \underline{P}_{i}(t)
    \label{eq:discreteEqnsOfMotion-Translation} \\
    \ell^i\,J^i\,\dot{v}^i(t) & = 
    Q_{\mathrm{v}}^i(\underline{X}(t);\underline{U}(t))+Q^i(t)
    \textrm{,}
    \label{eq:discreteEqnsOfMotion-Rotation}
\end{align}
\end{subequations}
where $\underline{P}_{i}(t)$ and $Q^{i}(t)$ define the external
loading, $\tilde{m}_{i}$ is the vertex-based mass, defined as
the sum of half the mass of the segments adjacent to vertex 
$\underline{x}_{i}$,
\begin{subequations}
        \label{eq:vertexInertia}
\begin{equation}
    \tilde{m}_{i} = \sum_{j\in J_{i}} \frac{m^j}{2}
    \textrm{,}
    \label{eq:segmentMass}
\end{equation}
where the set of adjacent segments is $J_{j} = \{i-1,i\}$ if
$0<i<n+1$, $J_{0} = \{0\}$ and $J_{n+1} = \{n\}$.  Note that this
vertex mass $\tilde{m}_{i}$ could change over time if adaptation were used. In 
equation~(\ref{eq:discreteEqnsOfMotion-Rotation}), $J^i$ is the 
density of moment of inertia of the cylinder attached to
segment $\underline{T}^i$ in actual configuration, per unit length:
\begin{equation}
    J^i =
2\rho^i\,I^i = \frac{m^i}{V^i}\,\frac{(A^i)^2}{2\pi}
    \label{eq:rotationalInertiaDiscrete}
\end{equation}
\end{subequations}
and $\rho^i = m^i/V^i$ is the mass density of segment $i$.

The equations of motion can be written in compact form by introducing
the generalized viscous force $\underline{F}_{\mathrm{v}}$, the
generalized external force $\underline{F}$ and the mass matrix
$\underline{\underline{M}}$.  The latter are obtained by collecting
vertex and segment-based components with the same ordering convention
as in the generalized velocity $\underline{U}$:
\begin{subequations}
    \label{eq:generalizedStuff}
    \begin{align}
        \underline{F}_{\mathrm{v}} & = (\underline{P}_{0}^\mathrm{v},
	Q_{\mathrm{v}}^0,\underline{P}_{1}^\mathrm{v},\cdots
	Q_{\mathrm{v}}^n,\underline{P}_{n+1}^\mathrm{v})
        \label{eq:generalizedStuff-Fv}\\
        \underline{F} & = (\underline{P}_{0},
	Q^0,\underline{P}_{1},\cdots
	Q^n,\underline{P}_{n+1})
        \label{eq:generalizedStuff-F}\\
        \underline{\underline{M}} & = 
	\diag(\tilde{m}_{0}\,\underline{\underline{1}},
	\ell^0\,J^0,
	\tilde{m}_{1}\,\underline{\underline{1}},
	\cdots,
	\ell^n\,J^n,\tilde{m}_{n+1}\,\underline{\underline{1}})
	\textrm{,}
        \label{eq:generalizedStuff-M}
    \end{align}
\end{subequations}
where $\underline{\underline{1}}$ represents the unit matrix in 3
dimensions. The equations of motion~(\ref{eq:discreteEqnsOfMotion})
and~(\ref{eq:discreteViscousForcesMom}) can be rewritten as follows,
\begin{subequations}
    \label{eq:compactDiscretEquationsOfMotion}
    \begin{align}
        \underline{\underline{M}}\cdot \dot{\underline{U}}(t) &
	= \underline{F}_{\mathrm{v}}(\underline{X}(t),\underline{U}(t))
	+ \underline{F}(t)
        \label{eq:compactDiscretEquationsOfMotion-Dynamics}\\
        \underline{F}_{\mathrm{v}}(\underline{X},\underline{U}) 
	& = - 
	\left.\fp{\mathcal{D}(\underline{X},\underline{\hat{U}})}{\underline{\hat{U}}}
	\right|_{\underline{\hat{U}} = \underline{U}}
        \label{eq:compactDiscretEquationsOfMotion-ViscousForces}\\
        \dot{\underline{X}}(t) & = \underline{\underline{\Pi}}_{n}\cdot 
	\underline{U}(t)
	\textrm{.}
        \label{eq:compactDiscretEquationsOfMotion-PositionUpdate}
    \end{align}
\end{subequations}
The third equation is the definition of vertex velocities,
$\underline{u}_{i}(t) = \dot{\underline{x}}_{i}(t)$.  This equation
makes use of the projection operator $\underline{\underline{\Pi}}_{n}$
mapping the degrees of freedom associated with vertices from the
$\underline{X}$ representation, from which segments are absent, to
the $\underline{U}$ representation:
\begin{equation}
    \underline{\underline{\Pi}}_{n} = \sum_{i=0}^{n+1}\sum_{j=0}^{2}
    \underline{\delta}_{3i+j}\otimes \underline{\delta}_{4i+j}
    \textrm{.}
    \label{eq:defPin}
\end{equation}
Here $\underline{\underline{\Pi}}_{n}$ is a matrix of size
$(3n+5)\times(4n+7)$, defined with the convention that vector indices
start at 0.  The index $i$ runs over vertices, the index $j$ over
space directions, and $\underline{\delta}_{k}$ represents the vector
whose entries are all 0, except for the $k$-th entry whose value is 1.
The values $k=3i+j$ and $k=4i+j$ appearing in subscript are the
indices of the degrees of freedom for vertex $i$ in direction $j$ in
either representation.

Since the discrete dissipation potential is consistent in the smooth
limit $n\to\infty$, the discrete viscous thread model converges to the
smooth model in this limit: formally, the dynamical system in
equations~(\ref{eq:compactDiscretEquationsOfMotion}) becomes
equivalent to the smooth equations of
motion~(\ref{eq:equationsOfMotionStrong}) combined with the
expression~(\ref{eq:smoothFunctionalDerivativeOfRayleighPotential})
for the internal viscous forces.  This convergence is checked
numerically in section~\ref{sec:validation}.

\subsection{Surface tension and other forces}
\label{ssec:discreteForceExamples}

The weight of the thread is taken in account by setting
$\underline{P}_{i} = \underline{g}\,\tilde{m}_{i}$ and $Q^i=0$ in the equation
of motion~(\ref{eq:generalizedStuff-F}).  Here $\tilde{m}_{i}$ is the mass
attached to a vertex, defined in equation~(\ref{eq:segmentMass}), and
$\underline{g}$ the acceleration of gravity.  In our validation
experiments, the thread is formed by expelling fluid from a syringe,
and letting it fall onto the ground under its own weight.  The contact
forces with the syringe or the ground need not be computed, as we
treat contact using kinematical constraint and not penalty forces.
This is explained in the forthcoming section~\ref{ssec:nozzle}.

\begin{figure}
    \centering
    \includegraphics{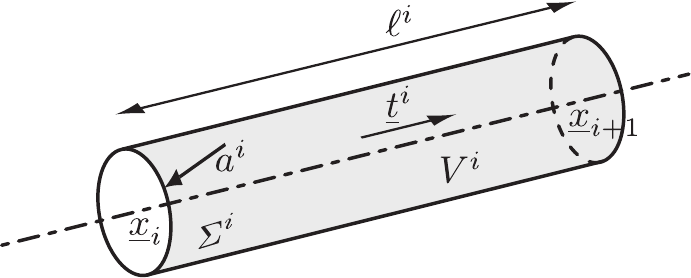}
    \caption{Surface tension is based on a cylindrical representation
    of the fluid attached to segments.}
    \label{fig:SurfaceTension}
\end{figure}

We shall now explain how surface tension is taken into account ---
this is the last type of forces that we shall need in our examples.
The present implementation of surface tension assumes that the
segments are cylinders, as in figure~\ref{fig:SurfaceTension}.  It is
significantly simpler than that based on truncated cones presented in
our conference
paper~\cite{Bergou-Audoly-EtAl-Discrete-Viscous-Threads-2010}.
The lateral area of the cylinder joining vertices $\underline{x}_{i}$
and $\underline{x}_{i+1}$ reads
\begin{equation}
    \Sigma^i = 2\pi\,a^i\,\ell^i = 2\,\sqrt{\pi\,V^i\,\ell^i}
    \textrm{,}
\end{equation}
as can be shown using $V^i= \pi\,(a^i)^2\,\ell^i$.  Its gradients with
respect to the position of its endpoints read
\begin{subequations}
    \label{eq:lateralSurfaceGradient}
    \begin{align}
	\underline{\nabla}_{\underline{x}_{i}}\Sigma^i &
	=
	- \left(\frac{\pi\,V^i}{\ell^i}\right)^{1/2}\,\underline{t}^i
	=
	-\pi\,a^i\,\underline{t}^i
	\label{eq:lateralSurfaceGradient-1}\\
	\underline{\nabla}_{\underline{x}_{i+1}}\Sigma^{i} & 
	=
	+\left(\frac{\pi\,V^i}{\ell^i}\right)^{1/2}\,\underline{t}^i
	=+\pi\,a^i\,\underline{t}^i
	\textrm{.}
	\label{eq:lateralSurfaceGradient-2}
    \end{align}
    \nonumber
\end{subequations}
Here we have used $\underline{\nabla}_{\underline{x}_{i}}\ell^i = 
\underline{\nabla}_{\underline{x}_{i}}|\underline{x}_{i+1} - 
\underline{x}_{i}| =
- 
\underline{t}^{i}$ and 
$\underline{\nabla}_{\underline{x}_{i+1}}\ell^i = +\underline{t}^{i}$.
Discrete surface tension forces are set up by means of the discrete
capillary energy,
\begin{equation}
    \mathcal{E}_{\gamma}(\underline{X}) = \sum_{i=0}^{n} 
    \gamma^i\,\Sigma^i(\underline{X})
    \textrm{.}
    \label{eq:discreteCapillaryEnergy}
\end{equation}
Here $\gamma^i$ is the fluid's surface tension at segment $i$.  In
this equation, we assume that the radius $a^i$ of the cylinder varies
over much longer length-scales that than the radius itself, and
neglect the longitudinal curvature of the lateral boundary in front of
its azimuthal curvature.  This is consistent with the thin thread
approximation used everywhere in this paper.  Let us note this
approximation is not suited to the analysis of the Rayleigh-Taylor
instability, whose critical wavelength is comparable to the radius.
This instability would have to be studied using the full equations for
3D viscous fluids anyway, and not the dimensionally reduced equations
for thin threads.

The expression for the discrete capillary forces at a vertex is given
by minus the gradient of the capillary
energy~(\ref{eq:discreteCapillaryEnergy}) with respect to vertex
positions,
\begin{subequations}
    \label{eq:discreteCapillaryForces}
    \begin{align}
        \underline{P}^{\gamma}_{i} & =
    \begin{cases}
        +n_{\gamma\mathrm{t}}^0\,\underline{t}^0 & \textrm{if $i=0$} \\
        -n_{\gamma\mathrm{t}}^n\,\underline{t}^n & \textrm{if $i=n+1$} \\
        +n_{\gamma\mathrm{t}}^i\,\underline{t}^i
	-n_{\gamma\mathrm{t}}^{i-1}\,\underline{t}^{i-1} & \textrm{if 
	$1\leq i\leq n$}
    \end{cases}
        \label{eq:discreteCapillaryForces-P}\\
        Q_{\gamma}^i & = 0
	\textrm{.}
        \label{eq:discreteCapillaryForces-Q}
    \end{align}
\end{subequations}
The second equation is a consequence of the fact that the capillary
energy depends only on vertex positions and not on the twist degree of
freedom.

These vertex forces $\underline{P}^{\gamma}_{i}$ are caused by an
longitudinal force $n_{\gamma\mathrm{t}}^i$, called the line tension,
acting along each segment.  Its magnitude is given by identification
with equation~(\ref{eq:lateralSurfaceGradient}),
\begin{equation}
    n_{\gamma\mathrm{t}}^i = \pi\,\gamma^i\,a^i
    \textrm{.}
    \label{eq:DiscreteLineTension}
\end{equation}
This expression for the line tension $n_{\gamma\mathrm{t}}^i$ can be
interpreted as resulting from the overpressure in the fluid caused by
the interface curvature $1/a^i$, according to the Young-Laplace law.
As it derives from energy proportional to the lateral area, the
capillary force tends to make the thread shorter and more compact, by
bringing the endpoints closer to each other and by flattening out
curved regions of the centerline.

The net capillary force on a vertex given in
equation~(\ref{eq:discreteCapillaryForces}) has two contributions that
almost cancel each other at each interior vertex (these contributions
are associated with each one of the two adjacent segments), but only
one contribution at the terminal vertices $\underline{x}_{0}$ and
$\underline{x}_{n+1}$.  This reflects the presence of Dirac
contributions at the endpoints in the smooth model, see
equation~(\ref{eq:smoothSurfaceTension-ForceDistribution}).  Note that
the area of the cap closing the cylindrical thread near its endpoints
is negligible, and the Dirac contributions are accurately captured even
though these caps are not taken into account.  Since the discrete
model for surface tension has been derived from the energy in
equation~(\ref{eq:discreteCapillaryEnergy}) which is a good
approximation of the capillary energy in
equation~(\ref{eq:smoothCapillaryEnergy}), the discrete capillary
forces converge to the smooth ones in the limit $n\to\infty$.  Our
discrete model for surface tension is validated in
section~\ref{ssec:validationOfSurfaceTension}.

\section{Time discretization, numerical implementation}
\label{sec:timeDiscrete}

\subsection{Representation of the dissipation potential by a band matrix}
\label{ssec:representationOfLinearForms}

The strain rate operators
$\mathcal{L}_\mathrm{s}^{i}(\underline{X};\underline{\hat{U}})$,
$\mathcal{L}_\mathrm{t}^{i}(\underline{X};\underline{\hat{U}})$ and
$\underline{\mathcal{L}}^\mathrm{b}(\underline{X};\underline{\hat{U}})$
defined in equations~(\ref{eq:DiscreteDissipation-LinearForm}) are
linear with respect to their virtual velocity argument
$\underline{\hat{U}}$.  The two first operators are real-valued: each
one can be represented as a vector, denoted
$\underline{\mathcal{L}}^i_{\mathrm{s}}(\underline{X})$ or
$\underline{\mathcal{L}}_{i}^{\mathrm{t}}(\underline{X})$, that acts
on $\underline{\hat{U}}$ by dot product.  The last operator is
vector-valued, and can be represented as a matrix
$\underline{\underline{\mathcal{L}}}_{i}^{\mathrm{b}} (\underline{X})$
acting on $\underline{\hat{U}}$:
\begin{subequations}
    \label{eq:DefinitionOfDiscreteLinearForm}
\begin{align}
    \mathcal{L}_\mathrm{s}^{i}(\underline{X}; 
    \underline{\hat{U}})
    & =
    \underline{\mathcal{L}}^i_{\mathrm{s}}(\underline{X})
    \cdot \underline{\hat{U}}
    \label{eq:DefinitionOfDiscreteLinearForm-LiS}
    \\  
    \mathcal{L}^\mathrm{t}_{i}
    (\underline{X};
    \underline{\hat{U}})
    & =
    \underline{\mathcal{L}}_{i}^{\mathrm{t}}
    (\underline{X})
    \cdot \underline{\hat{U}}
    \label{eq:DefinitionOfDiscreteLinearForm-LiT}
    \\
    \underline{\mathcal{L}}^\mathrm{b}_{i}
    (\underline{X};
    \underline{\hat{U}}) 
    & =
    \underline{\underline{\mathcal{L}}}_{i}^{\mathrm{b}}
    (\underline{X})
    \cdot \underline{\hat{U}}
    \textrm{.}
    \label{eq:DefinitionOfDiscreteLinearForm-LiB}
\end{align}
\end{subequations}
The tensors $\underline{\mathcal{L}}^i_{\mathrm{s}}(\underline{X})$,
$\underline{\mathcal{L}}_{i}^{\mathrm{t}} (\underline{X})$ and
$\underline{\underline{\mathcal{L}}}_{i}^{\mathrm{b}} (\underline{X})$
just introduced can be distinguished from the original functions as
they bear one additional bar below, and have a single argument.  They
can be calculated explicitly (\emph{i}) by
setting the virtual velocity $\underline{\hat{U}} =
\underline{\delta}_{j}$, \emph{i.\ e.}\ by canceling all entries of
$\underline{\hat{U}}$ except for the entry at index $j$ which is set
to one, (\emph{ii}) by evaluating the left-hand sides above using the
definitions~(\ref{eq:DiscreteDissipation-LinearForm}), (\emph{iii}) by
filling the entry or column with index $j$ in the target vector or
matrix, and (\emph{iv}) by iterating over the index $j$.
In other words, we build the tensors using
\begin{equation}
    \underline{\mathcal{L}}^i_{\mathrm{s}}(\underline{X})  
    = \sum_{j=0}^{4n+6}
    \mathcal{L}^i_{\mathrm{s}}(\underline{X};\underline{\delta}_{j})
    \otimes \underline{\delta}_{j}
    \textrm{,}
    \nonumber
\end{equation}
and similar formulas for $\underline{\mathcal{L}}_{i}^{\mathrm{t}}
(\underline{X})$ and
$\underline{\underline{\mathcal{L}}}_{i}^{\mathrm{b}}
(\underline{X})$. 

Since the linear forms
$\mathcal{L}_\mathrm{s}^{i}(\underline{X};\underline{\hat{U}})$,
$\mathcal{L}^\mathrm{t}_{i}(\underline{X};\underline{\hat{U}})$ and
$\underline{\mathcal{L}}^\mathrm{b}_{i}(\underline{X};\underline{\hat{U}})$
attached to a segment or a vertex $i$ depend only on the
virtual velocities of the neighboring segments or vertices, the
vectors and the matrix introduced in
equations~(\ref{eq:DefinitionOfDiscreteLinearForm}) are sparse.  This
sparse structure allows for efficient storage and manipulation.

\begin{figure}[tb]
    \centering
    \includegraphics[width=.8\textwidth]{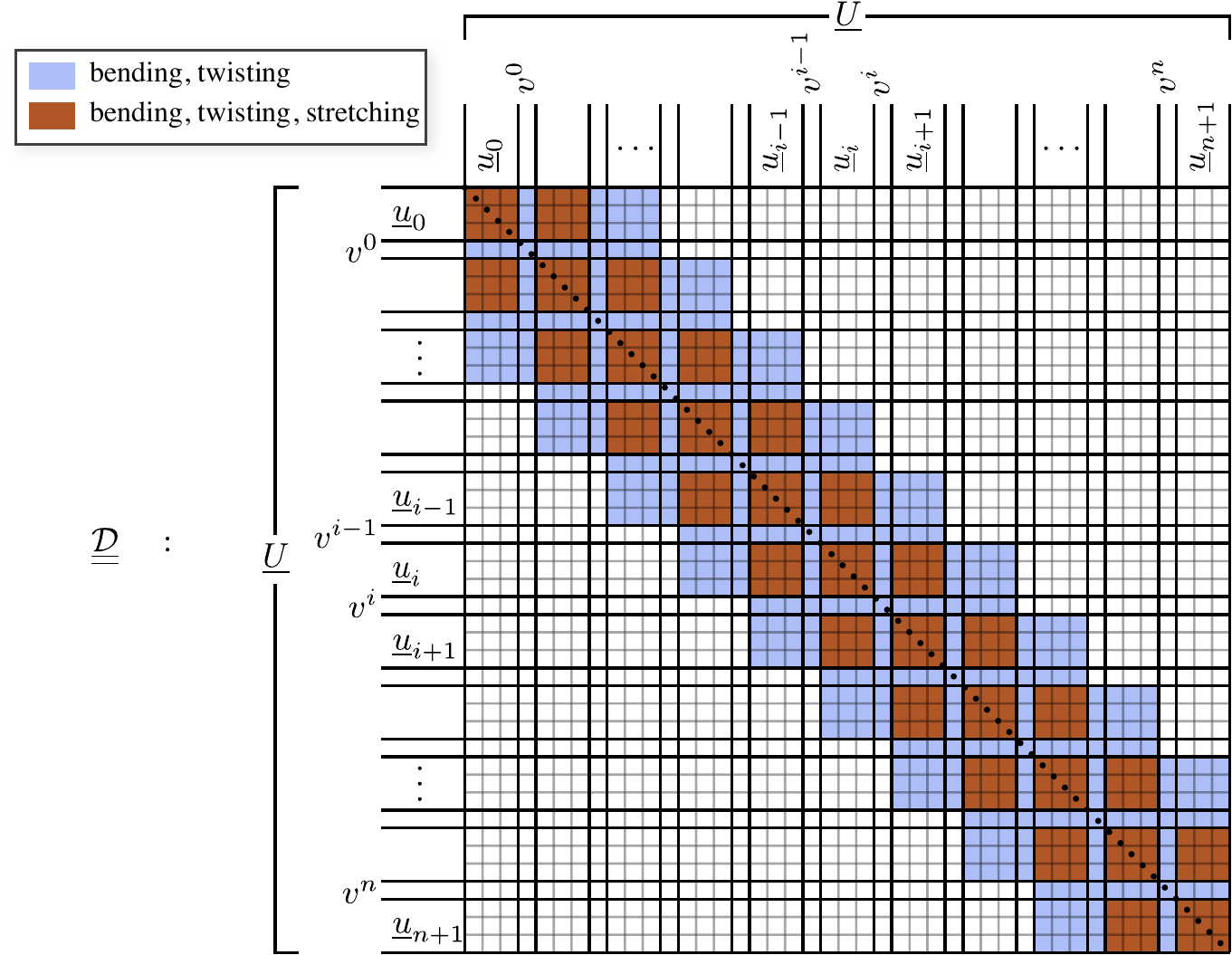}
    \caption{Band structure of the dissipation matrices
    $\underline{\underline{\mathcal{D}}}_{\mathrm{s}}(\underline{X})$,
    $\underline{\underline{\mathcal{D}}}_{\mathrm{t}}(\underline{X})$
    and
    $\underline{\underline{\mathcal{D}}}_{\mathrm{b}}(\underline{X})$
    for the stretching, twisting and bending modes.}
    \label{fig:SparsityPattern}
\end{figure}

Similarly, the dissipation potential is a quadratic
function of the virtual velocity $\underline{\hat{U}}$.  It can
therefore be represented by a symmetric matrix
$\underline{\underline{\mathcal{D}}}(\underline{X})$, such that for
any virtual velocity $\underline{\hat{U}}$
\begin{equation}
    \mathcal{D}(\underline{X};\underline{\hat{U}}) = 
    \frac{1}{2}\,\underline{\hat{U}}\cdot\underline{\underline{\mathcal{D}}}
    (\underline{X})
    \cdot \underline{\hat{U}}
    \textrm{.}
    \label{eq:EvaluationOfDissipationPotential-DiscreteSetting}
\end{equation}
This matrix is built up by combining the stretching, twisting and 
bending contributions,
\begin{equation}
    \label{eq:discreteDissipationMatrixDecomposition}
    \underline{\underline{\mathcal{D}}}(\underline{X}) = 
    \underline{\underline{\mathcal{D}}}_{\mathrm{s}}(\underline{X})
    + \underline{\underline{\mathcal{D}}}_{\mathrm{t}}(\underline{X})
    + \underline{\underline{\mathcal{D}}}_{\mathrm{b}}(\underline{X})
    \textrm{.}
\end{equation}
Explicit formulae for these contributions can be found by inserting
the representation of the linear operators in
equations~(\ref{eq:DefinitionOfDiscreteLinearForm}) into the
definitions~(\ref{eq:discreteDissipationInTermsOfLForms}) of the
discrete dissipation potentials:
\begin{subequations}
    \label{eq:D-MatrixContributions}
    \begin{align}
        \underline{\underline{\mathcal{D}}}_{\mathrm{s}}(\underline{X})
	&
	= 
	\sum_{0\leq i\leq n}
	D^i(\underline{X})
	\,\underline{\mathcal{L}}^i_{\mathrm{s}}(\underline{X})\otimes
	\,\underline{\mathcal{L}}^i_{\mathrm{s}}(\underline{X})
        \label{eq:D-MatrixContributions-Stretching} 
	\\
	\underline{\underline{\mathcal{D}}}_{\mathrm{t}}(\underline{X})
	&
	=
	\sum_{1\leq i\leq n}
	C_{i}(\underline{X})\,
	\underline{\mathcal{L}}_{i}^{\mathrm{t}}(\underline{X})\otimes
	\underline{\mathcal{L}}_{i}^{\mathrm{t}}(\underline{X})
	\label{eq:D-MatrixContributions-Twist}
	\\
        \underline{\underline{\mathcal{D}}}_{\mathrm{b}}(\underline{X})
	&
	=
	\sum_{1\leq i\leq n}
	B_{i}(\underline{X})\,
	\big(\underline{\underline{{\mathcal{L}}_{i}^{\mathrm{b}}}}(\underline{X})\big)^T\cdot
	\big(\underline{\underline{{\mathcal{L}}_{i}^{\mathrm{b}}}}(\underline{X})\big)
	\textrm{.}
	\label{eq:D-MatrixContributions-Bending} 
    \end{align}
\end{subequations}
Because the linear forms in the right-hand sides are represented by
sparse tensors, these symmetric matrices in the left-hand sides are
all band-diagonal.  Their band structure is shown in
figure~\ref{fig:SparsityPattern}.

Note that new vertices are created during the simulation, while others
collide with obstacles and are discarded.  As a result, the number of
vertices --- and therefore the dimension of the dissipation matrix ---
may vary from one time step to the next.

\subsection{Boundary conditions and kinematical constraints}
\label{ssec:kinematicalConstraints}

We consider the possibility that some degrees of freedom are
kinematically constrained.  Such constraints are typically used to
enforce boundary conditions.  For instance, in the case of a clamped
end, the velocities $\underline{u}_{0}$ and $\underline{u}_{1}$ of the
first two vertices and the angular spinning velocity of the first segment
$v^0$ are imposed by the motion of the clamp.

In the presence of kinematical constraints, the
the generalized velocity $\underline{U}_{t+\epsilon}$ at the end of the
time step is prescribed to be of the form
\begin{equation}
    \underline{U}_{t+\epsilon} = \underline{\underline{B}}\cdot 
    \underline{W}_{t+\epsilon} + 
    \underline{B}'
    \textrm{,}
    \label{eq:reducedDofs}
\end{equation}
where $t$ is the time at the beginning of the time step, $\epsilon$ is
the time step, and $\underline{W}_{t+\epsilon}$ is a reduced velocity
vector collecting independent degrees of freedom.  This vector
$\underline{W}_{t+\epsilon}$ is the unknown of the time-stepping
algorithm.  Its size, denoted $r$, may vary from one time step to the
next as constraints can be created or destroyed --- this happens for
instance when the thread makes its first contact with the ground.

The matrix $\underline{\underline{B}}$ is a $(4n+7)\times r$ matrix
that dispatches the independent degrees of freedom into the
generalized velocity vector $\underline{U}$.  Given a strictly
increasing numbering $b$ of the unconstrained degrees of freedom,
$b(0)< b(1)< \cdots <b(r-1)$, this matrix reads
\begin{equation}
    \underline{\underline{B}} = \sum_{i=0}^{r-1} 
    \underline{\delta}_{b(i)}\otimes \underline{\delta}_{i}
    \textrm{,}
    \label{eq:kinematicalConstraintsB}
\end{equation}
where $\underline{\delta}_{j}$ is the Dirac vector,
$(\underline{\delta}_{j})_{i} = 1$ for $i=j$, and 0 otherwise.

The last term in equation~(\ref{eq:reducedDofs}) is the vector
$\underline{B}'$ collecting the velocities of the constrained degrees
of freedom in the final state $\underline{U}_{t+\epsilon}$.  This
vector is typically filled using the motion of the bodies in contact
with the rod as explained in
sections~\ref{sec:interactionWithEnvironment}.  Entries of
$\underline{B}'$ corresponding to unconstrained degrees of freedom can
be set to zero by convention:
\begin{equation}
    \underline{\underline{B}}^T\cdot \underline{B}' = \underline{0}
    \textrm{.}
    \label{eq:supportOfBPrime}
\end{equation}
Indeed, allowing them to be non-zero would simply shift the
value of the unknown $\underline{W}_{t+\epsilon}$.

As an illustration, consider the case when there is no kinematical
constraints.  Then $\underline{r} = 4n+7$, $b(i)=i$ and
$\underline{B}' = 0$, which implies that $\underline{\underline{B}}$
is the identity matrix.  The more interesting case of a viscous
thread, both ends of which are attached to clamps having a motion of
pure translation with velocities $\underline{u}_{\mathrm{clamp}}^1$
and $\underline{u}_{\mathrm{clamp}}^2$ is handled by setting
\begin{subequations}
    \label{eq:clampClamp}
    \begin{align}
        r & = 4n + 7 - 2\times 7 = 4n - 7
        \label{eq:clampClamp-r}\\
        b(i) & = i-7
        \label{eq:clampClamp-b}\\
        \underline{B}' & = 
	(\underline{u}_{\mathrm{clamp}}^1,0,\underline{u}_{\mathrm{clamp}}^1,
	0,\underline{0},\cdots,\underline{0},0,
	\underline{u}_{\mathrm{clamp}}^2,0,\underline{u}_{\mathrm{clamp}}^2)
	\textrm{.}
	\label{eq:clampClamp-BPrime}
    \end{align}
\end{subequations}

In the presence of kinematical constraints, we discard the equations
of motion corresponding to the constrained degrees of freedom.  This
is achieved by left-multiplying both sides of
equation~(\ref{eq:compactDiscretEquationsOfMotion-Dynamics}) by
$\underline{\underline{B}}^T$, and the equation of motion becomes
\begin{equation}
    \underline{\underline{B}}^T\cdot
    \underline{\underline{M}}\cdot \dot{\underline{U}}(t)
	= \underline{\underline{B}}^T\cdot\Big(
	\underline{F}_{\mathrm{v}}(\underline{X}(t),\underline{U}(t))
	+ \underline{F}(t)
	\Big)
	\textrm{.}
    \label{eq:discreteDynamicsConstrained}
\end{equation}

\subsection{Time discretization and time-stepping}

The time step $\epsilon$ can be fixed or variable.  At each time step,
the updated position $\underline{X}_{t+\epsilon}$ and velocity
$\underline{U}_{t+\epsilon}$ must be determined from the actual
position $\underline{X}_{t}$ and velocity $\underline{U}_{t}$.  We
discretize equation~(\ref{eq:discreteDynamicsConstrained}) in time
using a linear implicit scheme.  The viscous force
$\underline{F}_{\mathrm{v}}$ is evaluated implicitly with respect
velocity but explicitly with respect to position:
\begin{equation}
    \underline{\underline{B}}^T\cdot
    \underline{\underline{M}}\cdot
    \frac{\underline{U}_{t+\epsilon}-\underline{U}_{t}}{\epsilon}
    = \underline{\underline{B}}^T\cdot 
    \left(
    \underline{F}_{\mathrm{v}}(\underline{X}_{t},\underline{U}_{t+\epsilon})+
    \underline{F}(t)
    \right)
    \textrm{.}
    \label{eq:ImplicitScheme}
\end{equation}
This choice combines good stability, as demonstrated by the validation
examples, and ease of implementation: only a linear solver is
required.  

The viscous force
$\underline{F}_{\mathrm{v}}(\underline{X}_{t},\underline{U}_{t+\epsilon})$
is indeed linear with respect to the unknown
$\underline{U}_{t+\epsilon}$ by
equations~(\ref{eq:compactDiscretEquationsOfMotion-ViscousForces})
and~(\ref{eq:EvaluationOfDissipationPotential-DiscreteSetting}),
\begin{equation}
    \underline{F}_{\mathrm{v}}(\underline{X};\underline{U}) = 
    -\fp{\mathcal{D}}{\underline{U}}
    = -\underline{\underline{\mathcal{D}}}(\underline{X})\cdot 
    \underline{U}
    \textrm{.}
    \label{eq:EvaluationOfViscousForces-DiscreteSetting}
\end{equation}
Inserting equations~(\ref{eq:reducedDofs})
and~(\ref{eq:EvaluationOfViscousForces-DiscreteSetting}) into the
equation of motion~(\ref{eq:discreteDynamicsConstrained}), we find
that the update rule for the velocity takes the form of a linear
equation to be solved for $\underline{W}_{t+\epsilon}$:
\begin{equation}
    \Big[
    \underline{\underline{B}}^T\cdot
        \left(
	\underline{\underline{M}}
	+\epsilon\,\underline{\underline{\mathcal{D}}}(\underline{X}_{t})
	\right)
    \cdot\underline{\underline{B}}
    \Big]\cdot \underline{W}_{t+\epsilon}
    =
    \underline{\underline{B}}^T\cdot
    (
    \epsilon\,(-\underline{\underline{\mathcal{D}}}(\underline{X}_{t})\cdot\underline{B}'
    + \underline{F}(t))
    - \underline{\underline{M}}\cdot(\underline{B}' - \underline{U}_{t})
    )
    \textrm{.}
    \label{eq:MatrixTimeStep-ForVT+1}
\end{equation}
In this equation, $\underline{\underline{B}}$ and $\underline{B}'$
encode kinematical constraints, $\underline{\underline{M}}$ is the
mass matrix, $\underline{\underline{\mathcal{D}}}$ the viscous
dissipation matrix representing internal stress, $\underline{F}(t)$ is
the external loading and $\underline{U}_{t}$ the velocity at the start
of the time step.  The matrix in square brackets in the left-hand side
is symmetric, positive definite for any value of the time increment
$\epsilon>0$.  As a result, this linear equation can be solved using
efficient and robust solvers.  

Note that the external force $\underline{F}(t)$, which includes in
particular the effect of capillary forces, is evaluated explicitly.
We have tried a linear implicit implementation of surface tension, but
have not observed any significant improvement in stability: an
explicit evaluation of surface tension force does not raise any
stability issue in all our tests.

Once the linear equation for $\underline{W}_{t+\epsilon}$ has been
solved, the generalized velocity $\underline{U}_{t+\epsilon}$ is
reconstructed by means of equation~(\ref{eq:reducedDofs}).  Positions
are then incremented using a discrete version of
equation~(\ref{eq:compactDiscretEquationsOfMotion-PositionUpdate}),
\begin{equation}
    \underline{X}_{t+\epsilon} = \underline{X}_{t} + 
    \underline{\underline{\Pi}}_{n}\cdot \underline{U}_{t+\epsilon}
    \textrm{.}
    \label{eq:positionUpdate}
\end{equation}

\subsection{Time-stepping: summary}
\label{eq:summaryOfImplementation}

The main task of the time stepping loop is to set up and solve
equation~(\ref{eq:MatrixTimeStep-ForVT+1}).
Algorithm~\ref{alg:timestep} explains how the various quantities
appearing in this equation are constructed, and provide an overview of
the implementation of the time step.
\begin{algorithm}[h]
  \begin{algorithmic}[1]
      \caption{Overview of a time step.\label{alg:timestep}}
      \REQUIRE{$\epsilon$}
      \COMMENT{time step}
    \REQUIRE $m^i$, $V^i$, $\mu^i$ 
    \COMMENT{fluid properties}
    \REQUIRE $\underline{x}_{i}(t)$, $\underline{u}_{i}(t)$,
    $v^i(t)$
    \COMMENT{initial positions and velocities}
    \REQUIRE{$\underline{\underline{B}}$, $\underline{B}'$}
    \COMMENT{kinematical constraints at boundaries, \S\ref{ssec:kinematicalConstraints} \S\ref{sec:interactionWithEnvironment}} 
    \REQUIRE{$\underline{F}_{T}$} 
    \COMMENT{external force, \S\ref{ssec:discreteForceExamples} and 
    eq.~(\ref{eq:generalizedStuff-F})} 
    \STATE{set $\ell^i$, $A^i$ and $\tilde{\ell}_{i}$}
    \COMMENT{geometry, \S\ref{ssec:discreteKinematics}Êand \S\ref{ssec:discreteIncompressibility}}
    \STATE{set $D^i$, $C_{i}$, $B_{i}$}
    \COMMENT{viscous moduli, eq.~(\ref{eq:discreteModuli})}
    \STATE{set $\underline{\mathcal{L}}^i_{\mathrm{s}}$,
    $\underline{\mathcal{L}}_{i}^{\mathrm{t}}$,
    $\underline{\underline{\mathcal{L}}}^\mathrm{b}_{i}$}
    \COMMENT{discrete strain rates, \S\ref{ssec:discreteStrainRateOperators} and \S\ref{ssec:representationOfLinearForms}}
    \STATE{set $\underline{\underline{\mathcal{D}}}$}
    \COMMENT{dissipation matrix, eqs.~(\ref{eq:discreteDissipationMatrixDecomposition}--\ref{eq:D-MatrixContributions})}
    \STATE assemble $\underline{X}_{t}$ and 
    $\underline{U}_{t}$ 
    \COMMENT{eqs.~(\ref{eq:positionalGeneralizedCoordinate})
    and~(\ref{eq:generalizedVelocity})}
    \STATE{set $\tilde{m}_{i}$ and $J^i$}
    \COMMENT{eqs.~(\ref{eq:vertexInertia}) 
    and/or~(\ref{eq:neglectDiscreteRotationalInertia})}
    \STATE{assemble $\underline{\underline{M}}$}
    \COMMENT{eq.~(\ref{eq:generalizedStuff-M})}
    \STATE{solve for $\underline{W}_{t+\epsilon}$}
    \COMMENT{eq.~(\ref{eq:MatrixTimeStep-ForVT+1})}
    \STATE{reconstruct $\underline{U}_{t+\epsilon}$}
    \COMMENT{eq.~(\ref{eq:reducedDofs})}
    \STATE{update $\underline{u}_{i}(t+\epsilon)$, $v^i(t+\epsilon)$}
    \COMMENT{eq.~(\ref{eq:generalizedVelocity})}
    \STATE{update $\underline{x}_{i}(t+\epsilon)$}
    \COMMENT{eq.~(\ref{eq:positionUpdate})}
  \end{algorithmic}
\end{algorithm}

Note that fluid properties such as the mass $m^j$, volume $V^j$ and
viscosity $\mu^j$ are stored in segments.  These properties are
constant in most of the examples shown here.  They may however depend
on time, either when adaptation is used (see
section~\ref{ssec:meshRefinement}), or when the dynamics of the thread
is coupled to another physical process, as for instance in the case of
heat transfer and temperature-dependent viscosity --- a separate
update rule is then required.  In addition segments store their
spinning velocity $v^j$.

By the argument of section~\ref{ssec:smoothQuasiStatic}, rotational
inertia can be neglected for a thin thread.  In all the examples shown
here, the rotational inertia was set to zero,
\begin{equation}
    J^j=0\textrm{.}
    \label{eq:neglectDiscreteRotationalInertia}
\end{equation}
We have compared simulations done using this approximation with
simulations based on the exact value of $J^j$ given in
equation~(\ref{eq:rotationalInertiaDiscrete}), and checked that the
results are indeed very close when the thread's radius is small.
%

\subsection{Adaptive mesh refinement}
\label{ssec:meshRefinement}

In the the experiments of
Morris~\cite{Morris-Dawes-EtAl-Meandering-instability-of-a-viscous-2008},
which we reproduce in section~\ref{ssec:sewingMachineExample}, gravity
stretches the thread by a factor which can be as large as 10 to 100.
In the absence of refinement, the segments in the bottom part of the
thread would be considerably longer than those at the top.  A good
spatial resolution is needed at the bottom, where the impact with the
hard surface forms a coils of small radius.  This makes simulations of
severely stretched threads extremely inefficient, unless spatial
refinement is used.

We have implemented adaptive mesh refinement as follows.  At the end
of every dynamic step detailed in algorithm~\ref{alg:timestep}, the
segments needing refinement are first marked according to some
user-defined criterion; next, segments that have been marked are
actually split: a new vertex is inserted in each segment undergoing
subdivision, and quantities such as mass, radius, position, velocity
etc.\ are calculated in the new vertices and segments, as explained
below.  The refined thread is then used as the initial state for the
next dynamic step.

The refinement criterion can be based on comparing length of the
segment to a maximum prescribed length.  Other criteria based on the
turning angle with respect to the neighboring segments have been
considered too, but not used in the examples.  The maximum allowed
length may be a function of the position of the midpoint of the
segment, to force refinement near a boundary for instance.  In each
case, the refinement criterion was adjusted manually to offer the best
compromise between efficiency and accuracy --- even if introduced in
the context of non-steady problems such as the viscous sewing machine,
refinement was always first validated in a steady coiling geometry, as
explained in section~\ref{sec:validation}.  

Whenever a segment has been marked, its subdivision is carried out as
follows.  A new vertex is inserted, two new segments are allocated and
the former segment is removed.  The position $\underline{x}_{i}$ and
velocity $\underline{u}_{i}$ of the new vertex are calculated by an
interpolation of order 4 based on the positions and velocities of its
neighboring vertices.  Vertices resulting from a subdivision
concomitant with that under consideration do not enter in this
interpolation, and so the result is independent of the order in which
marked segments are processed.  The mass $m^j$ stored in the original
segment is equally split among the two subsegments.  The viscosity
$\mu^j$, surface tension $\gamma^j$, spinning velocity $v^j$ of the
new subsegments are all computed by an interpolation of order 2 based
on the values of the former segment and of its neighbors.  Finally,
the volume $V^j$ of the subsegments is computed by first considering
an interpolation of the cross-sectional area $A^i$ at order 4, which
is then multiplied by the length of the subsegment.  This procedure
and the interpolation orders have been chosen in such a way that the
viscous twisting and bending forces, which depend on the derivatives
of order up to four of the positions, remain smooth upon subdivision.

\section{Interaction of the thread with other bodies}
\label{sec:interactionWithEnvironment}

Interaction with two types of rigid bodies are considered in our
validation examples: a fluid container, such as a syringe, delivering
fluid at a fixed volume rate, and a rigid impenetrable surface onto
which the thread impinges.  The interaction is taken care of using
kinematical constraints.  The vertices that have not yet been pushed
outside the container, as well as those that have collided with the
surface, are constrained.  Their positions and velocities are
prescribed by the motion of the body.  In this section, we explain how
the matrix $\underline{\underline{B}}$ and the vector
$\underline{B}'$, which represent these constraints in
algorithm~\ref{alg:timestep}, are calculated.  For the sake of
simplicity, we assume that the motion of the external bodies is
unaffected by the motion of the thread; two-way coupling with external
bodies could be enforced as a kinematical constraint, and would
typically be handled by a post-integration step using a manifold
projection
method~\cite{Hairer-Lubich-EtAl-Geometric-Numerical-Integration-2006}.

\subsection{Fluid container}
\label{ssec:nozzle}

In the experiments, the viscous fluid is fed from a container at a
constant volume rate $Q_{\mathrm{c}}$ through a circular opening of
diameter $d_{\mathrm{c}}$ by a syringe controlled by a step motor, see
figure~\ref{fig:nozzle}a.  Let $A_{\mathrm{c}} =
\pi\,\frac{{d_{\mathrm{c}}}^2}{4}$ be the area of the opening.  The
imposed volume rate sets an ejection velocity $U_{\mathrm{c}} =
Q_{\mathrm{c}} / A_{\mathrm{c}}$ of the thread relative to the container.
The velocity of the fluid
exiting from the container is the sum of the prescribed velocity of the
container, and this relative velocity $U_{\mathrm{c}}$.

At the exit from the container, the thread is clamped, \emph{i.~e.}
the centerline position, tangent, and the rotation are all prescribed.
This clamped boundary condition is implemented by blocking the
rotation of the segment joining the first two vertices, and
constraining the positions of these two vertices (this has the effect
of blocking the unit tangent vector parallel to the first segment).
These kinematical constraints are implemented as explained in
section~(\ref{ssec:kinematicalConstraints}).  Clamped boundary
conditions therefore require that two vertices lie inside the
container at all times.

\subsubsection{Simple container model}
\begin{figure}[tbp]
    \centering
    \includegraphics[width=.96\textwidth]{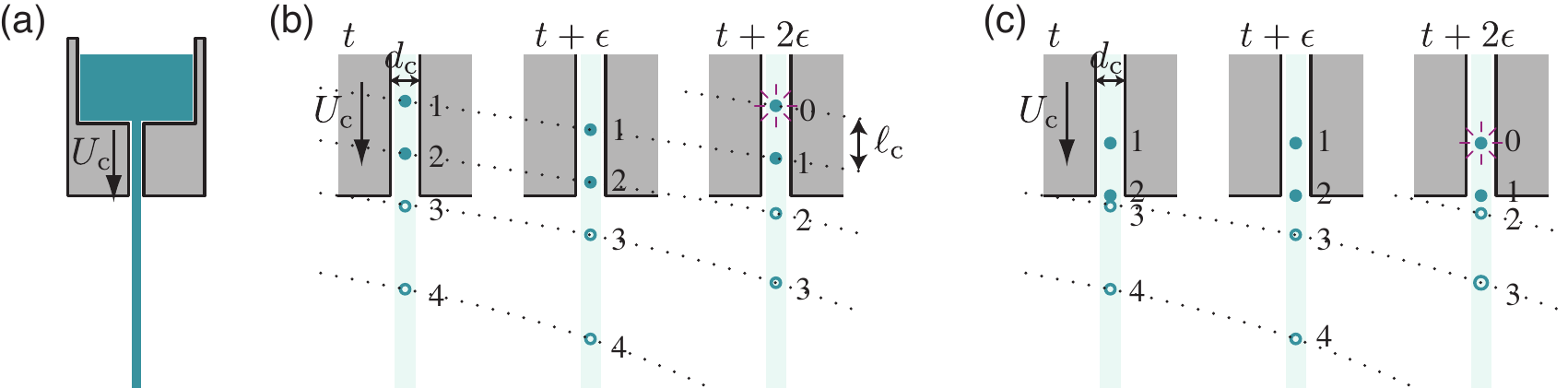}
    \caption{Discrete representation of the container delivering fluid
    through a circular opening of diameter $d_{\mathrm{c}}$ at a
    prescribed vertical velocity $U_{\mathrm{c}}$, shown in (a).
    Clamped boundary conditions are enforced by prescribing the
    position and velocities of the two topmost vertices, and blocking
    the rotation of the first segment joining them.  Filled disks
    represent constrained vertices (top part of the thread attached to
    the container), open circles represents unconstrained vertices
    (hanging part of the thread), dashed curves are the timelines of
    vertices.  (b) Simple implementation whereby the two first
    vertices move with prescribed vertical velocity $U_{\mathrm{c}}$.
    (c) In a refined implementation, the first two vertices are kept
    at a fixed position with respect to the container, and mass is
    continuously added into the second segment.  In both cases (b) and
    (c) a new vertex is periodically created from top (vertex labelled
    `0' appearing at time $t+2\epsilon$), and there are always two
    vertices inside the container.}
    \label{fig:nozzle}
\end{figure}

A simple discrete model for the container is sketched in
figure~\ref{fig:nozzle}b.
The two topmost vertices of the thread are located inside the
container and move with the velocity
$-U_{\mathrm{c}}\,\underline{e}_{z}$ relative to the container.  Here
$\underline{e}_{z}$ is the unit vector, usually vertical, which is
opposite to the direction of ejection.  At every time step, whenever
one of these two vertices goes past the opening of the container, it
is freed (its position constraint is discarded) and a new constrained
vertex is added on top of the thread, as happens in frame
$t+2\epsilon$ in figure~\ref{fig:nozzle}b; the new segment on top is
assigned a length $\ell_{\mathrm{c}}$, a volume
$(A_{\mathrm{c}}\,\ell_{\mathrm{c}})$, a mass
$(\rho\,A_{\mathrm{c}}\,\ell_{\mathrm{c}})$, and surface tension
$\gamma$.  Here, $\rho$ and $\gamma$ are the mass density and surface
tension of the fluid in the container, and the length
$\ell_{\mathrm{c}}$ is a discretization parameter chosen by
the user.

This simple implementation has a drawback: it induces oscillations of
the thread with a small amplitude and a large frequency.  Indeed the
effective fall height is determined by the position of the second
topmost vertex.  As a function of time, this height varies abruptly
every time a new vertex is added.  The amplitude of the discontinuity
is set by the spacing between vertices in the containers.  We found
that this small-amplitude, large-frequency forcing usually allowed
convergence of positions and velocities with finer and finer
discretization lengths $\ell_{\mathrm{c}}$, but induces unbounded
fluctuations of acceleration.

\subsubsection{Refined container model}
\label{ssec:NozzleImproved}

We found it necessary to suppress the oscillations to obtain
convergent and fully reproducible results, and to capture the subtle
patterns produced by the viscous sewing machine at large fall
heights~\cite{Brun-Ribe-EtAl-A-numerical-investigation-of-the-fluid-2012}.
To do so, we used an improved description of the container which is
sketched in figure~\ref{fig:nozzle}c.  The first two vertices do not
follow fluid particles but instead stay at a fixed position with
respect to the container.  As earlier, their velocity is constrained
to the value $-U_{\mathrm{c}}\,\underline{e}_{z}$ relative to the
container at the start of each dynamic step, but their position is
systematically reset after each dynamic step, so as to make the second
vertex coincide with the opening of the container and the first vertex
lie at distance $\ell_{\mathrm{c}}$ above it.  At each time step
$\epsilon$, an incremental volume of fluid
$(\epsilon\,Q_{\mathrm{c}})$ and the associated mass
$(\rho\,\epsilon\,Q_{\mathrm{c}})$ provided by the container are
assigned to the \emph{second} segment.  Whenever its volume exceeds
the target value $(A_{\mathrm{c}}\,\ell_{\mathrm{c}})$ fixed by the
discretization parameter $\ell_{\mathrm{c}}$, this segment is split, a
new vertex is inserted between the second and third vertices; the
fluid material in excess is then assigned to the new segment, as
sketched in figure~\ref{fig:nozzle}c.  In this improved implementation
of the boundary conditions, the fall height varies smoothly as a
function of time.  We found that this was sufficient to make the
acceleration converge smoothly in the limit of a fine discretization,
and produce stitching patterns that are both reproducible and
consistent with the sewing machine experiments.  All examples shown in
Sections~\ref{sec:validation} and~\ref{sec:final} make use of this
refined model.

\subsection{Collisions on a hard surface}
\label{ssec:descriptionOfFloor}

We now consider collisions with a rigid obstacle.  When the thread
hits the obstacle, it sticks to it and gets carried away by the
obstacle.  The obstacle may be at rest, as in the case of steady
coiling of section~\ref{sec:validation}, or in motion, as happens with
the moving belt in the sewing machine experiment of
section~\ref{ssec:sewingMachineExample}.  In the latter case, the
motion of the obstacle is prescribed.  Detection of collisions
involves comparing the distance of vertices to the obstacle, to the
radius $a^i$ of the adjacent segments defined in
equation~(\ref{eq:discreteRadiusAndAreaReconstruction}).

\subsubsection{`Capture and continue' mode}

In a straightforward implementation, called the `capture and continue'
mode, we detect collisions with the obstacle at the end of every
dynamic step, and mark the colliding vertices as being captured: in
all subsequent dynamic steps, their positions and velocities are
constrained based on the motion of the obstacle and on their relative
position to the obstacle when they were captured.  Slip-free
conditions are considered: the rotation of segments joining any pair
of captured vertices is blocked as well.  We found that this simple
description of the collisions was another source of large, spurious
fluctuations in acceleration.  The oscillations can be interpreted by
the fact that the vertical momentum resulting from a collision is not
transferred to the thread until the following time step, when the
position constraints take effect.  Another drawback of the method,
which is partly responsible for oscillations, is that the thread
penetrates into the obstacle by a small but very irregular depth,
roughly proportional to the time step duration.

\subsubsection{`Time roll-back' mode}
\label{sssec:rollback}

The oscillations were removed by using a 
\emph{roll-back}, which allows for a more accurate handling of collisions.  
A roll-back discards any dynamic time step ending up in unexpected
collisions.  The time step is recomputed
with the motion of the colliding vertices constrained in such a way
that they come exactly in contact with the obstacle at the end of the
time step. Roll-back can be viewed as an iterative 
predictor-corrector or iterative constraint refinement 
method~\cite{Otaduy:2009tla}: a new unexpected
collision may take place during the second tentative time step, inducing
a third attempt, etc. A list of expected
collisions is kept and updated after unsuccessful time step. 

Roll-back removes the two main limitations of the straightforward
`capture and continue' implementation: it suppress the delay in
transferring momentum from the obstacle to the thread, and makes the
thread land exactly at the surface of the obstacle, removing the
unwanted rugosity produced by the `capture and continue' mode.
We found that roll-back does indeed suppress the spurious oscillations
very effectively, bringing about benefits similar to those of the
improved nozzle implementation discussed in
section~\ref{ssec:NozzleImproved}.  All the validation examples shown
in the following sections make use of the time roll-back method.

If many collisions occur during a single time step, the roll-back
method does not work well and spurious fluctuations reemerge in
the acceleration as a function a time.  The reason is that we work in
the context of a linearized implicit scheme, in which the equations of
motion are always linearized using the solution at the start of the
time step.  The successive iterations of a given time step all start
from the same configurations.  As a result, they all make use of the
same set of linearized equations.  If too many collisions occur before
the end of the time step, this linearization becomes a poor
approximation of the actual motion.  To work around this limitation,
we combined roll-back with time adaptation: if a new collision is
detected before the end of the tentative time step, the next tentative
time step is shortened and scheduled to end at the collision time
estimated from the previous iteration.  This time adaptation reduces
the time step dynamically, in such a way that there is at most one
collision per time step.  The drawback is the increased complexity in
implementation, and the fact that the simulation time increases with
the rate of collisions.  
%

\section{Validation in a steady coiling geometry}
\label{sec:validation}

We proceed to validate our discrete model and verify our
implementation, by checking convergence in the smooth limit.  We
consider the steady coiling motion of a viscous thread stretched by
gravity and impinging on a surface at rest, as shown in
figure~\ref{fig:coilingValidation}b.  Our simulation results are
compared to reference solutions kindly provided by N.~Ribe, which are
based on numerical continuation of the time-independent problem
expressed in the co-rotating
frame~\cite{Ribe-Coiling-of-viscous-jets-2004}, and solved using the
AUTO
software~\cite{Doedel-Champneys-EtAl-AUTO97:-continuation-and-bifurcation-2002}.

\subsection{Validation of bending, stretching, gravity, inertia and collisions}

The following set of parameters are used for validation and
verification: the fluid's dynamical viscosity $\mu=0.2$ and mass
density $\rho=5\,10^{-4}$, the acceleration of gravity $g=9.81$, the
area $A_{\mathrm{c}}= 6.44\,10^{-3}$ of the circular outlet of the
container and the fluid's volume rate $Q_{\mathrm{c}} =
3.96\,10^{-3}$.  The surface tension $\gamma$ is set to zero until we
validate surface tension later in
section~\ref{ssec:validationOfSurfaceTension}.

Three dimensionless groups characterize the properties of the fluid
and the container~\cite{Ribe-Habibi-EtAl-Stability-of-liquid-rope-2006}:
\begin{equation}
	\label{defs:piparameters}
	\Pi_1 = \left(\frac{\nu^5}{g\,{Q_{\mathrm{c}}}^3}\right)^{1/5}\,,\qquad
	\Pi_2 = \left(\frac{\nu\,Q_{\mathrm{c}}}{g\,{d_{\mathrm{c}}}^4}\right)^{1/4}\,,\qquad
	\Pi_3 = \frac{{d_{\mathrm{c}}}^2 \,\gamma }{Q_{\mathrm{c}}\, 
	\mu }\,\textrm{,}
\end{equation}
where $\nu=\mu/\rho$ is the kinematical viscosity; the diameter of the
container's circular opening $d_{\mathrm{c}}$ and the extrusion
velocity $U_{\mathrm{c}} = Q_{\mathrm{c}}/A_{\mathrm{c}}$ have been
defined in section~\ref{ssec:nozzle}.  The numerical values of the
dimensionless groups are $\Pi_{1} = 7000$, $\Pi_{2} = 7$ and $\Pi_{3}
= 0$ for the set of parameters listed above.  This corresponds to
significant, but not extreme, stretching: the radius decreases by a
factor of order 2 during the course of the descent for the range of
heights considered below.

Based on the acceleration of gravity $g$ and on the kinematical
viscosity $\nu$, one can define a natural length scale $L^*$
and a natural time scale $T^*$ 
by
\begin{equation}
    L^*
    = \left(\frac{\nu^2}{g}\right)^{1/3},
     \qquad
     T^* =
     \left(\frac{\nu}{g^2}\right)^{1/3}
    \textrm{.}
    \label{eq:naturalScale}
\end{equation}
Numerically, $L^* = 25.36$ and $T^* = 1.61$ using the above set of
parameters.  These scales are used to make the simulation results
dimensionless when comparing to the reference solution.

Two additional discretization parameters are needed in the simulation:
the initial segment length $\ell_{\mathrm{c}}$, introduced in
section~\ref{ssec:nozzle}, and the time step $\epsilon$.  Unless
otherwise specified, their values are set to $\ell_{\mathrm{c}}=0.025$
and $\epsilon=0.02$.  Note that the average number of particles
emitted per time step is $(\epsilon\,U_{\mathrm{c}}/\ell_{\mathrm{c}}) =
0.49$.  A good trade-off between accuracy and efficiency requires that
this number is neither very small nor very large.

The reference solution of
N.~Ribe~\cite{Ribe-Coiling-of-viscous-jets-2004} takes the form of
data for the coiling radius or the coiling frequency, as a function of
the fall height.  To produce simulation data which can be compared to
this reference solution, a range of fall heights $H(t)$ is swept in a
single simulation run.  To this end, the motion of the container is
prescribed in a sequence of up to three phases in our simulations.  At
initial time $t=0$, it is placed at height $H_0$.  It is left still
until time $t=t_1$, when the steady coiling is established.  Then, the
container is moved upwards at a prescribed velocity $V_\mathrm{c}$
that is much smaller than the extrusion velocity $U_{\mathrm{c}}$,
until time $t=t_2$.  In a last phase, from time $t=t_2$ to the end of
the simulation, $t=t_\mathrm{f}$, the container is moved slowly downwards
at the same velocity $V_\mathrm{c}$.  Unless otherwise specified, we
set $t_{1} = 30$, $t_{2} = 1530$, $t_{\mathrm{f}} = 3030$ and 
$V_{\mathrm{c}} = 0.02$, which is 30 times slower than the extrusion 
velocity.

\begin{figure}
    \begin{center}
	\includegraphics[width=.80\columnwidth]{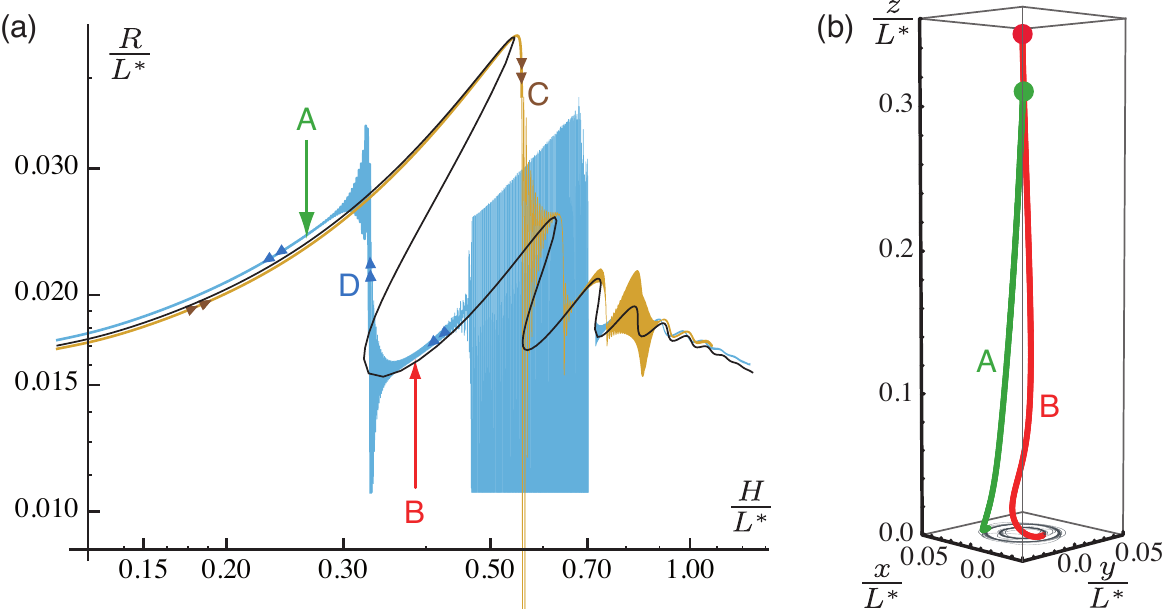}
	\caption{Validation in a steady coiling geometry, with
	$\Pi_{1} = 7000$, $\Pi_{2}= 7$ and $\Pi_{3} = 0$ (no surface
	tension).  (a) The coiling radius $R=(x^2+y^2)^{1/2}$ is recorded continuously
	as the fall height $H$ is varied, by slowly moving the
	container upwards (brown curve) and then downwards (blue
	curve).  It is compared against the reference solution of
	N.~Ribe obtained by numerical continuation (black), after
	rescaling by the length scale $L^*$.  Note the abrupt
	changes in the coiling radius at places where the reference
	solution folds back onto itself, as expected.  Filled regions correspond to
	a rapidly varying radius, either because the thread goes to a
	steady folding mode, as happens around $H=.60$, or because of
	a transient regime following a branch jump.  (b) Two typical
	configurations of the thread, for different fall heights.
	Note that the fundamental mode of a hanging viscous string is
	excited in `$\mathrm{A}$', while its first harmonic is excited
	in `$\mathrm{B}$': the red curve has a node near
	$z/L^*=.05$.}
	\label{fig:coilingValidation}
	\end{center}
\end{figure}

Validation results are shown in figure~\ref{fig:coilingValidation}.
Most of the time, the simulation lays down a thin curve (brown and
blue) in the plane $(H,R)$, where $R=(x^2+y^2)^{1/2}$ is the distance
of the point of contact of the thread with the floor, to the axis
passing through the nozzle.  This indicates a steady coiling regime.
When the container is moved up (brown curve), the simulated radius $R$
follows closely the reference solution, until the latter folds back
onto itself.  The portion of the reference curve immediately past the
fold point is known to be
unstable~\cite{Ribe-Habibi-EtAl-Stability-of-liquid-rope-2006}.  The
simulation then goes to a transient regime, labeled `$\mathrm{C}$' in
figure~\ref{fig:coilingValidation}a and shown in
figure~\ref{fig:transientInCoiling}.  It then settles to a different
branch of solutions having a smaller radius.  A series of
such bifurcations is observed in the simulation, a behavior that has
been observed in experiments
too~\cite{Ribe-Habibi-EtAl-Stability-of-liquid-rope-2006}.  A similar
sequence of transitions is observed when the container is moved down.
As multiple coiling solutions are in competition, a hysteretic
behavior is observed: transition from the branch corresponding to the
largest coiling radius to the second largest occurs at a height
`$\mathrm{C}$' when the container moves up, which is larger than the
transition height `$\mathrm{D}$' observed on the way
back.  We observe the occurrence of a folding mode in the interval
$.72> H/L^*> .47$ when the container moves down, but will not comment 
further on it as little
is known on the competition between the coiling and folding modes.
The small gap between the blue and brown curves in the left part of
figure~\ref{fig:coilingValidation}a can be attributed to the fact that
the velocity of the container $V_{\mathrm{c}}$ is small but finite.
Overall, the simulation is in good agreement with the reference curve,
and reproduces the details of its meandering shape.  This
validates the various physical ingredients that affect the reference
curve, namely viscous bending and stretching, gravity, inertia and
contact with the floor. 

A detailed run such as the one shown in
figure~\ref{fig:coilingValidation}, corresponding to a total
simulation time $t_{\mathrm{f}} = 3030$ and a total number of time
steps $t_{\mathrm{f}}/\epsilon \approx 150\,10^3$, runs in about
$30~\mathrm{min}$ on a $2.6~\mathrm{Ghz}$ Intel Core i7 processor
using $8~\mathrm{GB}$ of memory.  The maximum number of vertices, when
the fall height is maximum, is 460, corresponding to approximately
1800 degrees of freedom.

\subsection{Analysis of convergence}

\begin{figure}[tbp]
    \centerline{\includegraphics[width=.7\columnwidth]{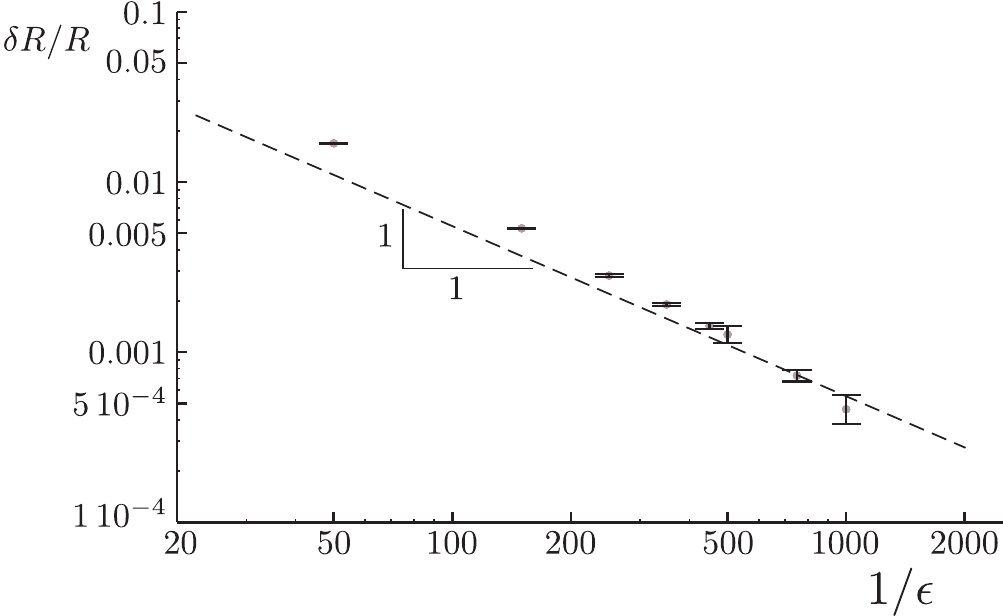}}
    \caption{Analysis of convergence towards the reference solution
    in steady coiling geometry, showing the relative error on the coiling
    radius $\delta R/R$ as a function of mesh size and time step.  The
    same parameters are used as in figure~\ref{fig:coilingValidation},
    but the time step and the discretization length are varied in the
    intervals $.02 \leq \epsilon \leq .2 $ and $.025\leq
    \ell_{\mathrm{c}}\leq .25$, their ratio being kept constant.  Fall
    height is fixed to $H=1.01$.}
    \protect\label{fig:ConvergenceRate}
\end{figure}

Convergence of the solution towards the reference solution of N.~Ribe
is shown in figure~\ref{fig:ConvergenceRate} as a function of the
discretization parameters.
Convergence is challenging in the presence of collisions on the floor,
and we found it necessary to use the refined models for the floor and
container described in sections~\ref{ssec:NozzleImproved}
and~\ref{sssec:rollback}.
In our simulation the coiling radius $R$ is measured after a time that
is long enough for the initial transient to disappear, and then
averaged over several periods.  The simulation is repeated for
different values of the time step $\epsilon$ and discretization length
$\ell_\mathrm{c}$, their ratio being kept constant.  Convergence of
our numerical method is confirmed by the fact that the residual error
goes to zero.  The convergence appears to be linear with the
discretization parameters.

\subsection{Validation of surface tension}
\label{ssec:validationOfSurfaceTension}

\begin{figure}[tbp]
    \centering
    \includegraphics[width=.55\textwidth]{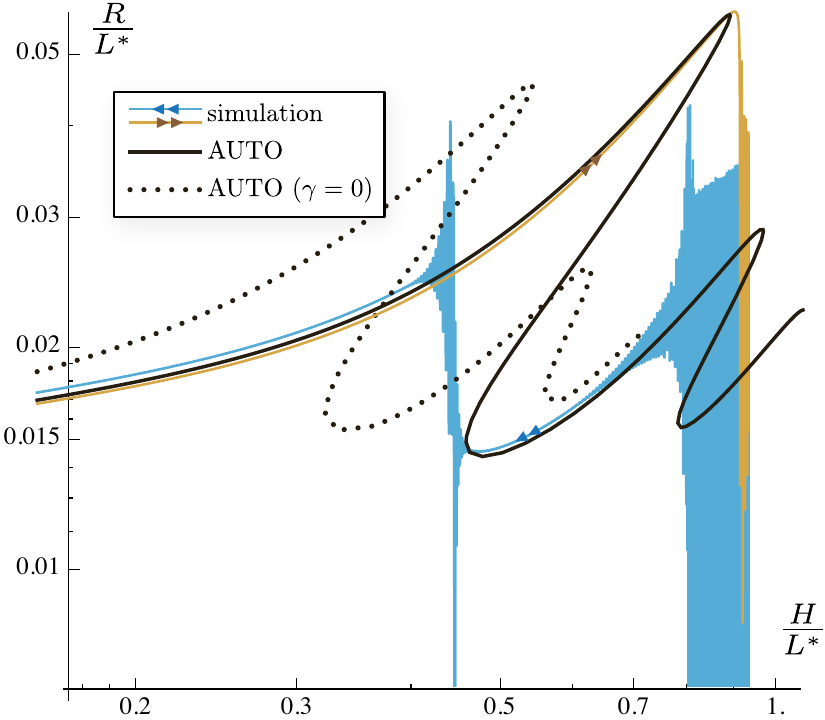}
    \caption{Validation of surface tension using steady coiling
    ($\Pi_{1} = 7000$, $\Pi_{2} = 7$, $\Pi_{3} = 10.3\,10^{-3}$).  The
    same parameters are used as in figure~\ref{fig:coilingValidation},
    except for the non-zero surface tension $\gamma= 10^{-3}$.  A good
    agreement is obtained with the reference curve that takes into
    account surface tension (solid black curve).  For reference, the
    reference solution with zero surface tension is shown (dotted
    curve).}
    \label{fig:valid_surftens_runb}
\end{figure}

With the aim to validate our discrete model for surface tension, we
repeat the validation shown in figure~\ref{fig:coilingValidation}
using the same set of parameters, except for the surface tension
coefficient, now set to $\gamma=10^{-3}$.  The corresponding
dimensionless parameter is $\Pi_{3} = 10.3\,10^{-3}$.  Surface tension
has a marked effect on the coiling radius, as shown by comparison of
the solid and dashed black curves in
figure~\ref{fig:valid_surftens_runb}.
We obtain a good agreement between the simulation and the new
reference curve.  This validates our discrete surface tension model.

\section{Discussion}
\label{sec:final}

\subsection{Transient regimes}

\begin{figure}[tbp]
    \centering
    \includegraphics[width=.99\textwidth]{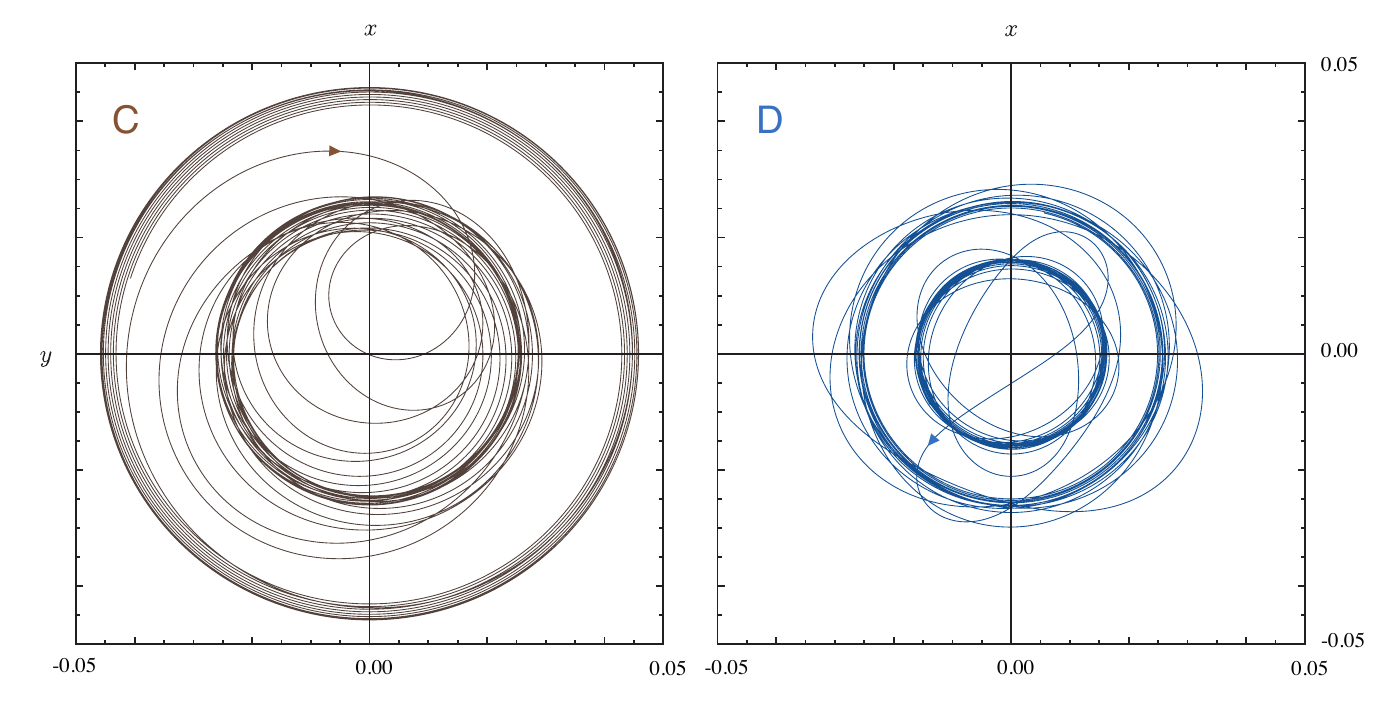}
    \caption{Trace laid out by the thread in the simulation, in the
    transient regimes following the jumps labelled `$\textrm{C}$' and
    `$\textrm{D}$' in figure~\ref{fig:coilingValidation}.  In
    `$\textrm{C}$', the fall height is slowly increased and the
    thread jumps to a solution having a smaller radius; in
    `$\textrm{D}$', the fall height is slowly decreased, and the
    thread jumps to a solution having a larger radius. Same 
    simulation parameters as in figure~\ref{fig:coilingValidation}.}
    \label{fig:transientInCoiling}
\end{figure}

Even though the steady coiling geometry provides a convenient set-up
for validation and verification, our numerical method can solve the
non-steady dynamics of viscous threads.  As an illustration, transient
regimes following jumps from one branch of steady coiling solutions to
the next are shown in figure~\ref{fig:transientInCoiling}.

\subsection{The viscous sewing machine}
\label{ssec:sewingMachineExample}

The fluid-mechanical sewing machine is an extension of the steady
coiling problem to the case of a moving substrate.  To the best of our
knowledge, our numerical method is the first one that can simulate
this non-steady phenomenon.  In fact the possibility to
set up numerical simulations of the experiments described in
references~\cite{Chiu-Webster-Lister-The-fall-of-a-viscous-thread-2006,%
Morris-Dawes-EtAl-Meandering-instability-of-a-viscous-2008} acted as
an incentive for us to develop the present simulation method.

\begin{figure}[tbp]
    \centering
    \includegraphics[width=.99\textwidth]{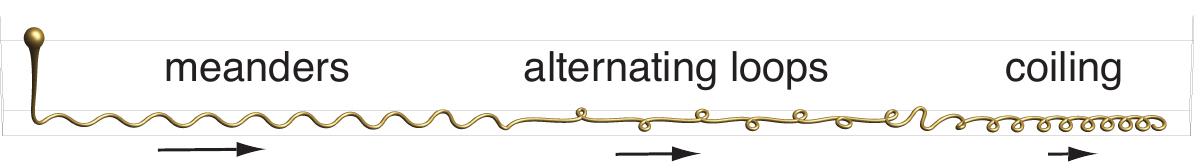}
    \caption{Simulation of the so-called numerical sewing machine.
    The thread is poured from a constant height onto a moving belt.
    As the velocity of the belt is increased, the pattern laid out by
    the thread undergoes a series of bifurcations, similar to those
    that have been reported in the experiments: for this particular
    value of the fall height, translated coiling, alternated loops and
    meanders are successively obtained.  Arrows indicate the velocity
    of the belt as the pattern was formed: the belt moves to the right
    and more ancient patterns, corresponding to a slower belt, are
    located on the right hand side.  Simulation parameters are the
    same as in
    reference~\cite{Morris-Dawes-EtAl-Meandering-instability-of-a-viscous-2008}
    and are provided in main text.}
    \label{fig:sewingMachine}
\end{figure}

A numerical investigation of the fluid-mechanical sewing machine is
the topic of a separate detailed
paper~\cite{Brun-Ribe-EtAl-A-numerical-investigation-of-the-fluid-2012}.
In figure~\ref{fig:sewingMachine}, a simple demonstration run is
presented.  In this simulation, the velocity of the belt is steadily
increased.  The patterns must be read from right to left: this
corresponds to the order in which they were produced, and so to an
increasing belt velocities.  On the right-hand side, for a low
belt velocity, the translated coiling pattern is obtained.  Increasing
the belt velocity, two successive transitions are observed, leading to
the formation of alternated loops first, and to meanders next.  For
even larger belt velocities, the oscillations disappear and the
pattern becomes straight (not shown) as the hanging part of the thread
takes on a catenary-like shape.  The same sequence of patterns has
been observed in the experiments, and is typical of small fall
heights.  A variety of patterns, some of which are quite complex, are
obtained at larger fall heights both in the experiments and in the
simulations, see
reference~\cite{Brun-Ribe-EtAl-A-numerical-investigation-of-the-fluid-2012}
for details.

The parameters used to produce the results of
figure~\ref{fig:sewingMachine} are the same as in the experiments of
Morris~\cite{Morris-Dawes-EtAl-Meandering-instability-of-a-viscous-2008}.
We use a convenient set of units in which the fluid's dynamical
viscosity is $\mu=1$, its mass density is $\rho=1$, and the
acceleration of gravity is $g=1$.  With this natural set of units, the
quantities $L^*$ and $T^*$ defined earlier read $L^* =1$ and $T^* =1$.
The fall height is fixed to $H=0.865$; this corresponds to a physical
fall height of $3.7~\mathrm{cm}$ in the
experiments~\cite{Morris-Dawes-EtAl-Meandering-instability-of-a-viscous-2008}.
The area of the circular outlet at the bottom of the container is set
to $A_{\mathrm{c}}= 0.0275$, and the imposed volume flow rate is
$Q_{\mathrm{c}} = 2.29\,10^{-5}$, and there is no surface tension,
$\gamma = 0$.  The corresponding values of the dimensionless groups
read $\Pi_{1} = 608.5$, $\Pi_{2} = 0.369 $, $\Pi_{3} = 0$.  The
spatial discretization parameter is set to $\ell_{\mathrm{c}}=0.005$,
and the temporal one to $\epsilon=0.05$.  The stretching effect of
gravity is severe, and the mesh is refined adaptively to maintain a
good resolution near the bottom of the thread while keeping a
reasonable number of degrees of freedom, as explained in
section~\ref{ssec:meshRefinement}: a given segment is refined up to
six times in the simulation.  The belt is held still until time
$t_{1}=919$ to the initial transient relax.  From this time on, the
belt velocity is steadily accelerated until the end of the simulation,
occurring at time $t_{\mathrm{f}} = 1650$, its final velocity being
$0.02$.

\subsection{Limitations and perspective}
\label{ssec:limitationsSemiImpliciteVsFullyImplicit}

For the simulation results to be reproducible, we found it critical to
fine-tune the discretization parameters in a steady coiling geometry,
before attempting to simulate non-steady behaviors.  Checking the
convergence of positions, typically by comparing the coiling radius to
the reference solution, appeared not to be sufficient.  We had to
suppress any fast residual oscillation in the computed acceleration,
by using the refined description of obstacles described in
section~\ref{sec:interactionWithEnvironment}, and by carefully
choosing the discretization parameters.

The present paper presents a linear implicit scheme: at each time
step, the velocity is updated using a linear expression for the
viscous forces, which is accurate in the vicinity of the previous
configuration, see
equations~(\ref{eq:discreteDissipationMatrixDecomposition}--\ref{eq:D-MatrixContributions})
and~(\ref{eq:MatrixTimeStep-ForVT+1}); the position is then updated
separately using this velocity, see
equation~(\ref{eq:positionUpdate}).  Compared to an explicit scheme,
this method is more difficult to implement but vastly superior in
terms of stability.  In a previous paper, we have explored a fully
implicit
scheme~\cite{Bergou-Audoly-EtAl-Discrete-Viscous-Threads-2010}, by
retaining the non-linear dependence of the viscous forces on
positions: in that case, each time step requires a non-linear
root-finding.  The benefit of the non-linear implicit approach is that
it preserves conservation laws associated with the symmetries of the
problem, such as the conservation of the angular momentum, when it
applies.  By contrast the method presented here displays the usual
dissipation of angular momentum.  This is a minor drawback as viscous
thread are rarely free-standing in practical applications.  In all the
demonstration examples shown earlier, neither the linear nor the
angular momentum of the thread are actually conserved, because of the
contact forces with the floor.

Simulations of viscous threads and elastic rods can be
unified~\cite{Bergou-Audoly-EtAl-Discrete-Viscous-Threads-2010}: in
that treatment all the code is shared by both rods and threads, except
that in the viscous case the undeformed configuration is updated at
each time step.  This makes the combined approach very appealing.
Nevertheless, a benefit of the specialized implementation presented
here is that it saves the burden of implementing the Hessian matrix
for naturally curved elastic rods.

In future work, it would be interesting to extend the present
numerical model to thin threads governed by more general constitutive
laws, such as visco-elastic
filaments~\cite{Bechtel-Cao-EtAl-Practical-application-of-a-higher-1992,%
Tabuteau-Mora-EtAl-Microscopic-Mechanisms-of-the-Brittleness-2009}
which can exhibit a complex and poorly understood
behavior~\cite{Majmudar-Varagnat-EtAl-Nonlinear-Dynamics-of-Coiling-2010}.
To this end, the discrete geometrical model exposed in the present
paper can be reused and combined with different constitutive laws.  It
would also be interesting to couple our thin thread model with 
existing simulation methods for 3D flows with free boundaries~\cite{%
Bonito-Picasso-EtAl-Numerical-simulation-of-3D-viscoelastic-2006,%
Rafiee-Manzari-EtAl-An-incompressible-SPH-method-for-simulation-2007,%
Oishi-Tome-EtAl-An-implicit-technique-for-solving-2008,%
Batty-Bridson-Accurate-Viscous-Free-2008}, in order to capture the
interaction of the thread with the slowly collapsing pile that forms
where it merges with the bath.

\subsection{Conclusion}

We have presented a numerical method for simulating the dynamics of
thin viscous threads.  In contrast with existing numerical methods, it
captures the combined effects of stretching, bending and twisting
forces, inertia and large rotations.  It is not restricted to steady
flows.  The method has been derived by writing the smooth equations of
motion for thin threads into a Lagrangian form, and using a careful
spatial discretization.  In particular, a discrete notion of twist has
been used, which is based on the geometry of parallel transport.  The
internal stress representing the internal viscous stress has been
derived from variational principles, using a Rayleigh potential.  All
the relevant physical quantities, such as strain rates and internal
stress, have been identified in the discrete setting.  Using a formal
convergence argument, we have shown equivalence of the discrete
equations with the classical smooth formulation.  The method has been
validated against reference solutions available for steady coiling.
Demonstration examples in the non-steady case have been shown.

We would like to thank Neil Ribe for getting us interested into the
fascinating dynamics of thin threads, and for sharing his continuation
data which enabled us to validate our code.

\appendix

\section{Equivalence with the constitutive equations of Ribe}
\label{app:RibeCompatibility}

In a classical paper, Ribe analyzed the helical coiling of viscous
jets falling on a plane~\cite{Ribe-Coiling-of-viscous-jets-2004}.  In
the frame rotating with the jet, the shape of the centerline is
stationary.  The equations for the shape of the jet are expressed as a
set of non-linear ordinary differential equations with boundary
conditions at both ends.  This non-linear boundary value problem was
solved by numerical continuation techniques, using the
AUTO~\cite{Doedel-Champneys-EtAl-AUTO97:-continuation-and-bifurcation-2002}
software.  These solutions, which corresponds to steady
configurations, have been used to validate our dynamical code.

We show below that the constitutive laws derived by Ribe from the Stokes
equations in 3D, are equivalent to our
equations~(\ref{eq:constitutiveEqnsEulerian}). However, the formalism used
by Ribe is different from ours, and we then need to reword his analysis.

Ribe introduces an Eulerian twist-curvature vector
$\underline{\pi}^\mathrm{E}$ which satisfies an equation similar to our
equation~(\ref{eq:DefMaterialCurvature}) for its Lagrangian variant
$\underline{\pi}$, where the derivative is taken with respect to $s$
instead of $S$, that is:
\begin{equation}
    \fp{\underline{d}_{i}'}{s} = \underline{\pi}^\mathrm{E} \times 
    \underline{d}_{i}\textrm{.}
    \label{eq:RibeTwistCurvatureVector}
\end{equation}
The components of $\underline{\pi}^\mathrm{E}$ are denoted
\begin{equation}
    \underline{\pi}^\mathrm{E} = \left[
    \begin{array}{c}
	\kappa_{1} \\ \kappa_{2} \\ \kappa_{3}
    \end{array}
    \right]
    \textrm{,}
    \label{eq:RibePiE}
\end{equation}
where the square brackets indicate a decomposition in the material frame
$(\underline{d}_{1}, \underline{d}_{2}, \underline{d}_{3})$, which is a
`moving' frame, \emph{i.~e.}\ this frame varies with the arc length 
parameter.

As in
equation~(\ref{eq:BinormalLagrangianCurvatureAndMaterialCurvature}),
the Eulerian kinematical twist $\tau^\mathrm{E}$ and binormal
curvature vector $\underline{K}^\mathrm{E}$ are obtained by a
decomposition of the twist-curvature vector
$\underline{\pi}^\mathrm{E}$ into transverse and tangential
components,
\begin{equation}
    \tau^\mathrm{E} = \kappa_{3},\qquad
    \underline{K}^\mathrm{E} = 
    \left[
    \begin{array}{c}
	\kappa_{1} \\ \kappa_{2} \\ 0
    \end{array}
    \right]
    \textrm{.}
    \label{eq:RibeTauK}
\end{equation}
In the frame rotating with the coils, the material rotation
$\underline{\omega}_{\mathrm{R}}$ is written in a form similar to
equation~(\ref{eq:OmegaDecomposition}),
\begin{equation}
    \underline{\omega}_{\mathrm{R}}
    =
    \underline{t}\times \dot{\underline{t}}_{\mathrm{R}}
    +
    \omega_{3}\,\underline{d}_{3}
	\textrm{,}
    \nonumber
\end{equation}
where $\underline{\dot{t}}_{R}$ is the time derivative evaluated in 
the rotating frame. Since the shape of the thread is steady in this 
frame, we have 
\begin{equation}
    \underline{\dot{t}}_{\mathrm{R}} 
    = 
    U\,\fp{\underline{\underline{\omega}_{\mathrm{R}}}}{s}
    =
    U\,\underline{\pi}^\mathrm{E}\times \underline{t}
    \textrm{,}
    \nonumber
\end{equation}
where $U= \underline{u}_{\mathrm{R}}\cdot \underline{t}$ is the axial
velocity of the fluid in the rotating frame. The expression for the
material angular velocity can be obtained by combining the above equations,
\begin{equation}
    \underline{\omega}_{\mathrm{R}} = 
        \left[
    \begin{array}{c}
	U\,\kappa_{1} \\ U\,\kappa_{2} \\ \omega_{3}
    \end{array}
    \right]
	\textrm{.}
    \label{eq:omegaRibe}
\end{equation}
The material angular velocity in the laboratory frame follows from the
composition of velocities, that is $\underline{\omega} =
\underline{\omega}_{\mathrm{R}} + \Omega\,\underline{e}_{z}$ where
$\Omega$ is the frequency of coiling, as well as the relative angular
velocity of the rotating frame with respect to the laboratory.

As a result, the gradient of rotation reads
\begin{equation}
    \fp{\underline{\omega}}{s} =
    \fp{}{s} 
            \left[
    \begin{array}{c}
	U\,\kappa_{1} \\ U\,\kappa_{2} \\ \omega_{3}
    \end{array}
    \right]
    =
    \left[
    \begin{array}{c}
	(U\,\kappa_{1})' \\ (U\,\kappa_{2})' \\ \omega_{3}'
    \end{array}
    \right]
    +
    \underline{\pi}^\mathrm{E} \times 
            \left[
    \begin{array}{c}
	U\,\kappa_{1} \\ U\,\kappa_{2} \\ \omega_{3}
    \end{array}
    \right]
    \textrm{,}
    \nonumber
\end{equation}
where the last term comes from the fact that the material frame is a 
moving frame, see equation~(\ref{eq:RibeTwistCurvatureVector}).

With the help of equations~(\ref{eq:RibePiE})
and~(\ref{eq:omegaRibe}), this vector reads
\begin{equation}
     \fp{\underline{\omega}}{s} =
        \left[
    \begin{array}{c}
	(U\,\kappa_{1})'+\kappa_{2}\,(\omega_{3}-U\,\kappa_{3})
	\\
	(U\,\kappa_{2})'-\kappa_{3}\,(\omega_{3}-U\,\kappa_{3})
	\\
	\omega_{3}'
    \end{array}
    \right]
    \nonumber
\end{equation}
When this vector is inserted into the constitutive
law~(\ref{eq:constitutiveEqnsEulerian-moment}), we obtain for the bending
moment $\underline{m}$ the same expressions as those derived by Ribe, who
used a lubrication-type of approximation in the Stokes equations in 3D.

\bibliographystyle{elsarticle-num}



\end{document}